\pdfoutput=1
\documentclass[a4paper,fleqn,usenatbib]{mnras}
\usepackage{amsmath}
\usepackage{tablefootnote}
\usepackage{natbib}
\usepackage{bm}
\usepackage{verbatim}
\usepackage{graphicx}
\usepackage[T1]{fontenc}
\usepackage{lmodern}
\usepackage{amssymb}
\usepackage{lipsum}
\usepackage{flushend}
\makeatletter
\patchcmd{\NAT@citex}
  {\@citea\NAT@hyper@{%
     \NAT@nmfmt{\NAT@nm}%
     \hyper@natlinkbreak{\NAT@aysep\NAT@spacechar}{\@citeb\@extra@b@citeb}%
     \NAT@date}}
  {\@citea\NAT@nmfmt{\NAT@nm}%
   \NAT@aysep\NAT@spacechar\NAT@hyper@{\NAT@date}}{}{}

\patchcmd{\NAT@citex}
  {\@citea\NAT@hyper@{%
     \NAT@nmfmt{\NAT@nm}%
     \hyper@natlinkbreak{\NAT@spacechar\NAT@@open\if*#1*\else#1\NAT@spacechar\fi}%
       {\@citeb\@extra@b@citeb}%
     \NAT@date}}
  {\@citea\NAT@nmfmt{\NAT@nm}%
   \NAT@spacechar\NAT@@open\if*#1*\else#1\NAT@spacechar\fi\NAT@hyper@{\NAT@date}}
  {}{}

\makeatother

\defcitealias{lazarian2000velocity}{LP00}
\defcitealias{lazarian2004velocity}{LP04}
\defcitealias{lazarian2006studying}{LP06}
\defcitealias{lazarian2012statistical}{LP12}
\defcitealias{0004-637X-818-2-178}{LP16}
\defcitealias{kandel2016extending}{KLP16}
\defcitealias{Kandel21012017}{KLP17}
\defcitealias{goldreich1995toward}{GS95}
\defcitealias{esquivel2011velocity}{EL11}
\defcitealias{burkhart2014measuring}{BX14}
\defcitealias{zhuravleva2012constraints}{ZX12}
\defcitealias{dickman1985largescale}{DK85}
\defcitealias{lazarian2003statistics}{LE03}
\defcitealias{arthur2016turbulence}{AMH16}
\title[Effects of dust on turbulence studies]{Effects of dust absorption on spectroscopic studies of turbulence}

\author[Kandel, Lazarian \& Pogosyan]{
D. Kandel,$^{1,3}$\thanks{E-mail: dkandel@ualberta.ca}
A. Lazarian$^{2}$
and D. Pogosyan$^{1,3}$
\\
$^{1}$Physics Department, University of Alberta, Edmonton, T6G 2E1, Canada\\
$^{2}$Department of Astronomy, University of Wisconsin, 475 North Charter Street, Madison, WI 53706, USA\\
$^{3}$CNRS and UPMC, UMR 7095, Institut d'Astrophysique de Paris, F-75014, Paris, France\\
}

\date{Accepted 2017 May 31. Received 2017 May 30; in original form 2016 November 29}

\pubyear{2017}

\begin{document}
\label{firstpage}
\pagerange{\pageref{firstpage}--\pageref{lastpage}}
\maketitle

\begin{abstract}
We study the effect of dust absorption on the recovery velocity and density spectra as well as on the anisotropies of magnetohydrodynamic turbulence using the Velocity Channel Analysis (VCA), Velocity Coordinate Spectrum (VCS) and Velocity Centroids. The dust limits volume up to an optical depth of unity. We show that in the case of the emissivity proportional to the density of emitters, the effects of random density get suppressed for strong dust absorption intensity variations arise from the velocity fluctuations only. However, for the emissivity proportional to squared density, both density and velocity fluctuations affect the observed intensities. We predict a new asymptotic regime for the spectrum of fluctuations for large scales exceeding the physical depths to unit optical depth. The spectrum gets shallower by unity in this regime. In addition, the dust absorption removes the degeneracy resulted in the universal $K^{-3}$ spectrum of intensity fluctuations of self-absorbing medium reported by Lazarian \& Pogosyan.  We show that the predicted result is consistent with the available HII region emission data. We find that for sub-Alfv\'enic and trans-Alfv\'enic turbulence one can get the information about both the magnetic field direction and the fundamental Alfv\'en, fast and slow modes that constitute MHD turbulence.

\end{abstract}

\section{Introduction}
It is now an accepted fact that the interstellar medium (ISM) is magnetized and turbulent (see \citealt{elmegreen2004interstellar}; \citealt{2007ARA&A..45..565M}; \citealt{mac2004control}). The Reynold's number in the ISM  may exceed $10^8$; in such case flows are expected to be turbulent. Moreover, analysis of scattering and scintillations observations shows turbulent spectra of electron density ranges many orders of magnitude, the so-called Big Power Law in the sky (see \citealt{armstrong1995electron}; \citealt{chepurnov2010extending}). The fluctuations of synchrotron emissivity and polarization also reflect the underlying magnetised turbulence in the ISM (see \citealt{lazarian2012statistical, 0004-637X-818-2-178} and references therein). 

In this paper, we are interested in using non-thermally broadened spectral lines in order to study the underlying turbulence statistics. The corresponding analytical description of statistics of spectral line intensity is provided in a series of our papers (\citealt{lazarian2000velocity, lazarian2004velocity, lazarian2006studying, lazarian2008studying}; \citealt{kandel2016extending, Kandel21012017}). These papers describe various aspects of the statistics of the position-position-velocity (PPV) cubes that can be available from observations\footnote{We mention parenthetically that the PPV cubes do not need to be fully sampled. For instance, the studies of turbulence with emission and absorption lines are possible for just a few lines passing through the volume (see Chepurnov \& Lazarian 2010).}. 

The VCA was introduced in \citet[hereafter \citetalias{lazarian2000velocity}]{lazarian2000velocity} and was extended for self-absorbing media in \citet[hereafter \citetalias{lazarian2004velocity}]{lazarian2004velocity}. The analytical expressions obtained predict how the intensity fluctuations in the velocity slice of PPV depend on the underlying turbulent velocity and density spectra with the change of the velocity slice thickness. This allows one to obtain the velocity and density spectra separately. 

While the VCA technique has been successfully applied to study turbulence spectrum in our galaxy and the Small Magellanic Cloud (SMC) (see \citetalias{lazarian2000velocity}; \citealt{stanimirovic2001velocity}; \citealt{2009SSRv..143..357L} and references therein),  the studies of anisotropies using VCA have so far only been performed with the synthetic data. When combined with the study of synchrotron fluctuations as suggested in \citet{lazarian2012statistical, 0004-637X-818-2-178}, VCA allows studies of the variation of turbulence properties from the cold to hot media. 

Another nobel way of obtaining information on turbulence was identified in \citet[hereafter \citetalias{lazarian2006studying}]{lazarian2006studying}. There it was suggested to study fluctuation of intensity along the velocity axis, and analytical relations between these fluctuations and the underlying turbulence velocity and density spectra were established. The technique was termed Velocity Coordinate Spectrum (VCS), and it has been successfully applied for studies of turbulence in molecular clouds \citep{padoan2009power}, 
for the HI turbulence at high galactic latitudes \citep{chepurnov2010velocity}, and for the HI in the SMC. 

The Velocity Centroid technique is perhaps the oldest technique for studying turbulence (see \citealt{munch1958internal}). Using the synthetic data obtained with simulation of MHD turbulence, \citet{esquivel2005velocity} showed that centroids are able recover spectra of subsonic turbulence as well as to trace anisotropies arising from the anisotropic nature of MHD turbulence. The direction of the anisotropy was shown to trace the direction of magnetic field. \citet{esquivel2011velocity} have shown that the degree of anisotropy is related to the media magnetization that can be characterized by the Alfv\'en Mach number $M_A=V_L/V_A$, where $V_L$ is the turbulence velocity at the injection scale $L$, while $V_A$ is the Alfv\'en velocity. The analytical treatment of velocity centroids that capitalized on the aforementioned PPV studies was presented in \citet[hereafter \citetalias{Kandel21012017}]{Kandel21012017}, where it was shown how to use velocity centroids to study the Alfv\'en, slow and fast MHD modes observationally.

The aforementioned techniques are complementary. The VCA was shown to be efficient for studies of supersonic turbulence, or subsonic turbulence using the heavier species of a turbulent flow. On the contrary, velocity centroids are shown to be much less affected by the thermal Doppler broadening and therefore applicable to the studies of subsonic turbulence using the main ingredients of the flow, e.g. atomic hydrogen. However, studies \citep{esquivel2005velocity} suggest that centroids are not reliable for the studies of supersonic turbulence.

All the studies that were mentioned above do not account for the absorption by dust, however. This is acceptable for studies of turbulence using radio lines, e.g. 21 cm HI line, CO lines. However, dust absorption affects optical and UV lines. In fact, this was a problem that the earlier researchers were aware of (see \citealt{munch1958internal}). This paper extends the description of the PPV statistics by accounting for the effect of the dust absorption for the PPV studies and shows how the interpretation of results using VCA, the VCS and velocity centroid technique change   in the presence of dust absorption.

In what follows, we discuss the radiative transport in the presence of dust in Sec. \ref{srad}, and introduce the power law statistics of density and velocity in Sec \ref{svelden}. The effects of dust on the recovery of the velocity and density spectra using the VCA, VCS and velocity centroids are studied in Sec. \ref{vcadust}, \ref{svcs} and \ref{centroidsdust}, respectively. The studies of anisotropies using both the VCA and the centroids are described in Sec. \ref{sanist}. In Sec. \ref{sec:collext}, applicability of VCA, VCS and centroids for collisionally excited lines is discussed.  Comparison of theory with observational data is carried out in Sec. \ref{obs}, while the discussion of the results and the summary are presented in Sec. \ref{sdiscuss} and Sec. \ref{ssumm}, respectively.

\section{Radiative transport in the presence of dust}\label{srad}
In the presence of dust, the standard equation of radiative transfer is  \citep{spitzer2008physical} 
\begin{equation}\label{radtrans}
\mathrm{d}I_\nu=-g_\nu I_\nu\mathrm{d}z+j_\nu\mathrm{d}z-\sigma_\nu I_\nu\mathrm{d}z,
\end{equation}
where the term $-\sigma_\nu I_\nu\mathrm{d}z$ denotes extinction due to dust. In the case of self-absorbing emission in spectral lines that is proportional to first power of density
\begin{align}
g_\nu &=\alpha\rho_g(z)\Phi_\nu(z)~,\\
j_\nu &=\epsilon\rho_g(z)\Phi_\nu(z)~,\\
\sigma_\nu &=\kappa'\rho_d(z)~,
\end{align}
where $\rho_g$ is the density of emitting gas, $\rho_d$ is the density of dust, $\Phi_\nu(z)$ is the frequency distribution of emitters along the LOS, and $\kappa'$ is the dust extinction coefficient, which is in general frequency dependent. For the purpose of our analysis, we assume it to be constant within the line width of a Doppler broadened emission line. 
The solution to the Eq. \eqref{radtrans} takes the following form 
\begin{align}\label{basic}
I_\nu=&\epsilon\int_0^{S}\mathop{\mathrm{d}z}\rho_g(\bm{x})\Phi_\nu(\bm{x})\nonumber\\
&\times\exp\left[-\int_0^z\mathop{\mathrm{d}z'}\left(\alpha\rho_g(\bm{X}, z')\Phi_\nu(\bm{X}, z')+\kappa'\rho_d(\bm{X}, z')\right)\right]~.
\end{align}
In the case when dust absorption is negligible, i.e. when $\kappa'\rightarrow 0$, the spectral intensity reduces to 
\begin{equation}
I_{v_1}(\bm{X})=\frac{\epsilon}{\alpha}\left[1-\mathrm{e}^{-\alpha\rho_s(\bm{X},v_1)}\right]~,
\end{equation}
where $\rho_s$ is the density of emitters in PPV space, and is given by
\begin{equation}
\rho_s(\bm{X},v_1)=\int_0^S\mathop{\mathrm{d}z}\rho(\bm{x})\Phi_v(\bm{x})~.
\end{equation}
In writing the above equations, we have used that fact that the Doppler-shifted frequency information can be used to deduct LOS velocity of the emitters, and hence all frequency dependence  in radiative transfer equation can be equivalently designated as  velocity dependence $\nu\rightarrow v$. 

The study of turbulence using emission from self-absorbing medium was studied in detail in \citet{lazarian2004velocity} and \citetalias{Kandel21012017}. In this paper, we focus on the studies of turbulence using emission from a dusty medium.

In the case when self-absorption is negligible, i.e. $\alpha\rightarrow 0$, Eq. \eqref{basic} reduces to
\begin{equation}\label{eq:dustgeneq}
I_\nu(\bm{X})=\epsilon\int_0^{S}\mathop{\mathrm{d}z}\rho_g(\bm{x})\Phi_v(\bm{x})\exp\left[-\int_0^z\mathrm{d}z\,\kappa'\rho_d(\bm{X}, z')\right]~,
\end{equation}
where $\Phi_v(\bm{x})$ is given
\begin{equation}
\Phi_v(\bm{x})=\frac{1}{\sqrt{2\pi\beta}}\exp\left[-\frac{(v-u(\bm{x})-v_c(\bm{x}))^2}{2\beta}\right]~,
\end{equation} 
where $u(\bm{x})$ is the LOS component of turbulent velocity and $v_c(\bm{x})$ is the LOS component of coherent velocity that is not part of turbulent cascade. In this paper, we assume that $v_c(\bm{x})$ is small, else it has to be modelled independently. 
Clearly, the main effect of dust is it cuts out signals from optical depths larger than 1, thus attenuating intensity signals. In the subsequent analysis, unless explicitly stated, we will consider the case when self-absorption is negligible. Before starting further derivations, we assume that the dust density is directly proportional to the gas density\footnote{ This assumption has been shown to hold good in various papers \citep{lada1994dust}.}, so that
\begin{equation}
\kappa'\rho_d(\bm{x})=\kappa\rho_g(\bm{x})\equiv \kappa\rho(\bm{x})~,
\end{equation}
where $\kappa$ is the redefined dust absorption coefficient, which takes into account the proportionality factor between density of gas and density of dust.

\section{Velocity and density fields}\label{svelden}
In the first part of this paper, we are interested in the spectra of density and velocity fields, which are fluctuating in a turbulent media. Two important statistical measure of fluctuations $\delta\mathcal{R}(\bm{x})$ of a random quantity $\mathcal{R}(\bm{x})$  are correlation function 
\begin{equation}
\xi_{\mathcal{R}}(\bm{x}_1,\bm{x}_2)=\langle\mathcal{R}(\bm{x}_1)\mathcal{R}(\bm{x}_2)\rangle~,
\end{equation}
and structure function 
\begin{equation}
D_{\mathcal{R}}(\bm{x}_1,\bm{x}_2)=\langle\left(\mathcal{R}(\bm{x}_1)-\mathcal{R}(\bm{x}_2)\right)^2\rangle~.
\end{equation}
We assume turbulence to be homogeneous in which case, the correlation function and structure function can be represented purely as a function of spatial separation, so that $\xi_{\mathcal{R}}(\bm{x}_1,\bm{x}_2)=\xi_{\mathcal{R}}(\bm{r})$ and $ D_{\mathcal{R}}(\bm{x}_1,\bm{x}_2)=D_{\mathcal{R}}(\bm{r})$, where $\bm{r}=\bm{x}_2-\bm{x}_1$.

Scaling of power distribution of random fields are of two types: steep and shallow. For a steep field, power of fluctuations is dominated by moderate to large scale fluctuations, while for shallow field, power is dominated by small scale fluctuations. Depending upon whether fluctuations are steep or shallow, it is appropriate to either use structure function or use correlation function. A major difference between correlation function and structure function is that, while the structure function at scale $r$ is determined by the integrated power of fluctuations over scales smaller than $r$, the correlation function is determined by the integral of the power over scales larger than $r$. Therefore, the correlation function is more appropriate for shallow spectra, while structure function is more appropriate for steep spectra.

Observations show that velocity spectra in the ISM is steep, while density can be either steep or shallow. Leaving discussion of angular dependence of velocity and density statistics to Sec. \ref{sanist}, we take LOS velocity structure function as
\begin{equation}\label{velocitymodel}
D_z(\bm{r})=D_z(S)\left(\frac{r}{S}\right)^\nu~,\qquad r<S
\end{equation}
where $\nu>0$. In the above equation, $S$ is the injection scale of turbulence, and for isolated clouds, we assume that $S$ is comparable to the size of the cloud.  Modelling of the structure function at $r>S$ requires modelling of the injection process. The correlation length of velocity field is comparable to the energy injection scale. 

Similarly, the structure function of steep density field is taken to be
\begin{equation}\label{steepdend}
D_\rho(\bm{r})\approx 2\sigma_\rho^2\left(\frac{r_c}{r}\right)^{-\nu_\rho}~,\quad \nu_\rho<0~, \quad r<r_c~,
\end{equation}
while for shallow density spectrum, the density correlation is modelled as
\begin{equation}\label{shallowdenx}
\xi_\rho(\bm{r})\approx \rho_0^2+\sigma_\rho^2\left(\frac{r_c}{r}\right)^{\nu_\rho}~,\quad \nu_\rho>0~, \quad r>r_c~,
\end{equation}
where $\rho_0$ is the mean density, $\sigma_\rho$ the density dispersion, and $r_c$ the correlation length of the density field. In the case steep density field, $r_c$ is comparable to the size of injection scale, i.e. $r_c \sim S$, and the power law scaling regime lies at $r < r_c$, while for shallow density, $r_c$ is small, and the 
power law scaling is exhibited at $r > r_c$. In the subsequent parts of this paper, we use Eq. \eqref{steepdend} when $\nu_\rho<0$ and Eq. \eqref{shallowdenx} when $\nu_\rho>0$.

Here, as in our previous papers (see \citetalias{lazarian2000velocity}),  velocity fluctuations are assumed to be Gaussian. In \citetalias{lazarian2000velocity}, no assumptions on density field were made in the formulation of VCA for optically thin emission lines. However, as we shall see in the next section, one needs an explicit model of density field in order to carry out further analysis in the presence of dust absorption. In the case when density dispersion is smaller than the mean density (which is so for the steep density), it is safe to assume density fluctuations as Gaussian random field. However, if the density dispersion is larger than the mean density (which can be so for the  shallow spectra), Gaussian distribution is not a good model for density field, as it violates an important constraint that $\rho>0$ (see \citealt{fry1986nonlinear}). Thus, for a shallow spectra, we take density fluctuations to obey log-normal distribution. This distribution is physically reasonable as it always satisfies the constraint $\rho>0$, and is computationally convenient \citep{barrow1983limits}. This assumption also fits with recent observations, which suggest that density probability distribution function of the diffuse gas in the ISM is lognormal (see \citealt{berkhuijsen2008density}).

\section{Effects of dust on VCA}\label{vcadust}
Velocity channel analysis  (see \citetalias{lazarian2000velocity})
concerns with studies of the correlations of line intensities measured
in velocity channels 
\begin{equation}\label{eq:windowinten}
I_{v_i}(\bm{X_1};\Delta V_i) = \int_{-\infty}^\infty d v_1 \; 
I_{v_1}({\bm X}) \; W(v_1-v_i;\Delta V_i)
\end{equation}
Here $v_i$ is the central velocity (frequency) of the channel ``i'', and
$ W(v_1-v_i;\Delta V_i)$ is the channel profile of the width $\Delta V_i$ 
(such as $ W(v_1-v_i;\Delta V_i) \sim 0$ when  $ |v_1-v_i| > \Delta V_i/2$).
VCA's focus is on intensity
correlations between pairs of LOS as function of LOS 
separation and, importantly, as a function of velocity channel width.
Relevant statistical measures are the correlation function
\begin{equation}
\xi(\bm{R},v_i,\Delta V) = 
\left\langle I_{v_i}(\bm{X}_1;\Delta V)
I_{v_i}(\bm{X}_2, \Delta V) \right\rangle
\end{equation}
and the structure function
\begin{equation}\label{vstrucd}
\mathcal{D}(\bm{R},v_i,\Delta V) = 
\left\langle \left (I_{v_i}(\bm{X}_1,\Delta V)
- I_{v_i}(\bm{X}_2,\Delta V) \right)^2 \right\rangle~.
\end{equation}

There are two important regimes: {\it thin} velocity slice regime,
where information
about velocity spectra can be obtained, and {\it thick} velocity slice regime,
where density spectra can be obtained. Which regime is realized depends on
the comparison between the characteristic difference of turbulence velocities 
between two LOS and the velocity channel width. 
The slice is {\it thin} if $\Delta V < D_z^{1/2}(R)$ 
(\citetalias{lazarian2000velocity}). Thus, to measure velocity effects 
in VCA, one either needs a narrow velocity channel or the measurements of correlations at
sufficiently separated LOS. 
In practice, the minimal width of channels is determined by the spectral
resolution of the instrument, while wider channels are obtained synthetically
by coadding signal of individual high resolution channels
(assuming for simplicity that channels are co-adjacent). This gives VCA
the ability to separate velocity and density effects, if thin channel regime is
available, by varying the slice thickness in the analysis. Here we note that
thermal broadening interferes with the ability to carry out thin slice measurements by
effectively increasing the channel thickness as 
$\Delta V^2 \to \Delta V^2 + \beta_{\text{T}}$. Thus, VCA is most effective for 
supersonic turbulence or when used with lines from massive species with 
slow thermal velocities.

For asymptotic analysis, {\it thick} slice regime corresponds formally to
the channel profile $W \sim 1$, i.e. integration over the whole line.
The explicit expression for correlation function is then 
\begin{equation}
\xi(\bm{R}) \sim \int_{-\infty}^\infty \!\! dv_1
\int_{-\infty}^\infty \!\! dv_2 
\left\langle I_{v_1}(\bm{X}_1) I_{v_2}(\bm{X}_2) \right\rangle
\end{equation}
\textit{Thin} slice asymptotics is reproduced for $W(v_1 - v_i) \approx 
\Delta V_i \delta_D(v_1-v_i)$ and
\begin{equation}
\xi(\bm{R}) \sim \Delta V^2
\left\langle I(\bm{X}_1) I(\bm{X}_2) \right\rangle
\end{equation}

Previous papers on VCA did not consider interstellar dust extinction.
In this section, we consider the effect of dust on VCA statistics, and try to
answer the following question: can velocity and density spectra be recovered
using VCA even in the presence of significant dust absorption? In
Sec.~\ref{sec:thickslice} and
\ref{sec:thin}, we will consider optically thin emission lines, and in Sec.~\ref{sec:selfab} we will
discuss the effect of self-absorption. The full expression for intensity
correlation in the presence of dust absorption is presented in Appendix \ref{fullexp}.

We start with discussing main qualitative effects. The optical depth due 
to dust absorption 
\begin{equation}
\tau(\bm{X},z) \equiv \kappa \int_0^z \rho(\bm{X},z^\prime) dz^\prime ~,
\end{equation}
and the LOS distance $\Delta(\bm{X})$ at which optical depth becomes unity,
\begin{equation}
\kappa\int_0^{\Delta(\bm{X})}\mathrm{d}z\rho(\bm{X},z)=1~,
\quad \tau(\bm{X},\Delta(\bm{X})) \equiv 1, 
\end{equation}
play the central role in determining the relevant statistical properties of intensity correlation. In the absence of the turbulent velocity effects, or
more specifically, when spectral line is integrated over 
$\Phi_v$ distribution,
Eq.~\eqref{eq:dustgeneq} becomes
\begin{align}\label{eq:Inovel}
I(\bm{X})\propto 1-\mathrm{e}^{-\tau(\bm{X},S)}.
\end{align}
This is the case applicable for \textit{thick} velocity slices, 
further detailed in Sec.~\ref{sec:thickslice}.
Main observation here is that when optical depth through the volume
exceeds unity along all LOS, $\tau(\bm{X},S) > 1$, the intensity
fluctuations due to density fluctuations are exponentially suppressed.
This comes from compensation between density variations along the line of 
sight and corresponding variations of the physical LOS distance 
$\Delta(\bm{X})$ at which $\tau$ 
reaches unity.   This can be seen by approximating exponential cutoff in LOS integral by a
step $\Theta$ function
\begin{equation}
e^{-\tau(\bm{X},z)} \approx \Theta\left(\Delta(\bm{X})-z \right), 
\end{equation}
after which Eq.~\eqref{eq:dustgeneq} gives constant intensity
\begin{equation}\label{eq:Idensat}
I \approx \epsilon\int_0^\Delta\mathrm{d}z\rho(\bm{x}) = 
\frac{\epsilon\tau(\bm{X},\Delta(\bm{X}))}{\kappa} \equiv \frac{\epsilon}{\kappa}
\end{equation}
This suppression of density fluctuation is not present when densities of
emitters and dust do not follow each other (in particular, when distribution
of dust is taken to be uniform), or when intensity is 
proportional to the square of the emitters density.
It is also absent when absorption is weak, namely optical
depth through the cloud is less than unity, $\tau(\bm{X},S) < 1$.

Let us see what happens when velocity effects are present, as, for instance,
when intensity is measured in \textit{thin} velocity slices. Applying 
the step approximation to the full Eq.~\eqref{eq:dustgeneq} we have
\begin{equation}
I_v(\bm{X}) \approx \int_0^{\Delta(\bm{X})} \!\! dz \;
\rho(\bm{X},z) \Phi_v(\bm{X},z)
\end{equation}
which highlights two main effects of the strong absorption.  
Strong absorption means small $\Delta(\bm{X})$ which suggests an estimate
\begin{equation}
\label{eq:Icrude}
I_v(\bm{X}) \sim \Delta(\bm{X})
\rho(\bm{X},0) \Phi_v(\bm{X},0) 
\approx \frac{1}{\kappa} \Phi_v(\bm{X},0)~.
\end{equation}
Thus, firstly, 
the signal is reduced to the nearest 2D surface of the cloud rather than
being integrated along LOS through 3D volume, and, secondly, the
density fluctuations are compensated away, leaving only the velocity
induced intensity fluctuations. However, 
correlation studies introduce additional scales besides $S$, that can be much 
shorter, namely the density correlation length $r_c$ 
and the separation between pairs of LOS, $R$ which can be taken
arbitrarily small (limited by the instrument resultion). 
Thus we must consider whether $\Delta/R$ and
$\Delta/r_c$ are small or large before concluding whether
$\Delta(\bm{X})$ is sufficiently small or not.

The estimate Eq.~\eqref{eq:Icrude} seemingly reguires density to be near
constant
through $[0,\Delta(\bm{X})]$ LOS interval, but to what extend it is 
essential to the described picture?
We can develop a more precise estimate rewriting Eq.~\eqref{eq:dustgeneq}
by integrating by parts 
\begin{align}\label{eq:thinintpart}
I_v(\bm{X})=\frac{1}{\kappa}\bigg(\left[\Phi_v(\bm{X},0)
-\Phi_v(\bm{X}, S)\mathrm{e}^{-\tau(\bm{X},S)}\right]\nonumber\\
+\int_0^S\mathrm{d}z\,\frac{\mathrm{d}\Phi_v(\bm{x})}{\mathrm{d}z}\mathrm{e}^{-\kappa\int_0^z\mathrm{d}z'\,\rho(z')}\bigg)~.
\end{align}
to eliminate density pre-factor before applying $\Theta$-function approximation
for the exponential term. This gives
\begin{align}\label{eq:thinintpartmain1}
I_v(\bm{X}) \approx \frac{1}{\kappa}\left[\Phi_v(\bm{X},\Delta(\bm{X}))-\Phi_v(\bm{X}, S)\mathrm{e}^{-\tau(\bm{X,S})}\right]~.
\end{align} 
This expression is valid only for strong absorption, $\tau(\bm{X},S) > 1$,
so the exponential term can be neglected. We leave this exponentially suppressed term here just to show
that Eq.~\eqref{eq:thinintpartmain1} is consistent with Eq.~\eqref{eq:Inovel}
when the velocities in the spectral line are integrated over. Otherwise, 
we may just write
\begin{align}\label{eq:thinintpartmain}
I_v(\bm{X}) \approx \frac{1}{\kappa} \Phi_v(\bm{X},\Delta(\bm{X})),
\quad \tau(\bm{X},S) > 1
\end{align} 
This expression explicitly neglects the width
of the region to $\Delta(\bm{X})$,  but does not, however, assume 
anything about the density uniformity. Thus, the conclusion about 2D
surface nature of the signal at high absorption holds independently on
density properties.  Density fluctuations are introduced only
through fluctuations of the LOS positions of this surface, and
one can expect that if $\Delta(\bm X)$ does not fluctuate much,
contribution of density fluctuations is small. This is the manifestation of
density fluctuation suppression discussed above.

In the following sub-sections, we discuss with more rigour two important VCA regimes: thin velocity slice and thick velocity slice regimes.

\subsection{Thick velocity slice}\label{sec:thickslice}
We are first interested in a simpler case of thick velocity channel. In this case, all emission along the LOS is collected independent of the velocity of emitters, which can be seen using Eqs. \eqref{eq:dustgeneq} and \eqref{eq:windowinten} with $W(v)=1$, which gives
\begin{equation}\label{simradtrans}
I(\bm{X})=\frac{\epsilon}{\kappa}\left[1-\mathrm{e}^{-\kappa\mathcal{N}(\bm{X})}\right]~,
\end{equation}
where $\mathcal{N}(\bm{X})$ is the column-density, and is given by 
\begin{equation}
\mathcal{N}(\bm{X})=\int_0^S\mathrm{d}z\rho(\bm{x})~, 
\quad \tau(\bm{X},S)=\kappa \mathcal{N}(\bm{X})~.
\end{equation}

Eq. \eqref{simradtrans} shows that dust absorption leads to exponential suppression of density contribution. Thus, making measurements in thick slices in the presence of dust is informative only when absorption is not too large (more accurate criteria for extent absorption is discussed further on). Due to the appearance of density dependence in exponential fashion, in the presence of dust one has to carry out further analysis with an explicit model of density field, unlike in \citetalias{lazarian2000velocity}, where no assumption about the model of density field had to be made. 

Statistics of the column density $\mathcal{N}(\mathbf{X})$ is discussed in the
Appendix~\ref{appcolumn}. For density field with steep spectrum of
fluctuations, the most relevant case is when density dispersion is less than
the mean density\footnote{For steep density, perturbations of largest amplitude
have wavelength comparable to the size of the cloud. If we determine mean
density over the size of the cloud, then small scale perturbations naturally
have amplitude less than the mean.}.
Therefore, we can model density fluctuations $\delta\rho$ as a Gaussian random
field, with $\langle\delta\rho\rangle=0$. This translates to the Gaussian
behaviour of the column density as well.

On the other hand, for shallow density field, a more appropriate and general
model for the density field is log-normal field.  Numerical calculations
present in Appendix \ref{appcolumn} show that column density $\mathcal{N}$
for a shallow density still tends to be Gaussian if the integration length $S$
along the LOS is much larger than the correlation length $r_c$ of density field
(which can be easily satisfied for shallow spectra, as $r_c$ in this case is
very small). This is because, outside of the correlation length, the
fluctuations of density are essentially independent (as correlation decays
outside of correlation length), and the sum of uncorrelated random fluctuations
should tend to Gaussian quantity according to the Central limit theorem.
The near Gaussian nature of the column-density is clearly illustrated in
Fig.~\ref{pdf}.
Appendix~\ref{appcolumn} also defines and discusses the column density
variance $\sigma_\mathcal{N}^2$ and correlation $\xi_\mathcal{N}(\mathbf{R})$
and structure $d_\mathcal{N}(\mathbf{R})$ functions and their ranges of utility.

With this, we are now ready to proceed further on. In configuration space, to obtain correlation function of fluctuations of intensity, one needs to have knowledge about the mean intensity.  From Eq. \eqref{simradtrans}, we have the mean intensity as
\begin{equation}
\bar{I}=\frac{\epsilon}{\kappa}\left[1-\left\langle\mathrm{e}^{-\kappa\mathcal{N}(\bm{X})}\right\rangle\right]~,
\end{equation}
and correlation of fluctuations of intensity $I(\mathbf{X}) - \bar I$ as
\begin{align}\label{flucint}
\xi(\bm{R})=\frac{\epsilon^2}{\kappa^2}\bigg\langle\left(\left\langle\mathrm{e}^{-\kappa\mathcal{N}(\bm{X}_1)}\right\rangle-\mathrm{e}^{-\kappa\mathcal{N}(\bm{X}_1)}\right)\nonumber\\
\left(\left\langle\mathrm{e}^{-\kappa\mathcal{N}(\bm{X}_2)}\right\rangle-\mathrm{e}^{-\kappa\mathcal{N}(\bm{X}_2)}\right)\bigg\rangle.
\end{align} 
With the assumption of homogeneity, we have
$\left\langle\mathrm{e}^{-\kappa\mathcal{N}(\bm{X}_1)}\right\rangle=
\left\langle\mathrm{e}^{-\kappa\mathcal{N}(\bm{X}_2)}\right\rangle$,
which allows us to simplify Eq.~\eqref{flucint} to
\begin{equation}\label{flucint1}
\xi(\bm{R})=\frac{\epsilon^2}{\kappa^2}\left[\left\langle\mathrm{e}^{-\kappa\left(\mathcal{N}(\bm{X}_1)+\mathcal{N}(\bm{X}_2)\right)}\right\rangle-\left\langle\mathrm{e}^{-\kappa\mathcal{N}(\bm{X}_1)}\right\rangle^2\right]~.
\end{equation}
For a Gaussian $\mathcal{N}$, the first term in the Eq.~\eqref{flucint1} can be written as
\begin{equation}\label{gaussshallow}
\left\langle\mathrm{e}^{-\kappa\left(\mathcal{N}(\bm{X}_1)+\mathcal{N}(\bm{X}_2)\right)}\right\rangle=\left\langle\mathrm{e}^{-\kappa\mathcal{N}(\bm{X}_1)}\right\rangle^2\mathrm{e}^{\kappa^2\xi_{\mathcal{N}}(R)}~,
\end{equation}
where $\xi_{\mathcal{N}}(\bm{R})=\left\langle\left(\mathcal{N}(\bm{X}_1)
-\bar{\mathcal{N}}\right)\left(\mathcal{N}(\bm{X}_2)
-\bar{\mathcal{N}}\right)\right\rangle$.
Thus, using Eqs. \eqref{flucint1} and \eqref{gaussshallow} we finally have
\begin{align}\label{shalcor}
\xi(R)=\frac{\epsilon^2}{\kappa^2}\left\langle\mathrm{e}^{-\kappa\mathcal{N}(\bm{X}_1)}\right\rangle^2\left(\mathrm{e}^{\kappa^2\xi_{\mathcal{N}}(R)}-1\right)~.
\end{align}
For $\kappa^2\xi_{\mathcal{N}}(R)<1$, the above Eq. \eqref{shalcor} reduces to
\begin{equation}\label{thickshallowcorr}
\xi(R)=\epsilon^2\left\langle\mathrm{e}^{-\kappa\mathcal{N}(\bm{X}_1)}\right\rangle^2\xi_{\mathcal{N}}(R)~.
\end{equation}
Note that $\kappa^2 \xi_\mathcal{N}(0) = \langle \delta\tau^2 \rangle $, the variance of the optical depth through the volume. 
Thus, the expression Eq.~\eqref{thickshallowcorr} holds everywhere, and one can recover density statistics, if absorption due to dust is sufficiently uniform, $ \langle \delta\tau^2\rangle < 1$. However, it also holds for sufficiently large $R$ even if dust absorption is less uniform.

As discussed in Appendix~\ref{appcolumn}, correlation function approach
is useful if density spectrum is shallow with $\nu_\rho > 1$. In this case
Eq.~\eqref{thickshallowcorr} leads to 
a very interesting conclusion that one can recover density scaling as in
usual thick slice VCA even
in the presence of dust if one studies correlation of intensity at large lags
$R$. The useful scale range is given by the condition $\kappa^2\xi_{\mathcal{N}}(R)<1$, which for $\nu_\rho > 1$
sets a correlation length for the observed intensities at 
\begin{equation}
\label{eq:rc*1}
r_c^*\approx r_c\left\langle\delta\tau^2\right\rangle^{\frac{1}{\nu_\rho-1}}~,\quad \nu_\rho>1~.
\end{equation}
Thus, for $\nu_\rho > 1$ one can obtain density scaling measuring the
correlation function at lags
$R>\mathrm{max}(r_c^*,r_c)$. This condition is more and more limiting
as fluctuations of the optical depth grow beyond
$\langle\delta \tau^2\rangle=1$,
but may still be satisfied for shallow
density which has small correlation length $r_c$.

For $\nu_\rho < 1$ one should focus on measuring the structure
function of intensity.  Using Eq. \eqref{shalcor}, it can be written as 
\begin{align}
\mathcal{D}(\bm{R})&=2\left(\xi(0)-\xi(\bm{R})\right)\nonumber\\
&=\frac{2\epsilon^2}{\kappa^2}\left\langle\mathrm{e}^{-\kappa\mathcal{N}(\bm{X}_1)}\right\rangle^2\left(\mathrm{e}^{\kappa^2\xi_{\mathcal{N}}(0)}-\mathrm{e}^{\kappa^2\xi_{\mathcal{N}}(R)}\right)~,
\end{align}
which can be further simplified to 
\begin{equation}\label{eq:dustthickmain}
\mathcal{D}(\bm{R})=\frac{2\epsilon^2}{\kappa^2}\left\langle\mathrm{e}^{-\kappa\mathcal{N}(\bm{X}_1)}\right\rangle^2
\mathrm{e}^{\kappa^2\sigma_{\mathcal{N}}^2}
\left(1-\mathrm{e}^{-\frac{\kappa^2}{2}d_{\mathcal{N}}(R)}\right)~,
\end{equation}
At sufficiently small lags $R$, the usual result of the structure function of optically thin emission lines is recovered, i.e.
\begin{equation}
\mathcal{D}(\bm{R})\propto d_{\mathcal{N}}(\bm{R})~,
\end{equation} 
while at larger lags $R$, the structure function is saturated.
The saturation scale, which is, in effect, the correlation length of intensity,
is estimated by setting $\frac{\kappa^2}{2}d_{\mathcal{N}}(r_c^*)\sim 1$,
which now gives
\begin{equation}
\label{eq:rc*2}
r_c^*\sim r_c\left(\frac{1}{\langle\delta\tau^2\rangle}\right)^{\frac{1}{1-\nu_\rho}}~,\quad \nu_\rho<1~.
\end{equation}
where we note that Eq.~\eqref{eq:rc*1} and Eq.~\eqref{eq:rc*2} are equivalent.
Thus, by measuring structure function of density in thick 
slices, one can recover density spectra at scales 
$R<\mathrm{min}(r_c^*,r_c)$.
For steep spectra with $\nu_\rho < 0$ and $r_c \sim S$
the critical scale is less than the size of the cloud, i.e. $r_c^*<S$
only when dust optical fluctuations are exceeding unity,
while for weak absorption fluctuations it
is comparable to the size of the cloud, i.e. $r_c^*\sim S$.
Therefore, in the case when dust inhomogeneity is strong, the
recovery of steep density statistics in thick velocity slices may be limited
by the spatial resolution of telescope.
The situation is the least advantageous for $0 < \nu_\rho < 1$ range of shallow
spectra where we should use the structure function measurements but need to
resolve $r_c$ scale which now can be small.

We stress that the amplitude of correlations in both
Eq.~\eqref{thickshallowcorr}  and Eq.~\eqref{eq:dustthickmain}
will be exponentially suppressed if optical depth of the cloud is much larger than unity, and is challenging to measure if dust absorption is strong and instrumental noise is not properly factored out.

\subsection{Thin velocity slice}\label{sec:thin}
To facilitate comparison between the case of presence and absence of dust
absorption, we first review the main results
of VCA in thin slices in the absence of dust.

In the absence of dust, Eq. \eqref{eq:dustgeneq} is reduced to 
\begin{equation}
I_v(\bm{X})=\epsilon\rho_s(\bm{X},v)=\epsilon\int_0^S\mathrm{d}z\,\rho(\bm{x})\Phi_v(\bm{x})~,
\end{equation}
i.e. the spectral intensity of optically thin emission lines is  directly given by PPV density of emitters. In this case, the correlation of spectral intensity is given by
\begin{align}\label{intencorr}
\xi(\bm{R}, v_1, v_2)=\epsilon^2\int_0^S\mathrm{d}z_1\,\int_0^S\mathrm{d}z_2\,\langle\rho(\bm{x}_1)\rho(\bm{x}_2)\rangle\nonumber\\
\langle\Phi_{v_1}(\bm{x}_1)\Phi_{v_2}(\bm{x}_2)\rangle~,
\end{align}
where we have assumed uncorrelated density and velocity field. Assuming velocity fluctuations to be Gaussian random quantities, Eq. \eqref{intencorr} can be used to finally obtain correlation of thin slice (see \citetalias{lazarian2004velocity})
\begin{equation}\label{corrthinin}
\xi(\bm{R})\propto S \int_0^S\left(1-\frac{z}{2S}\right)\frac{\bar{\rho}^2+\xi_{\rho}(\bm{R},z)}{\sqrt{D_z(\bm{R},z)}}~,
\end{equation}
thus the structure function is 
\begin{equation}\label{strthinw}
\mathcal{D}(\bm{R})\propto S \int_0^S\left(\frac{\bar{\rho}^2+\xi_{\rho}(0,z)}{\sqrt{D_z(0,z)}}-\frac{\bar{\rho}^2+\xi_{\rho}(\bm{R},z)}{\sqrt{D_z(\bm{R},z)}}\right)
\end{equation}
For steep density spectra, the small $R$ asymptote of $\mathcal{D}(R)$ is $R^{1-\nu/2}$, while for shallow spectra, it is $R^{1-\nu_\rho-\nu/2}$. 
The latter asymptotics actually describes the scaling of $\mathcal{D}(R)$ when density index $\nu_\rho < 1- \nu/2$. This presents a unique way
to obtain velocity spectra of the turbulence.

We now study thin velocity slice limit in the presence of dust absorption.
For that, we start with Eq.~\eqref{eq:thinintpartmain}. We remind the reader
that although the representation in Eq.~\eqref{eq:thinintpartmain} is able
to describe regimes where dust absorption is strong, it {\it cannot} describe
the regimes where dust absorption is weak. The weak absorption regime will be
studied numerically later in this section. Strong absorption requirement
for validity of Eq.~\eqref{eq:thinintpartmain} means short
LOS distance $\Delta(\mathbf{X})$ to $\tau=1$ relative to all scales
of the problem, namely $\Delta(\mathbf{X}) < R \le S $.

We now carry out explicit analytical calculations for the correlation of intensity in the presence of strong dust absorption. Using Eq.~\eqref{eq:thinintpartmain}, the correlation of intensity in the {\it thin} velocity slice and
in the strong dust absorption regime can be written as
\begin{align}\label{eq:xivelden}
\xi(\bm{R})=\frac{1}{\kappa^2}\langle\Phi(\bm{X}_1,\Delta(\bm{X}_1))\Phi(\bm{X}_2,\Delta(\bm{X}_2))\rangle,
\end{align}
where $\langle..\rangle$ denotes statistical averaging in both velocity field and density field. Under the assumption of uncorrelated density and velocity field, this averaging can be performed independently, and thus we first carry out averaging with respect to velocity field by assuming Gaussian velocity field. This allows us to write Eq. \eqref{eq:xivelden} as
\begin{align}\label{xibwin}
\xi(\bm{R})\approx\frac{1}{\kappa^2}\left\langle\frac{1}{\sqrt{D_z\left(R,\Delta(\bm{X}_1)-\Delta(\bm{X}_2)\right)}}\right\rangle~.
\end{align}
where the remaining averaging is with respect to $\Delta(\bm{X})$ as a random
fluctuating quantity. Using Eq. \eqref{velocitymodel}, correlation given by Eq. \eqref{xibwin} can be explicitly written as
\begin{align}\label{correshallowd}
\xi(\bm{R})\approx \frac{S^{\nu/2}}{\kappa^2 D_z^{1/2}(S)}
\left\langle
\left(R^2+\left(\Delta(\bm{X}_1)-\Delta(\bm{X}_2)\right)^2\right)^{-\frac{\nu}{4}}
\right\rangle,
\end{align}
where we have restored all the relevant dimensionalities. A crucial point that needs to be explained at this point is the regime of validity of Eq. \eqref{correshallowd}. As a part of explanation, we take the case of constant density. In this case, on one hand, our approximate expression Eq. \eqref{correshallowd} gives a scaling of intensity correlation as $\sim R^{-\nu/2}$. On the other hand, an exact expression of intensity correlation can be obtained by using Eq. \eqref{corrthinin}, with $\xi_\rho$ set to zero, and $S$ set to $\Delta_0$. This expression gives a correlation scaling of $R^{-\nu/2}$ only at $R>\Delta_0$. We stress that density effects are suppressed by dust absorption, and therefore it is reasonable to deduce the regime of validity of Eq. \eqref{correshallowd} by considering the constant density case. With all these discussion,  we suggest the statistically averaged criterion for the validity of Eq. \eqref{correshallowd}
\begin{equation}
\label{eq:Delta0}
R > \Delta_0 \equiv 
\sqrt{\langle \Delta^2 \rangle}
\end{equation}

Appendix~\ref{deltacutstat} discusses statistics of
$\Delta$ in various regimes of density perturbations,
in particular gives expressions for $\langle \Delta^2 \rangle$
in Eqs.~(\ref{eq:Deltasquare1}, \ref{eq:Deltasquare2}).  For small perturbations,
$\Delta_0 \approx \langle \Delta \rangle \approx 1/(\kappa \bar{\rho})$.
At lags $R$ satisfying Eq.~\eqref{eq:Delta0}, we obtain from
Eq.~\eqref{correshallowd} the density insensitive scaling of the 
intensity correlation function
\begin{equation}
\label{eq:largeRasymp}
\xi(R) \sim  \frac{1}{\kappa^2 D_z^{1/2}(S)} \left(\frac{R}{S}\right)^{-\nu/2} ~ .
\end{equation}

To evaluate averaging in Eq. \eqref{correshallowd} accurately,
one needs to properly model
the statistics of $\Delta$, in particular the probability distribution function
of $\Delta$.  To give specific example, let us take dust density to be Gaussian.
This is reasonable when the density dispersion is less than the mean density.
In the case when density fluctuations are small in comparison to the mean
density (which is good for steep density spectra, but more restrictive for
shallow density spectra), the above equation reduces to 
\begin{align}\label{eq:thinmaincorr1}
\xi(\bm{R})\propto \left\langle\left(R^2+\frac{1}{\kappa^2\bar{\rho}^4}\left(\delta\rho(\bm{X}_2)-\delta\rho(\bm{X}_1)\right)^2\right)^{-\nu/4}\right\rangle~.
\end{align}
Performing averaging with Gaussian $\delta\rho$, one obtains
\begin{align}\label{modifiedxi}
\xi(\bm{R})\propto \bar{d}_{\rho}(R)^{-\nu/4} U\left(\frac{\nu}{4};\frac{2+\nu}{4};\frac{R^2}{2\bar{d}_{\rho}}\right)~,
\end{align}
where $U$ is the confluent Hypergeometric function, and  $\bar{d}_\rho(R)$ is the normalized structure function of density fluctuations
\begin{equation}
\bar{d}_\rho(R)=\frac{1}{\kappa^2\bar{\rho}^4}\left\langle\left(\delta\rho(\bm{X}_2)-\delta\rho(\bm{X}_1)\right)^2\right\rangle~.
\end{equation}
The asymptote of Eq. \eqref{modifiedxi} is $R^{-\nu/2}$ at $R^2 > \bar{d}(R)$.
Since $\bar{d}(R) \le 2 \sigma_\rho^2/(\kappa^2 \bar{\rho}^4) \approx 
\langle \Delta^2 \rangle \sigma_\rho^2/\bar{\rho}^2 $,
this asymptote matches the general result of 
Eq.~\eqref{eq:largeRasymp} over all its range of validity.

\begin{figure*}
\centering
\includegraphics[scale=0.5]{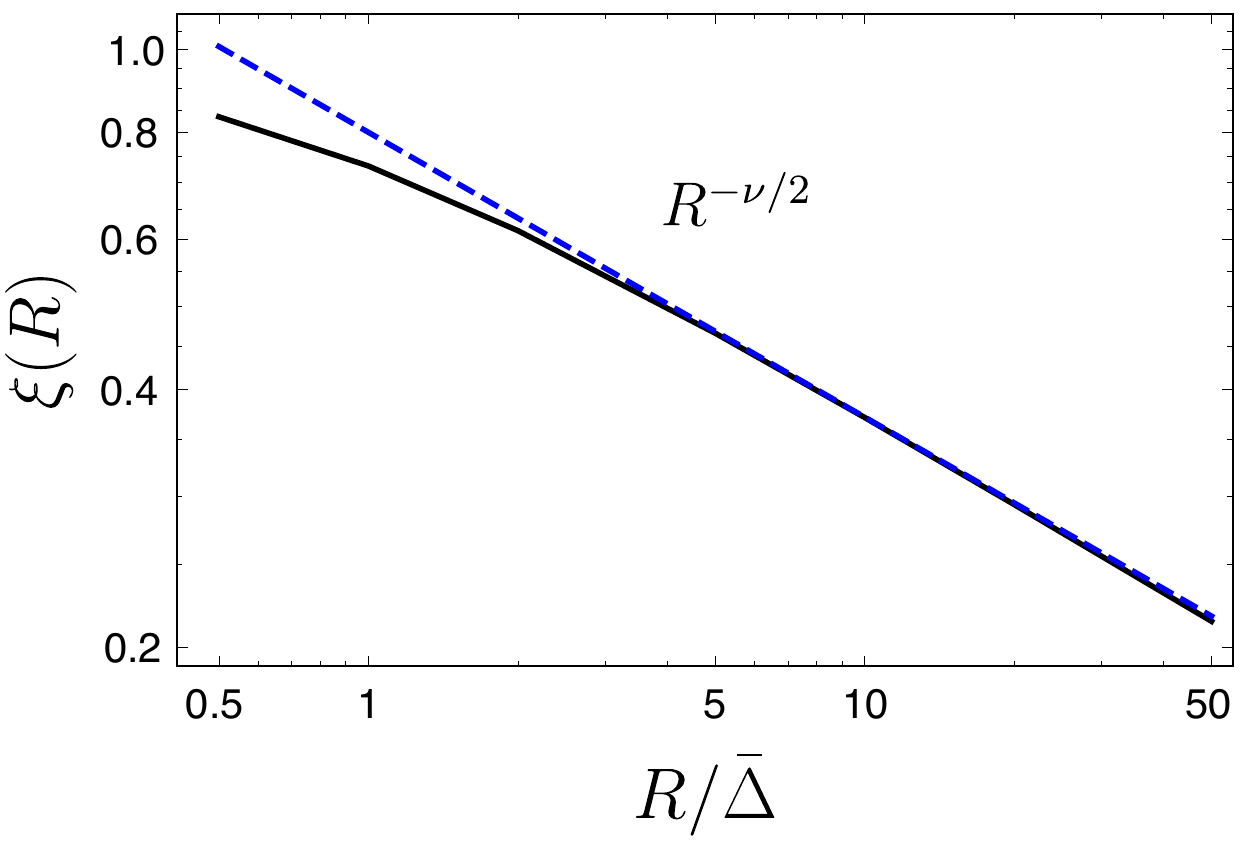}\hspace*{2cm}
\includegraphics[scale=0.5]{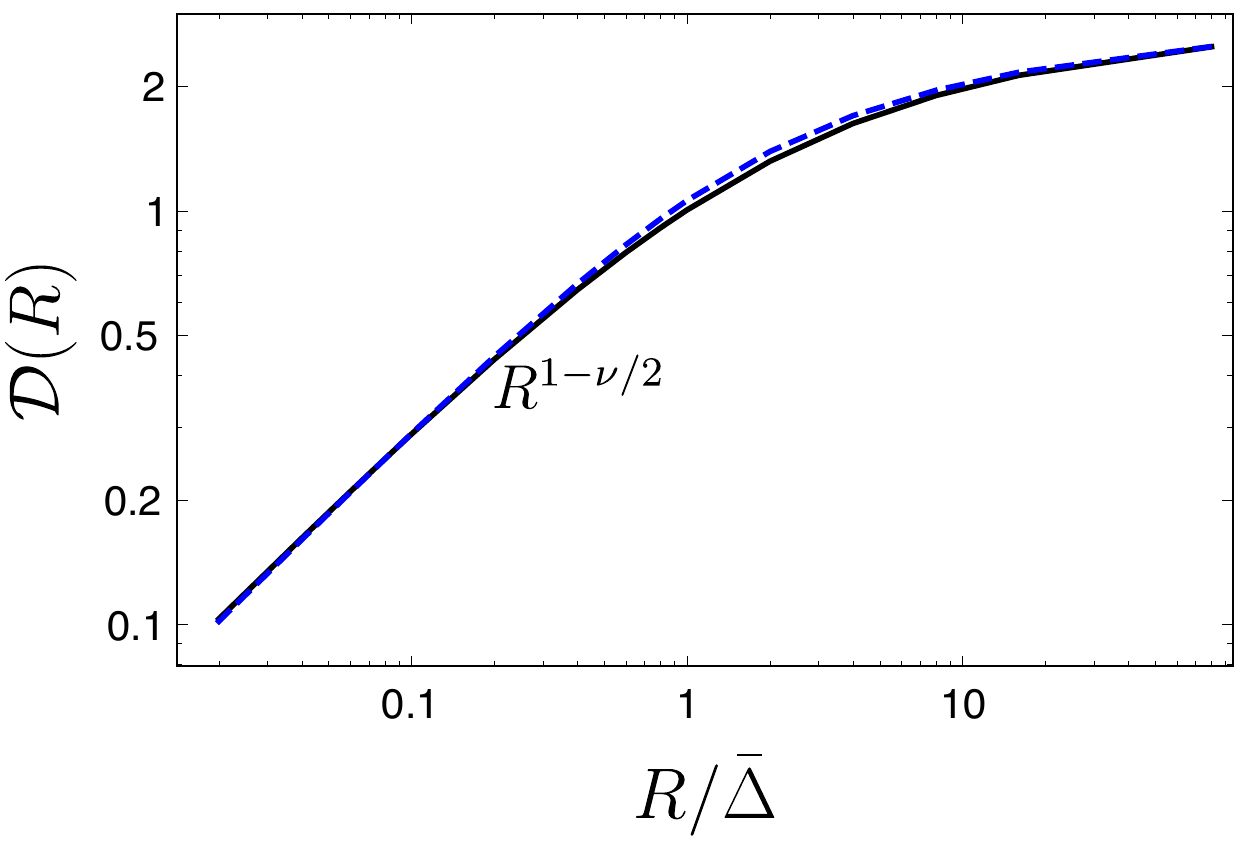}
\caption{Left: Intensity correlation function produced by full numerical integration for a thin slice case for steep density and velocity field, with spectral index $\nu=2/3$ and $\nu_\rho=1/3$. The parameters used are: $\sigma_\rho/\bar{\rho}=0.2, r_c/\bar{\Delta}=S/\bar{\Delta}=100$. Both velocity and density fluctuations are taken to following Gaussian distribution. Right: Intensity structure function at different scales for the same parameters. The solid curve is produced using exact numerics (see Appendix \ref{fullexp} and Eq. \eqref{vstrucd}), while the dotted curve is produced by using Eqs. \eqref{unified} and \eqref{vstrucd}, with $S/\Delta_0=66$. At short $R$, the asymptote is $R^{1-\nu/2}$, while at larger lags, the structure function becomes saturated. }
\label{steep}
\end{figure*}

At scales $R < \Delta_0$ the thickness of the transparent layer at
the front of the cloud cannot be considered negligible, but more distant
regions do not contribute.
In other words, in place of Eq.~\eqref{corrthinin} we can write
\begin{align}\label{unified}
\xi(\bm{R}) \sim \Delta_0 
\int_0^{\Delta_0}\!\!\!\mathrm{d}z\,\left(1-\frac{z}{2\Delta_0} \right)
\frac{\bar{\rho}^2+\xi_{\rho}(\bm{R},z)}{\sqrt{D_z(\bm{R},z)}}
\;,\quad R<\Delta_0\;.
\end{align}
Thus, dust absorption does not modify VCA predictions in this regime.
This has been represented in Fig \ref{steep}, where for
$R\gtrsim \Delta_0$, the asymptote of correlation function is clearly
$R^{-\nu/2}$, whereas for $R < \Delta_0$, the asymptote of structure function
follows the usual VCA scaling. Eq.~\eqref{unified} also obviously 
covers the regime of weak absorption when $\Delta_0 \gtrsim S$, in which case $\Delta_0$ should be replaced by $S$.

A crucial point that needs to stressed at this point is that Eq. \eqref{unified} is applicable only when dust absorption is weak or moderate, i.e. $\Delta_0\sim S$. If dust absorption is strong, i.e. $\Delta_0\ll S$, density effects are erased, and the intensity correlation is given by
\begin{align}\label{unified1}
\xi(\bm{R}) \sim \Delta_0 
\int_0^{\Delta_0}\!\!\!\mathrm{d}z\,\left(1-\frac{z}{2\Delta_0} \right)
\frac{\bar{\rho}^2}{\sqrt{D_z(\bm{R},z)}}
\;,\quad R<\Delta_0\;.
\end{align}

\begin{figure}
\centering
\includegraphics[scale=0.5]{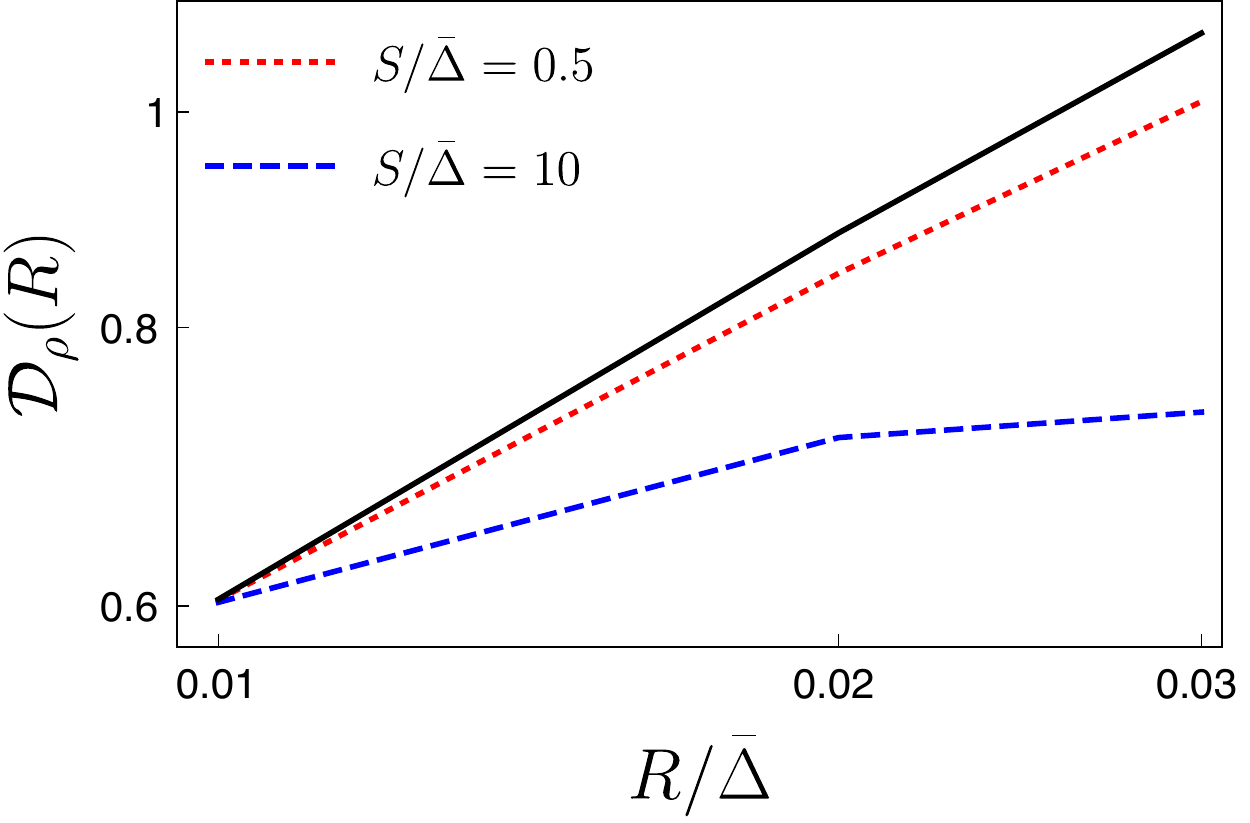}\hspace*{2cm}
\caption{The density part of structure function for different dust absorption extent for shallow density with $\nu_\rho=-1/2$, $r_c/\bar{Delta}=0.01$, $\sigma/bar{\Delta}=0.1$ and $\nu=2/3$. The solid line is produced using the simplified model presented in Eq. \eqref{unified}, with just the density part. As shown in the figure, the density contribution flattens as dust absorption becomes strong.}
\label{denflat}
\end{figure}

To test this, we carry out numerical integration with exact analytical expression presented in Appendix \ref{fullexp}. Expression presented in the Appendix \ref{fullexp} has its own limitation that it is not applicable if the density dispersion is larger than the mean density, as this becomes inconsistent with the assumption of Gaussianity. This poses challenge in deducing whether or not the density effects are erased due to strong dust absorption, as for small $\sigma$, the pure velocity part dominates while the density effects are sub-dominant regardless of the strength of dust absorption. To overcome this difficulty, we carry out numerical integration of just the density contribution part in the correlation given by Eq. \eqref{mainformcor}. As one might already see in Eq. \eqref{mainformcor}, the $\xi_\rho$ part is partly compensated by the differential part that comes with the negative sign. As shown in Fig. \ref{denflat}, the contribution of density part to intensity structure function flattens as dust absorption becomes strong, suggesting that density effects are unimportant in this regime. 

The validity of Eqs. \eqref{unified} and \eqref{unified1} is further tested by 
carrying out numerical integration of the full expression of intensity
correlation presented in Appendix \ref{fullexp}. The results are presented in
Fig.~\ref{steep} for steep density
spectra.  Left panel in Fig. \ref{steep} shows 
the correlation function of the intensity, confirming that large lag $R$
asymptotics follows Eq.~\eqref{eq:largeRasymp} while right panel shows
the structure function, emphasizing small $R$ scaling. Large $R$ 
correlation scaling is confirmed to be insensitive to the density fluctuations.

The dust cut-off depth $\Delta_0$ in the Eq. \eqref{eq:Delta0}
is the scale $R$ where the structure function starts to saturate, 
so it plays the role of the correlation length of the intensity fluctuations. 
It is close to $1/(\kappa \bar{\rho})$, 
though an accurate analytical determination
of this scale is not possible within our asymptotic analysis.
However, our study shows that measuring the intensity 
correlation length in observations of dusty regions
provides direct information about the level of dust absorption.

At small $R$ such that $D_z(R) < \Delta V^2$ the velocity slice thickness
$\Delta V$ can not be neglected and \textit{thin} slice approximation in
Eq.~\eqref{unified} is not applicable anymore. To cover this regime,
we should restore the velocity integral in the correlation function, which for strong dust absorption leads to 
\begin{align}\label{eq:unifiedwithvel}
\xi(\bm{R}) \sim \Delta_0 \int_{-\Delta V}^{\Delta V} \!\!\! \mathrm{d}v
\int_0^{\Delta_0}\!\!\!\mathrm{d}z\,\left(1-\frac{z}{2\Delta_0}
\right)\frac{\bar{\rho}^2}{\sqrt{D_z(\bm{R},z)}}\nonumber\\
\times\exp\left[-\frac{v^2}{2D_z(\bm{R},z)}\right]~,\quad R<\Delta_0~.
\end{align}

An apparent complication that might arise with our analysis is that while at small $R$ structure function is applicable, correlation function is used at larger $R$. An unified picture of the regimes with strong dust absorption and that with weak dust absorption can be obtained with the language of power spectrum. The power spectrum $P(K)$ of structure function of intensity is defined as
\begin{align}\label{powinten}
P(\bm{K})=-\frac{1}{2}\int\mathrm{d}^2\bm{R}\,\mathrm{e}^{\mathrm{i}\bm{K}\cdot \bm{R}}\mathcal{D}(\bm{R})
=\int\mathrm{d}^2\bm{R}\mathrm{e}^{\mathrm{i}\bm{K}\cdot \bm{R}}\xi(\bm{R})~,
\end{align}
which signifies that the power spectrum of structure function and correlation function are the same upto some constant factor. 

\begin{figure}
\centering
\includegraphics[scale=0.4]{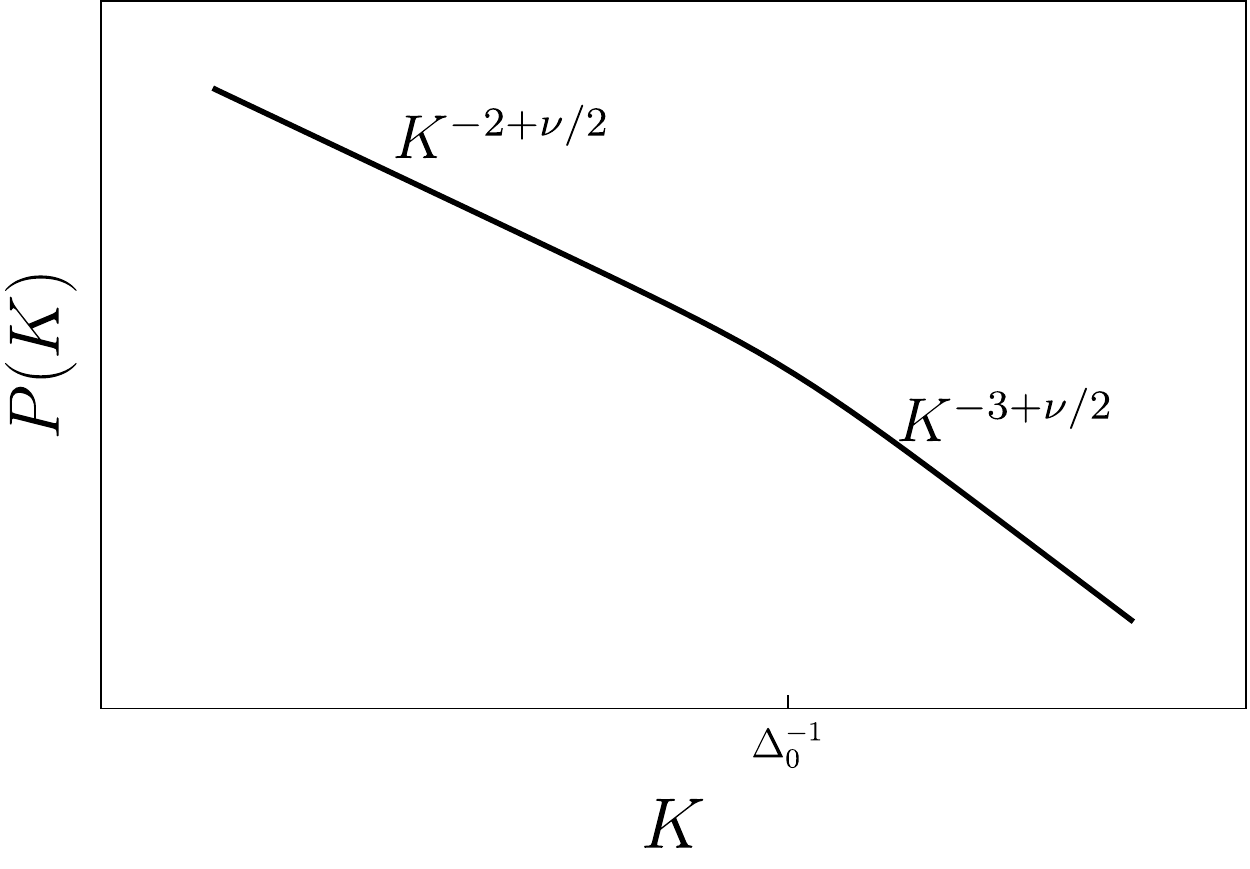}
\caption{Power spectrum of intensity fluctuation as a function of $K$ for thin velocity slice case for the case of strong dust absorption. Density effects are erased due to strong absorption, and there is a break in the slope of power spectrum at a scale characteristic to the dust cut-off.}
\label{powerspec}
\end{figure}

In the language of power spectrum, one can see the break in the slope of power spectrum as one goes from high $K$ to low $K$ in the thin slice limit. This has been clearly shown for both steep and shallow density in  Fig. \ref{powerspec}. Notice in Fig. \ref{powerspec} that the slope changes from  $-2+\nu/2$ to $-3+\nu/2$ for steep density and to $-3+\nu_\rho+\nu/2$ for shallow density with weak dust absorption.  Thus, it is useful to use correlation function at low $K$ and structure function at high $K$.We stress that the main effect is switching from integrated signal at small $R$ to just two dimensional slice in configuration space at large $R$. However, the non-trivial effect is at large $R$, or for strong dust absorption density effects are erased.

\subsection{Dust effect on self absorbing medium}\label{sec:selfab}
The study in the previous sections were carried out for optically thin emission lines, i.e. for the case when self-absorption is negligible. However, effect of self-absorption  can be important in various interstellar environments, for instance in molecular clouds. An important study on the effects of self-absorption in the intensity statistics was carried out in \citetalias{lazarian2004velocity}. Their study suggests that self-absorption introduces non linear effects on the statistics of intensity, and that the recovery of velocity and density spectrum might not be always possible. 

With the result of previous section, one can extend the results of previous section to account for the effect of both self-absorption and dust-absorption.  The main result of \citetalias{lazarian2004velocity} for structure function of intensity fluctuations can be expressed using $\xi_s(\bm{R}, v)$ as:
\begin{align}\label{abs1}
\mathcal{D}(\bm{R})=\int\mathop{\mathrm{d}v}\tilde{W}(v)\mathrm{e}^{-\frac{\alpha^2}{2}\tilde{d}_s(0,v)}\left[\xi_s(0,v)-\xi_s(\bm{R},v)\right]~.
\end{align}
The only difference between this result and the result of \citetalias{lazarian2004velocity} is the replacement of $S$ by $\Delta_0$ in all the integrals that are carried out over the LOS.  To quantify the effect of dust on self-absorption, we first start with the expression of $\tilde{d}_s(0,v)$ in Eq. \eqref{abs1} (see \citetalias{lazarian2004velocity})
\begin{align}\label{abwindow}
\tilde{d}_s(0,v)\propto \frac{\bar{\rho}^2\Delta_0}{D_z(S)}\int_{0}^{\Delta_0}\mathrm{d}z\,\frac{1}{|z|^{\nu/2}}\left[1-\exp\left(-\frac{v^2}{2|z|^m}\right)\right]~,
\end{align}
The difference between this expression and the one in \citetalias{lazarian2004velocity} is the replacement of $S$ by $\Delta_0$.
Eqs. \eqref{abs1} and \eqref{abwindow} physically imply that dust absorption decorrelates signals in physical space, while self-absorption decorrelates signals in velocity space. From Eqs. \eqref{abs1} and \eqref{abwindow}, one can immediately see that the effective window introduced by self-absorption decreases for any $\Delta_0<S$, thus diluting the effect of self-absorption. This is an interesting result: stronger dust absorption results in weaker self-absorption.

The results of \citetalias{lazarian2004velocity} suggest that in the case of negligible dust absorption, self-absorption leads to a universal regime where structure function or power spectrum of intensity fluctuations are independent of underlying velocity spectra, i.e. $\mathcal{D}(R)\sim R$. This extent of this regime is set by the strength of self-absorption, and is seen at scales
\begin{equation}
R<S\left(\frac{v_{\text{ab}}^2}{D_z(S)}\right)^{1/\nu}~,
\end{equation}
where $v_{ab}$ is the velocity scale at which $\alpha^2\tilde{d}_s(0,v_{\text{ab}})=1$ (c.f Eq. \eqref{abwindow}). As a simple approximation, in the presence of dust absorption, the scale at which this universal regime is seen is given by
\begin{equation}
R<\Delta_0\left(\frac{v_{\text{ab}}^2}{D_z(S)}\right)^{1/\nu}~.
\end{equation}
Clearly, the scales at which this regime is seen in the presence of dust absorption is reduced by a factor $\Delta_0/S<1$. Physically, this can be understood in the following sense:  strong dust absorption decorrelates signals from separations larger than the dust cut-off, and therefore the extent of self absorption a signal experiences is decreased.

\section{Effect of dust on VCS}\label{svcs}
Velocity coordinate spectrum (VCS, \citetalias{lazarian2006studying}) is another powerful technique to study supersonic turbulence. This technique measures power spectrum of intensity fluctuations along the velocity axis in PPV space. With VCS, one can study turbulence even in a spatially unresolved cloud. In this section, we discuss how dust absorption might affect studies of turbulence using VCS.

As it was shown in \citetalias{lazarian2000velocity}, the power spectrum of intensity fluctuations along the velocity axis is given by
\begin{equation}
P(k_v)=\int\mathrm{d}^2\bm{K}P(\bm{K},k_v)~,
\end{equation}
which is 
\begin{align}\label{powerVCS}
P(k_v)=\mathrm{e}^{-\beta k_v^2}\int\mathrm{d}^2\bm{R}B(\bm{R})\int_0^S\mathrm{d}z\,\xi_\rho(\bm{R}, z)\nonumber\\
\exp\left[-\frac{k_v^2D_z(\bm{R}, z)}{2}\right]~,
\end{align}
where $B(\bm{R})$ is the window of the telescope beam, which describes the resolution of the instrument. 
The two regime of interest are high spatial resolution (narrow beam) and poor spatial resolution. For a narrow beam, $B(\bm{R})=\delta(R)$, which gives asymptotic form of $P(k_v)$ at high $k_v$ for optically thin emission lines with negligible dust emission (see \citetalias{lazarian2006studying})
\begin{equation}\label{narrow}
P(k_v)\propto e^{-\beta k_v^2}\left[k_vD_z^{1/2}(S)\right]^{-2(1-\nu_\rho)/\nu}~,
\end{equation}
while for a wide beam the window is $B(\bm{R})=1$ (which upon integration over the entire image is equivalent to setting $\bm{K}=0$), and the asymptote is 
\begin{equation}\label{wide}
P(k_v)\propto e^{-\beta k_v^2}\left[k_vD_z^{1/2}(S)\right]^{-2(3-\nu_\rho)/\nu}~.
\end{equation}

\begin{figure}
\centering
\includegraphics[scale=0.5]{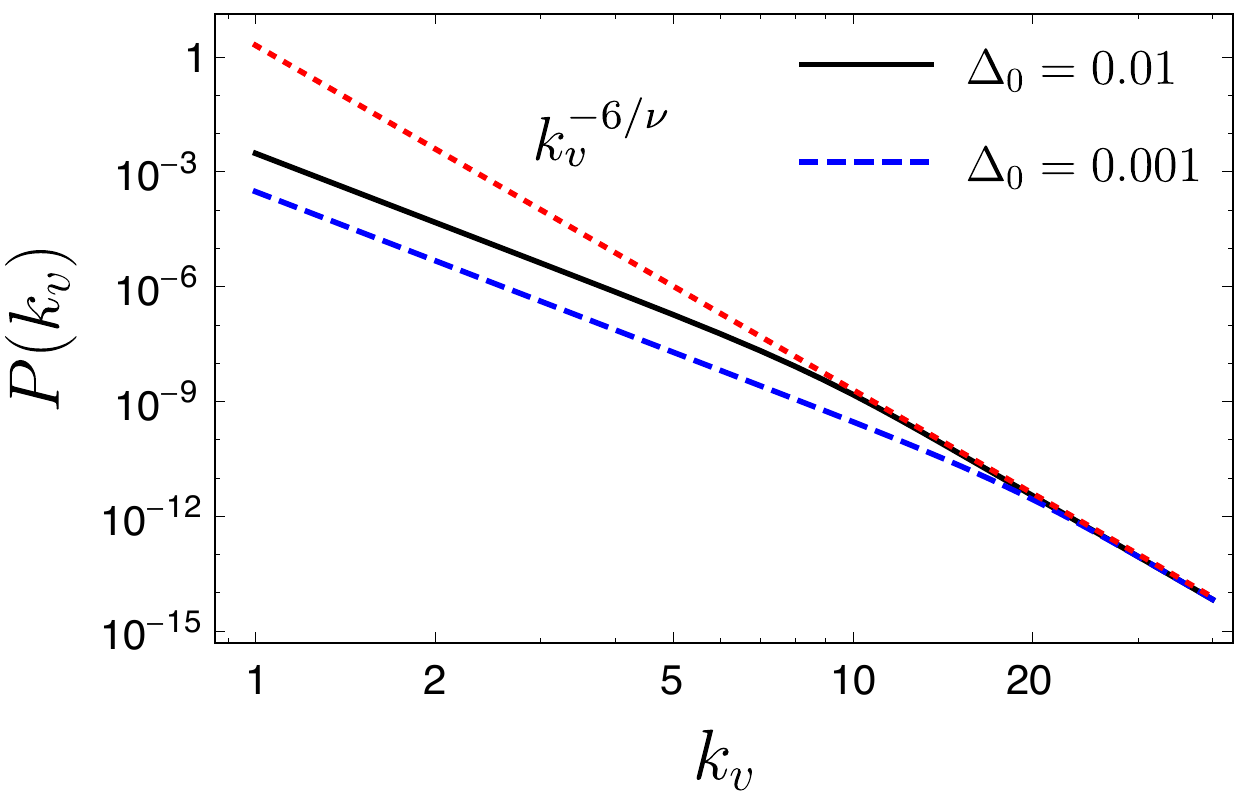}
\caption{Power spectrum of intensity fluctuations along $v$ axis, obtained numerically using Eq. \eqref{dustpowerVCS}. The velocity field is taken to be Kolmogorov, density to be constant and $\beta=0$.}
\label{fig:powerVCS}
\end{figure}

As was shown in previous sections, the main effect of dust absorption is the introduction of cut-off scales beyond which intensity signals are decorrelated. Thus, the dust cut-off scale $\Delta$ effectively acts as LOS extent $S$ of the turbulent cloud, and Eq. \eqref{powerVCS} can be written as
\begin{align}\label{dustpowerVCS}
P(k_v)\approx\mathrm{e}^{-\beta k_v^2}\int\mathrm{d}^2\bm{R}B(\bm{R})\int_0^{\Delta_{\text{av}}}\mathrm{d}z\,\xi_\rho(\bm{R}, z)\nonumber\\
\exp\left[-\frac{k_v^2D_z(\bm{R}, z)}{2}\right]~,
\end{align}
where $\Delta_{\text{av}}$ is some average of dust cut-off, and comes from the fact that the dust absorption cut-off along different LOS are different. In general, $\Delta_{\text{av}}\sim \Delta_0$. In the case when $k_v^2D_z(\Delta_0)\gg 1$, the LOS integration limit in Eq. \eqref{dustpowerVCS} can be extended to infinity, fluctuations of $\Delta$ along different LOS does not affect the averaging, allowing one to recover the usual results of VCS. Formally, this condition corresponds to
\begin{equation}\label{eq:vcsen}
k_v^{-1}<\sqrt{D_z(S)\left(\frac{\Delta_{\text{av}}}{S}\right)^\nu}~.
\end{equation}
If Eq. \eqref{eq:vcsen} is satisfied, the asymptotic forms represented by Eqs. \eqref{narrow} and \eqref{wide} for narrow and wide beam respectively will be valid even with the presence of dust absorption, as shown in Fig \ref{fig:powerVCS}, where the asymptote is established at some $k_v$ regardless of what the dust cut-off is. However, it is important to keep in mind that these asymptotic forms are applicable only at large wavenumbers, which depends on the value of $\Delta_{\text{av}}$ is. This is what Fig. \ref{fig:powerVCS} demonstrates, where it is shown how different dust cut-off $\Delta_{\text{av}}$ lead to asymptote at different $k_v$. 

An important scale is the velocity scale $V_{\Delta B}$ where the transition from poor to high resolution occurs
\begin{equation}\label{vscale}
V_{\Delta B}\equiv \sqrt{D_z(S)\left(\frac{\Delta B(S/\Delta_{\text{av}})}{S}\right)^{\nu}}~.
\end{equation}
It is clear from Eq. \eqref{vscale} that due to dust absorption the width of the telescope beam effectively increases from $\Delta B$ to $\Delta B(S/\Delta_{\text{av}})$, which implies that dust absorption makes spatial resolution poorer.
At scales 
\begin{equation}\label{nvscale}
k_v^{-1}>V_{\Delta B}~,
\end{equation}
the beam is narrow, while on shorter scales its width is important. In the case of negligible thermal broadening, the maximum value of $V_{\Delta_B}$ is $\sqrt{D_z(S)}$. Thus, a narrow beam requires $\Delta B<\Delta_0$, implying that it might not always be possible to achieve narrow beam especially if dust absorption is strong (i.e. $\Delta_0$ is small). 

The condition required for the asymptotes represented by narrow beam is more stringent than what Eq. \eqref{nvscale} gives. Combining Eqs. \eqref{eq:vcsen} and \eqref{nvscale}, one can see that there is a range of wavenumbers $k_v$ for which one can achieve narrow beam:
\begin{equation}\label{narrowrange}
\sqrt{D_z(S)\left(\frac{\Delta B}{\Delta_{\text{av}}}\right)^{\nu}}<k_v^{-1}<\sqrt{D_z(S)\left(\frac{\Delta_{\text{av}}}{S}\right)^\nu}~.
\end{equation}
An important result that can be inferred from the above equation is that while the above equation can be easily satisfied by large $k_v$ for large $\Delta_{\text{av}}$, the range of $k_v$ becomes narrower as $\Delta_{\text{av}}$ decreases. In particular, if $\Delta_{\text{av}}\lesssim \sqrt{S\Delta B}$, one cannot achieve the narrow beam.

It is also important to note that if dust absorption is strong, the power spectrum of fluctuations of intensity in velocity space is purely due to velocity effects, and the contribution due to density fluctuations are suppressed. In the weak dust absorption regime, the usual VCS statistics is restored.

\section{Centroids}\label{centroidsdust}
Centroid, which is velocity moment of intensity, is another powerful technique to study velocity field in a turbulent medium using Doppler broadened emission lines. Traditionally, centroid is defined as
\begin{equation}
C_N(\bm{X})=\frac{\int_{-\infty}^\infty\mathrm{d}v\,vI_v(\bm{X})}{\int_{-\infty}^\infty\mathrm{d}v\,I_v(\bm{X})}~.
\end{equation}
The normalisation in the definition of centroid makes analytical construction difficult. To overcome this, \citetalias{lazarian2003statistics} introduced a notion of unnormalized velocity centroids (UVC), which is defined as 
\begin{equation}
C(\bm{X})=\int_{-\infty}^\infty\mathrm{d}v\,vI_v(\bm{X})~.
\end{equation}
For optically thin emission lines with negligible dust-absorption, the above definition gives
\begin{equation}
C(\bm{X})=\int_0^S\mathrm{d}z\,\rho(\bm{x})u(\bm{x})~,
\end{equation}
and the structure function in the case of constant density is given by
\begin{equation}
\mathcal{D}(R)=\rho^2\int_0^S\mathrm{d}z_1\int_0^S\mathrm{d}z_2\,D_z(\bm{r})~,
\end{equation}
which at $R\ll S$ has an asymptotic form of
\begin{equation}
\mathcal{D}(R)\propto R^{1+\nu}~.
\end{equation}
Thus, by measuring the structure function of centroids, one can obtain the spectral slope of velocity field. 

We now focus on effects of dust absorption for the studies of turbulence using velocity centroids. The formalism developed in Sec. \ref{sec:thin} can be easily extended to  centroids. We start with the definition of centroids
\begin{equation}
C(\bm{X})=\int_{-\infty}^\infty\mathrm{d}v\,vI_v(\bm{X})=\int_0^S\mathrm{d}z\,\rho(\bm{x})u(\bm{x})\mathrm{e}^{-\kappa\int_0^S\mathrm{d}z'\,\rho(z')}.
\end{equation}

\begin{figure}
\centering
\includegraphics[scale=0.5]{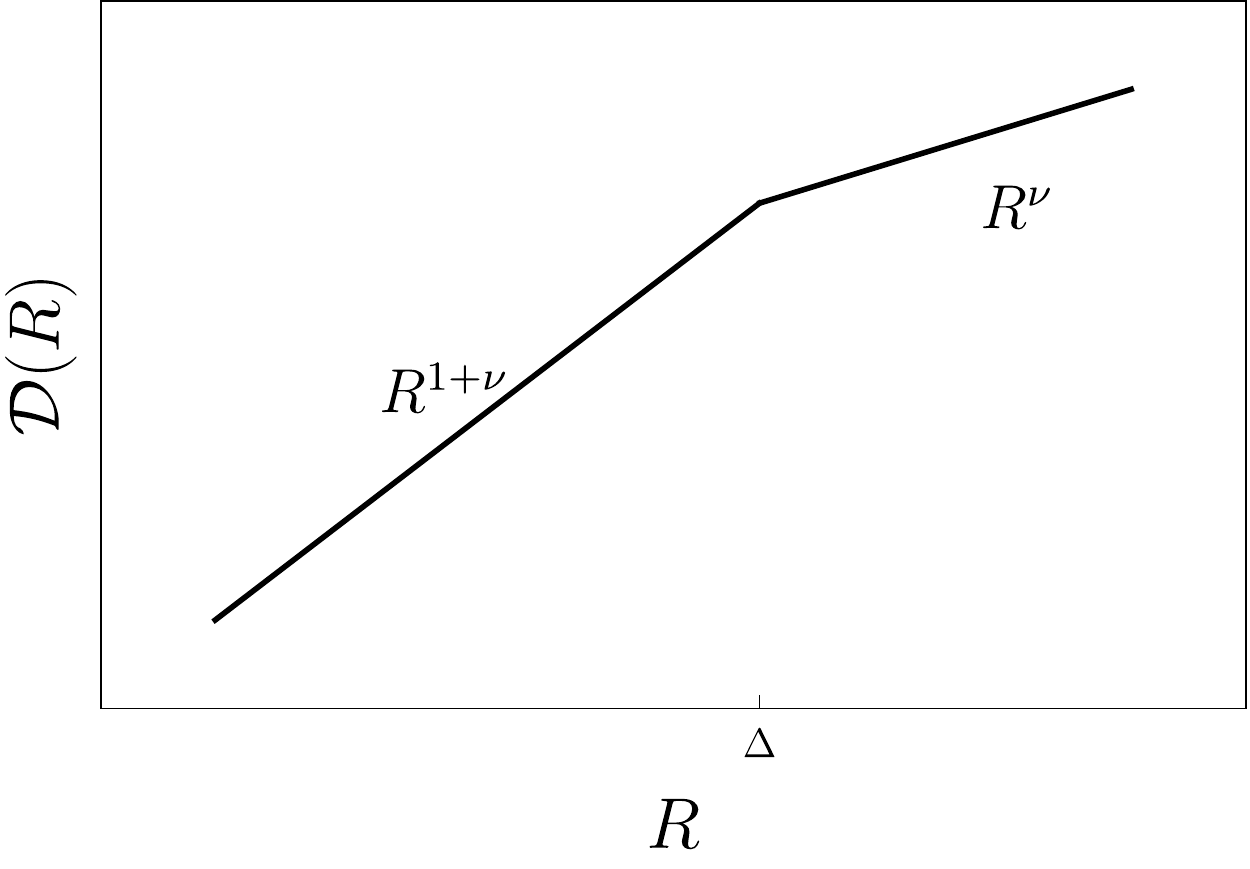}
\caption{Centroid structure function at different scales for Kolmogorov velocity field. At short scales $R$, structure function scaling is not affected by dust, while at large scales, dust flattens the spectra by an index of 1.}
\label{centroidbr}
\end{figure}

In the case of weak dust absorption, we do not expect much changes in the centroid statistics. However, for strong dust absorption, using similar idea as in Sec. \ref{sec:thin}, one can easily obtain
\begin{equation}
C(\bm{X})\approx\frac{1}{\kappa}u(\bm{X},\Delta(\bm{X})).
\end{equation}
Therefore, the structure function of centroids is given by
\begin{equation}
\mathcal{D}(R)=\frac{1}{\kappa^2}\langle D_z(R,\Delta(\bm{X}_1)-\Delta(\bm{X}_2))\rangle~,
\end{equation}
where the averaging is over $\Delta$ as a random fluctuation. The above approximation is valid for $R>\Delta_0$. For $R<\Delta_0$, one should obtain the usual centroid structure function. With this, one can write
\begin{align}\label{lcent}
\mathcal{D}(R)&\sim R^{\nu}\qquad \text{ for } R>\Delta_0~,\\
\label{scent}
\mathcal{D}(R)&\sim R^{1+\nu}\qquad \text{ for } R<\Delta_0~.
\end{align}
The above scalings are also represented in Fig. \ref{centroidbr}. 
Notice that unlike in VCA, dust does not change the statistics from steep at short scales to shallow at large scales. 

\begin{table*}
\centering
\caption{Summary of scalings of various statistical techniques in the presence of dust absorption.}
\begin{tabular}{l c c c c}  
\hline
\emph{Technique} & \emph{Scale} & \emph{Tool} & \emph{Scaling} & \emph{Equation}\\
\hline
thin slice VCA & $R<\Delta_0$ &structure function & $R^{1-\nu/2}$ & Eq. \eqref{strthinw}\\ 
&$R>\Delta_0$ & correlation function & $R^{-\nu/2}$ & Eq. \eqref{modifiedxi}\\
\hline
thick slice (steep density) & $R<R_c$ & structure function & $R^{1+\nu_\rho}$ & Eq. \eqref{eq:dustthickmain}\\
& $R>R_c$ & structure function & saturated  & Eq. \eqref{eq:dustthickmain} \\
& & & & \\
thick slice (shallow density)& $R<R_c$ & correlation function & saturated  & Eq. \eqref{shalcor}\\
& $R>R_c$ & correlation function & $R^{1-\nu_\rho}$  & Eq. \eqref{shalcor}\\
\hline
centroids & $R<\Delta_0$ & structure function & $R^{1+\nu}$  & Eq. \eqref{scent}\\
& $R>\Delta_0$ & structure function & $R^{\nu}$ & Eq. \eqref{lcent}\\
\hline
VCS & $k_v^{-1}>V_{\Delta B}$& power spectrum & $k_v^{-2(1-\nu_\rho)/\nu}$  & Eq. \eqref{narrow}\\
& $k_v^{-1}<V_{\Delta B}$& power spectrum & $k_v^{-2(3-\nu_\rho)/\nu}$  & Eq. \eqref{wide}\\
\hline
\end{tabular}
\label{tab:summary}
\end{table*} 

\section{Anisotropic MHD Turbulence}\label{sanist}
Our analysis in the previous sections focused on the spectra of density and velocity field, and thus we worked with the assumptions that the fields are isotropic and homogeneous. The interstellar medium is magnetised and thus there exists a preferred direction (along the direction of magnetic field), and therefore isotropy is not a good assumption any more. 
The properties of MHD turbulence depend on the degree of magnetization. An useful measure of the degree of magnetization is the Alfv\'en Mach number $M_{\text{A}}=V_L/V_A$, where $V_L$ is the injection velocity at the scale $L$ and $V_A$ is the Alfv\'en velocity. For $M_{\text{A}}\gg 1$ magnetic forces are not important at large scales and the cascade should be similar to ordinary hydrodynamic cascade in the vicinity of the injection scale. The modern theory of MHD turbulence was proposed in \citet[hereafter \citetalias{goldreich1995toward}]{goldreich1995toward},  where it was argued that turbulence eddies are strongly anisotropic, and are elongated towards the direction of {\it local} magnetic field. This theory was elaborated in further studies, in particular the concept of local system of reference (\citealt{lazarian1999reconnection}; \citealt{cho2000anisotropy}; \citealt{maron2001simulations}; \citealt{cho2002simulations}) was introduced. According to this concept, turbulent motions should not be viewed in the system of reference of the mean magnetic field as all earlier theories of MHD turbulence attempted to do, but in the system of reference of magnetic field comparable with the size of the eddies. However, from the point of view of the observational studies of the turbulence in a volume when the only available statistics are those averaged along the line of sight, the measurements should be carried out in the system of mean magnetic field, rather than the local system of reference. Therefore, one has to describe Alfv\'enic turbulence in the global system of reference  (see the discussions in \citealt{cho2002compressible}; \citealt{esquivel2005velocity}; \citetalias{lazarian2012statistical}). This modifies the available statistics. For instance, in the local system of reference \citetalias{goldreich1995toward} predicts the existence of two different energy spectra, namely, the parallel and perpendicular, in the global system of reference  only the spectrum of dominant perpendicular fluctuations is available. Similarly, while in the local system of reference the anisotropy increases with the decrease of size of the eddies, the anisotropy stays constant in the global system of reference.

MHD turbulence can be presented as a superposition of interacting fundamental modes, i.e. Alfv\'en, slow and fast. The first theoretical considerations in favour of this were given in \citetalias{goldreich1995toward} (see also \citealt{lithwick2001compressible}), which were extended and numerically tested in \cite{cho2002compressible, cho2003compressible} and in \cite{kowal2010velocity}. Because the compressible and incompressible modes weakly exchange their energy (\citealt{cho2002compressible}), it is possible to consider the modes separately. The details of velocity structure function of these fundamental modes are presented in  \citet[hereafter \citetalias{kandel2016extending}]{kandel2016extending}. 

In this section, we aim to discuss effects of dust absorption on the study of anisotropy. For that we will explicitly use some of the important results from \citetalias{kandel2016extending}. We first briefly review different fundamental MHD modes.

\subsection{Different MHD modes}
In a magnetised medium, turbulence can be described by decomposition of basic MHD modes : Alfv\'en, slow and fast. This decomposition is carried out with respect to the mean magnetic field, and is considered to be reasonable as long as perturbations of magnetic field are less than the mean magnetic field. In general, the Fourier component of velocity in a mode is given by $v(\bm{k}) = a_k\hat{\xi}_k$, where $\bm{k}$ is the wavevector, $a_k$ is the random complex amplitude of a mode and $\hat{\xi}$ is the direction of allowed displacement. With this, the power spectrum can be written as 
\begin{equation}
\langle v_i(\bm{k}v_j^*(\bm{k}')\rangle=\mathcal{A}(k, \hat{k}\cdot\hat{\lambda})(\hat{\xi}_k\otimes\hat{\xi}_k^*)_{ij}\delta(\bm{k}-\bm{k}')~,
\end{equation}
where $\mathcal{A}(k, \hat{k}\cdot\hat{\lambda})\equiv \langle \hat{a}_k\hat{a}_k^*\rangle $ is the power spectrum of amplitude fluctuations, and is in general anisotropic (as its angular dependence reflects), and  $(\hat{\xi}_k\otimes\hat{\xi}_k^*)_{ij}$ is the tensor structure of the mode, which is also in general anisotropic. 

In the subsequent subsections, we briefly review general properties and power spectrum of each of the fundamental MHD modes.
\subsubsection{Alfv\'en mode}
Alfv\'en modes are incompressible modes with most of the energy residing in them at around the wavenumbers where eddy turnover time is equal to the period of the Alfv\'en wave.  The displacement of this mode in a plasma is orthogonal to the wave-vector and the direction of magnetic field, i.e. $\hat{\xi}_k\propto \hat{k}\times\hat{\lambda}$, which leads to anisotropic tensor structure of Alfv\'en mode as
(see \citetalias{kandel2016extending})
\begin{align}\label{eq:alfventensor}
&\left(\hat{\xi}_{\bm{k}} \otimes \hat{\xi}^*_{\bm{k}}\right)_{ij}\nonumber\\
&=\left(\delta_{ij}-\hat{k}_i\hat{k}_j\right)-\frac{(\hat{\bm{k}}\cdot\hat{\lambda})^2\hat{k}_i\hat{k}_j+\hat{\lambda}_i\hat{\lambda}_j-(\hat{k}\cdot\hat{\lambda})(\hat{k}_i\hat{\lambda}_j+\hat{k}_j\hat{\lambda}_i)}{1-(\hat{\bm{k}}\cdot\hat{\lambda})^2}~.
\end{align}
In the global system of reference (which is the only reference system available for observations), the anisotropy of power spectrum is determined by the anisotropy at outer scale, and the power spectrum of amplitude fluctuations is given by
\begin{equation}
\mathcal{A}(k, \hat{k}\cdot\hat{\lambda})\propto k^{-11/3}\exp\left[-M_a^{-4/3}\frac{\left|\hat{\bm{k}}\cdot\hat{\lambda}\right|}{\left(1-(\hat{\bm{k}}\cdot\hat{\lambda})^2\right)^{2/3}}\right]~,
\end{equation}
where $\hat{\lambda}$ is the unit vector pointing towards the direction of mean magnetic field, and $M_A$ is the Alfv\'en Mach number. Clearly, if the LOS direction is aligned with the direction of magnetic field, then one can not see any anisotropic effect.

\subsubsection{Fast mode}
Fast modes are compressible modes and in a plasma with $\beta\equiv P_{\text{gas}}/P_{\text{mag}}\gg 1$, they behave like sonic waves, while in a low-$\beta$ plasma, they propagate with Alfv\'en speed due to the compressions of magnetic field. 
In the absence of collisionless damping, turbulent cascade of fast mode are expected to persist over time spans longer than that of Alfv\'en or slow modes, and they are expected to marginally interact with Alfv\'en modes \citep{cho2002compressible}. The power spectrum of amplitude fluctuations of this mode is isotropic, i.e.
\begin{equation}
\mathcal{A}(k)\propto k^{-7/2}~.
\end{equation}
The tensor structure of this mode can be derived by considering the allowed displacement of this mode in a plasma. In the high-$\beta$ regime, the displacement of fast mode is purely radial, i.e. parallel to wavevector $\hat{k}$, which leads to isotropic tensor structure
\begin{equation}
\left(\hat{\xi}_{\bm{k}} \otimes \hat{\xi}^*_{\bm{k}}\right)_{ij}\propto \hat{k}_i\hat{k}_j~,
\end{equation}
while in the low-$\beta$ regime, the displacement of this mode is orthogonal to the direction of magnetic field, i.e. $\hat{\xi}_k\propto \hat{\lambda}\times (\hat{\lambda}\times\hat{\bm{k}})$, and leads to tensor structure
\begin{equation}\label{fastcormode}
\left(\hat{\xi}_{\bm{k}} \otimes \hat{\xi}^*_{\bm{k}}\right)_{ij}=\frac{\hat{k}_i\hat{k_j}-(\hat{k}.\hat{\lambda})(\hat{k}_i\hat{\lambda}_j+\hat{k}_j\hat{\lambda}_i)+(\hat{k}.\hat{\lambda})^2\hat{\lambda}_i\hat{\lambda}_j}{1-(\hat{k}.\hat{\lambda})^2}~,
\end{equation}
which is clearly anisotropic. 

\subsubsection{Slow mode}
Slow modes in high-$\beta$ plasma are similar to pseudo-Alfv\'en modes, while in low-$\beta$ plasma they are density perturbations propagating with sonic speed parallel to magnetic field (see \citealt{cho2003compressible}). The power spectrum of this mode is anisotropic, and is in fact same as that of Alfv\'en mode. The tensor structure of this mode is derived by considering the allowed displacement of this mode in a plasma. In the high-$\beta$ regime, the displacement of this mode is perpendicular to the wavevector $\hat{k}$, i.e. $\hat{\xi}_k\propto \hat{\bm{k}}\times(\hat{\bm{k}}\times\hat{\lambda})$, leading to anisotropic tensor structure (see \citetalias{kandel2016extending})
\begin{equation}
\left(\hat{\xi}_{\bm{k}} \otimes \hat{\xi}^*_{\bm{k}}\right)_{ij}=\frac{(\hat{\bm{k}}\cdot\hat{\lambda})^2\hat{k}_i\hat{k}_j+\hat{\lambda}_i\hat{\lambda}_j-(\hat{k}\cdot\hat{\lambda})(\hat{k}_i\hat{\lambda}_j+\hat{k}_j\hat{\lambda}_i)}{1-(\hat{\bm{k}}\cdot\hat{\lambda})^2}~.
\end{equation}

On the other hand, in the low-$\beta$ regime, the displacement of this mode is parallel to the direction of magnetic field, i.e. $\hat{\xi}_k\propto \hat{\lambda}$, and leads to tensor structure
\begin{equation}
\left(\hat{\xi}_{\bm{k}} \otimes \hat{\xi}^*_{\bm{k}}\right)_{ij}=\hat{\lambda}_i\hat{\lambda}_j~,
\end{equation}
which is also clearly anisotropic.

\subsection{Effects of dust absorption on VCA and centroid anisotropy}
The intensity statistics of emission lines is dependent on the LOS velocity. In a magnetised medium, motions, and hence intensity statistics, are affected by the magnetic field. These effects are manifested through the anisotropic nature of correlation and structure function of intensity and centroids. The study of intensity anisotropy was carried out in \citetalias{kandel2016extending} and the study of centroids anisotropy was carried out in \citetalias{Kandel21012017}. In this section, we will discuss how the results in these paper would be affected by dust-absorption. 

Due to the cut-off introduced by dust absorption, it was shown in Section \ref{vcadust}  that the statistics of correlation function and structure function is modified at different lags $R$. In particular, for thin slice case, it was shown that intensity statistics changes from steep at small lags $R$ to shallow at large lags $R$. Similarly, it was shown that in the thick slice regime, intensity structure function saturates for steep density at $R$ larger some critical value, which depends on dust absorption coefficient. Thus, it is natural to expect different anisotropic behaviour of intensity correlation and structure function depending upon the lags $R$ under consideration.

Let us first start with thick velocity slice regime of optically thin emission lines. From Eq. \eqref{eq:dustthickmain}, it is clear that for steep spectra at large scales $R$, the intensity structure function saturates, while at small scales the usual VCA structure function is recovered. Thus, one can conclude that at small scales, the anisotropy will be the same as that predicted by VCA (see \citetalias{kandel2016extending}), while at large scales the statistics is isotropic. Thus, without a good spatial resolution, one cannot study anisotropies in the presence of dust absorption. On the other hand, shallow density is usually isotropic, and thus we do not expect anisotropies in correlation function of intensity in thick slice regime if density is shallow.

We now consider thin slice case. In this case, at short lags $R$, dust absorption does not affect power law scaling, and level of anisotropy is similar to that predicted in \citetalias{kandel2016extending}. On the other hand, results from Sec. \ref{sec:thin} show that the structure function is saturated at scales $R>\Delta_0$, and thus correlation function is better measure at these scales. Therefore, one can expect isotropization of intensity statistics at these scales if one uses structure function. On the other hand, anisotropies are manifested in the intensity correlation function at these scales, and the anisotropy of correlation function of intensity is directly related to the anisotropy of velocity structure function. It is clear from this discussion that the study of anisotropy in the thin slice regime is complicated by the fact that one has to use two different statistics at two different scale ranges. A way to unify the study of anisotropy is to study anisotropy of power spectrum of intensity fluctuations. An anisotropic power spectrum can be decomposed into multipoles
\begin{equation}
P(\bm{K})=\sum_{n=0}^{\infty}P_n\cos(n\phi_k)~,
\end{equation}
where $P_n$ is the multipole moments of power spectrum. An useful measure of level of anisotropy in the language of power spectrum is the quadrupole to monopole ratio $\mathcal{R}$ defined as
\begin{equation}
\mathcal{R}=\frac{P_2(K)}{P_0(K)}
\end{equation}
which depends on the scale $K$ and magnetisation of media. 

The anisotropy level of different MHD modes at small lags $R$ has already been presented in \citetalias{kandel2016extending}. Here, we proceed to compute this anisotropy level for lags $R>\Delta_0$. From Eq. \eqref{correshallowd}, and for sufficiently large lags, it is enough for our discussion to write the correlation function as
\begin{equation}\label{eq:xianisthin}
\xi(\bm{R})\approx 1/\sqrt{D_z(\bm{R},0)}.
\end{equation}

\begin{figure*}
\centering
\includegraphics[scale=0.5]{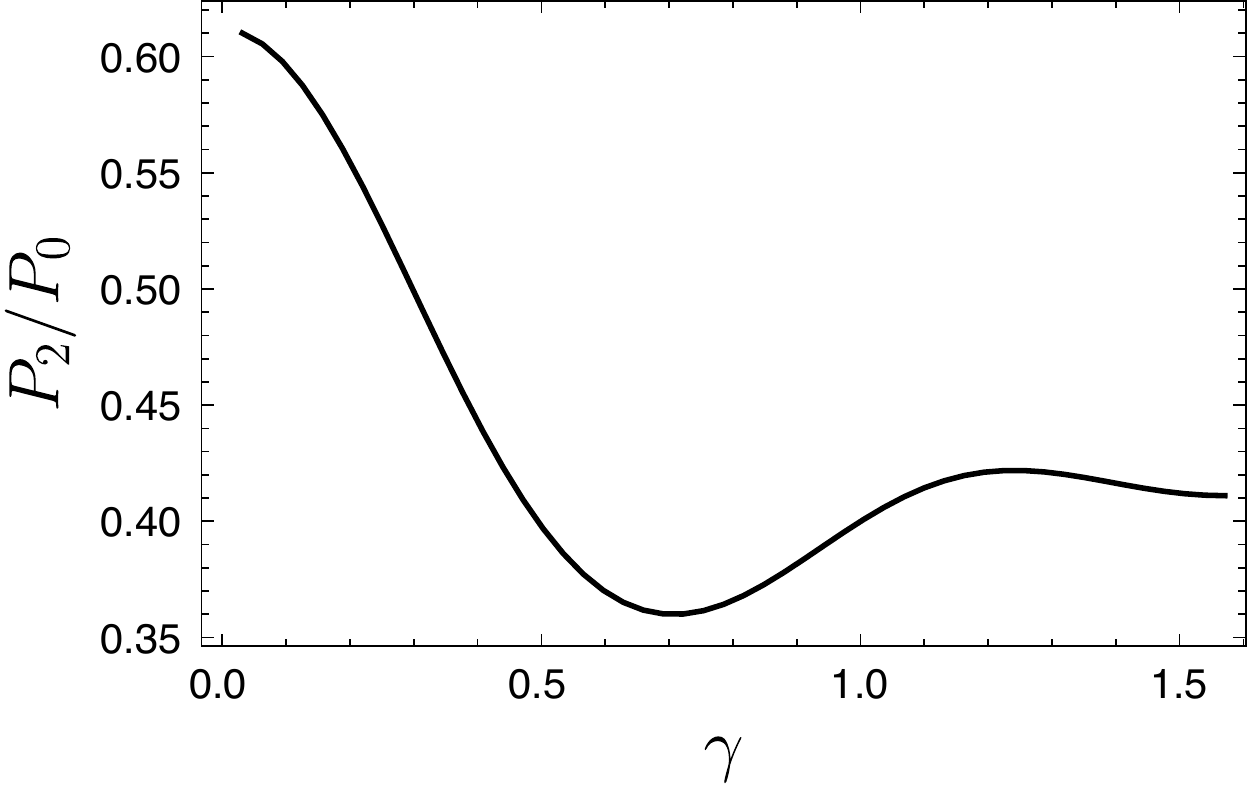}\hspace*{0.2cm}
\includegraphics[scale=0.5]{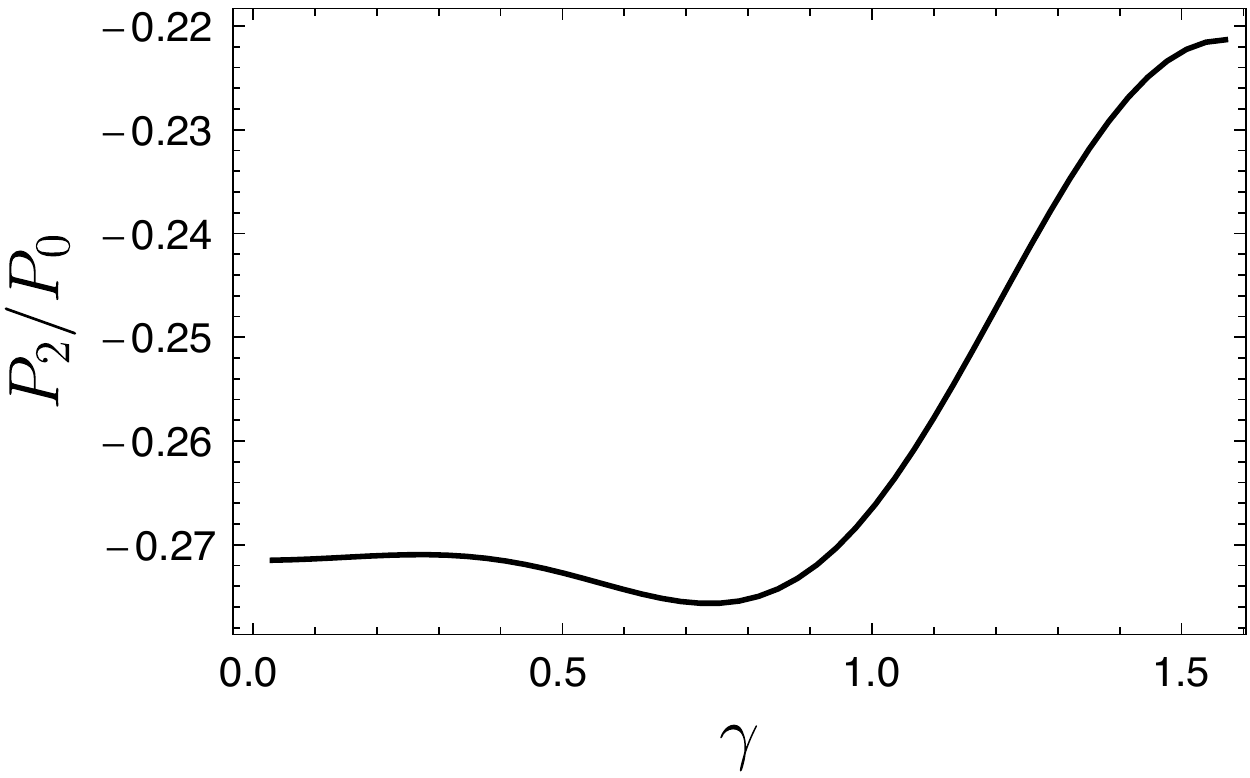}
\caption{Alfv\'en mode at $M_a=0.7$: quadrupole to monopole ratio of power spectrum of intensity fluctuations. Left-hand panel: the observed ratio in the case of negligible dust absorption or at scales $K\gtrsim \Delta_0^{-1}$. Right-hand panel: the observed ratio in the presence of dust absorption at scales $K\lesssim \Delta_0^{-1}$.}
\label{anisalfven}
\end{figure*}

\begin{figure*}
\centering
\includegraphics[scale=0.5]{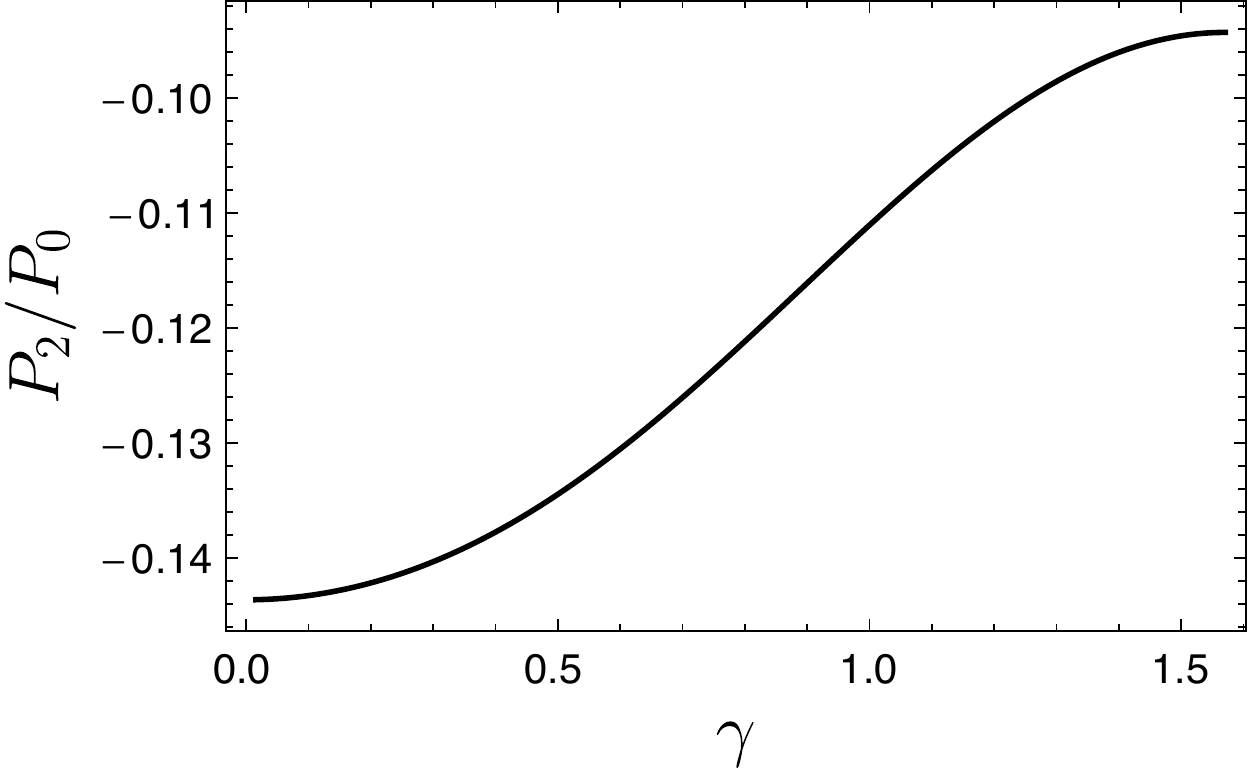}\hspace*{0.2cm}
\includegraphics[scale=0.5]{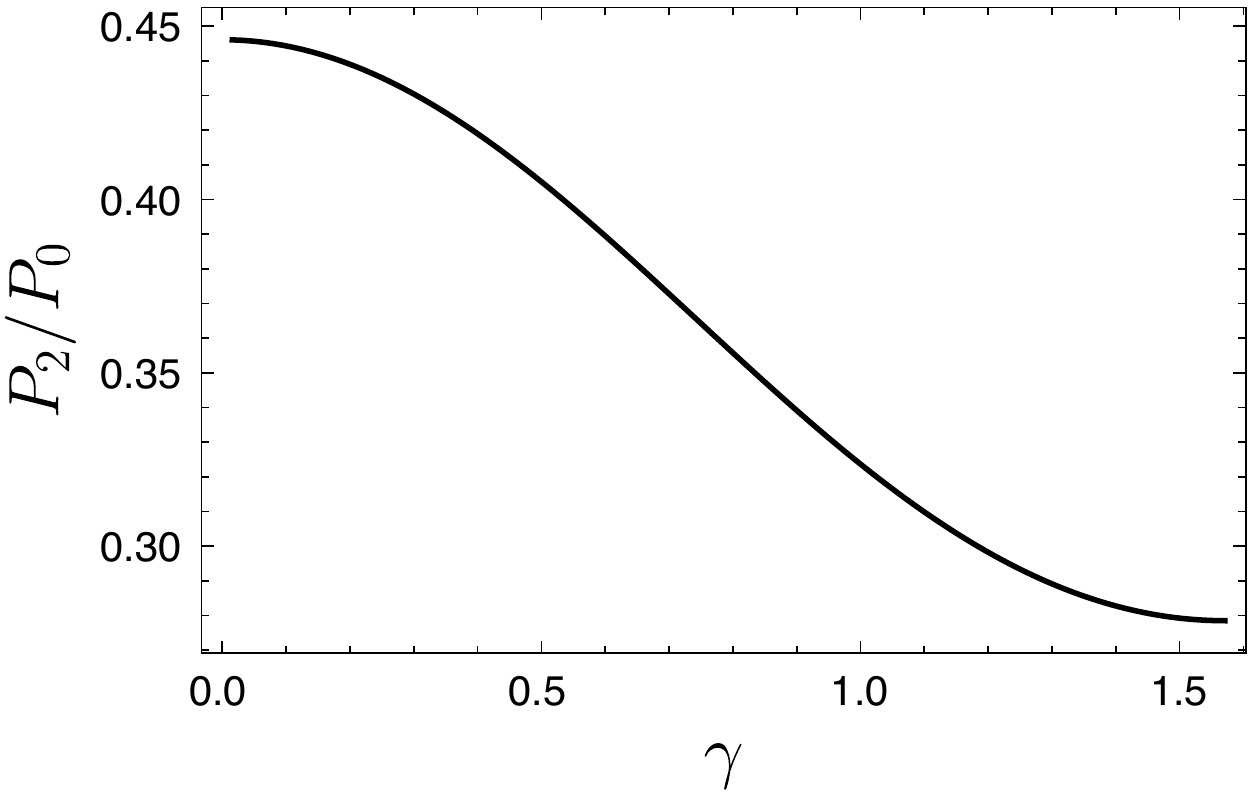}
\caption{Low-$\beta$ fast modes: quadrupole to monopole ratio of power spectrum of intensity fluctuations. Left-hand panel: the observed ratio in the case of negligible dust absorption or at scales $K\gtrsim \Delta_0^{-1}$. Right-hand panel: the observed ratio in the presence of dust absorption at scales $K\lesssim \Delta_0^{-1}$.}
\label{anisfast}
\end{figure*}

\begin{figure*}
\centering
\includegraphics[scale=0.5]{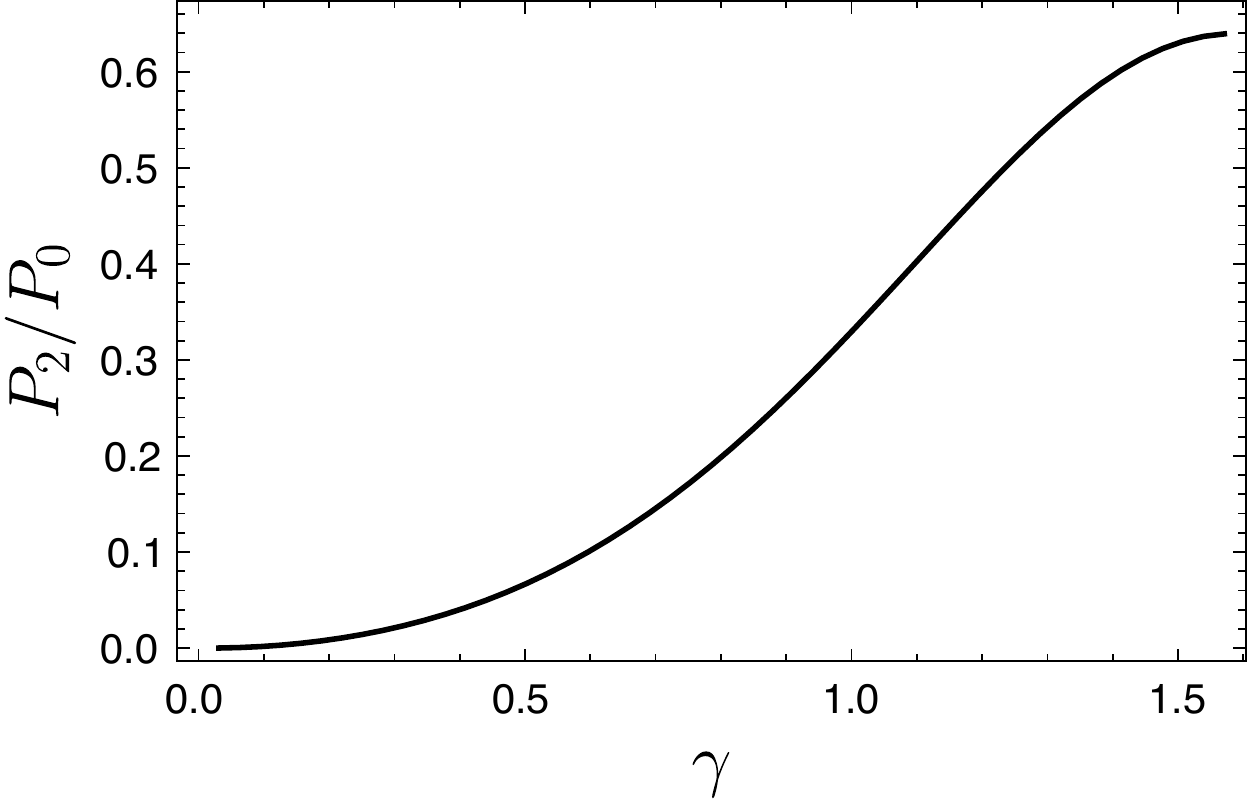}\hspace*{0.2cm}
\includegraphics[scale=0.5]{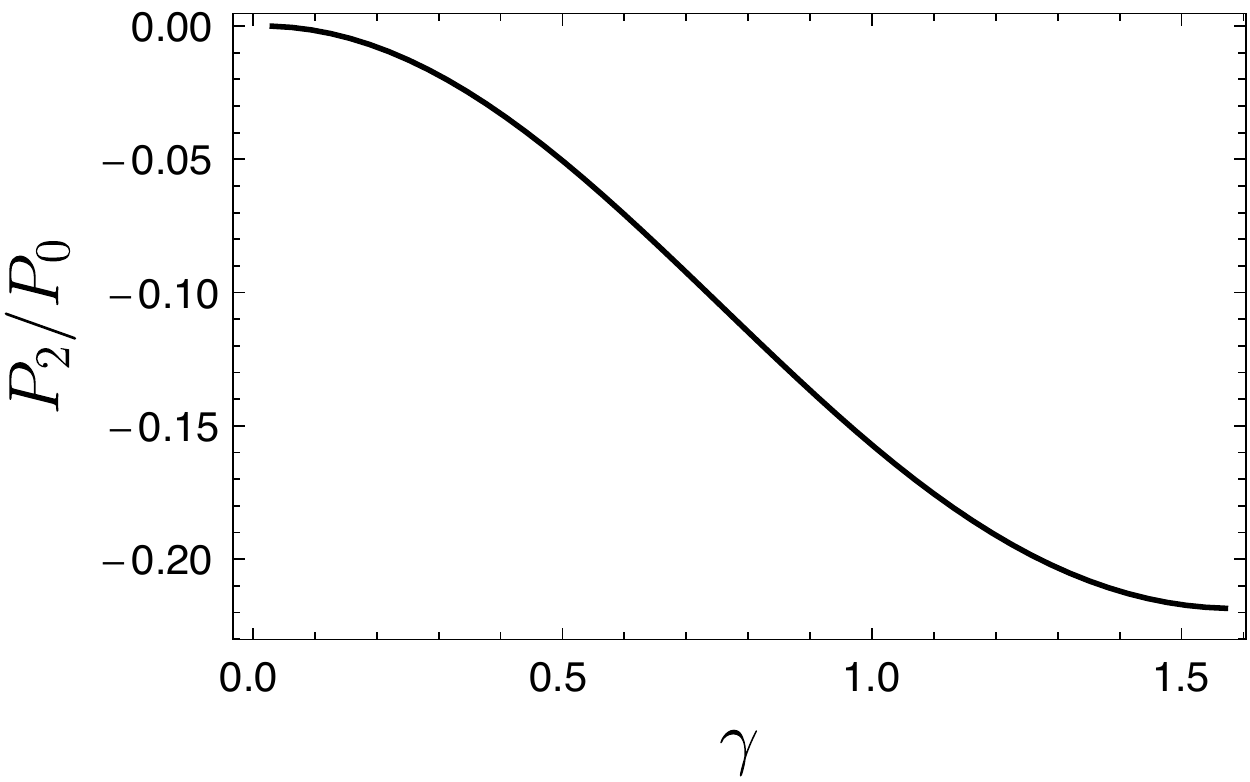}
\caption{High-$\beta$ slow modes at $M_A=0.7$: quadrupole to monopole ratio of power spectrum of intensity fluctuations. Left-hand panel: the observed ratio in the case of negligible dust absorption or at scales $K\gtrsim \Delta_0^{-1}$. Right-hand panel: the observed ratio in the presence of dust absorption at scales $K\lesssim \Delta_0^{-1}$.}
\label{anisslowhb}
\end{figure*}

\begin{figure*}
\centering
\includegraphics[scale=0.5]{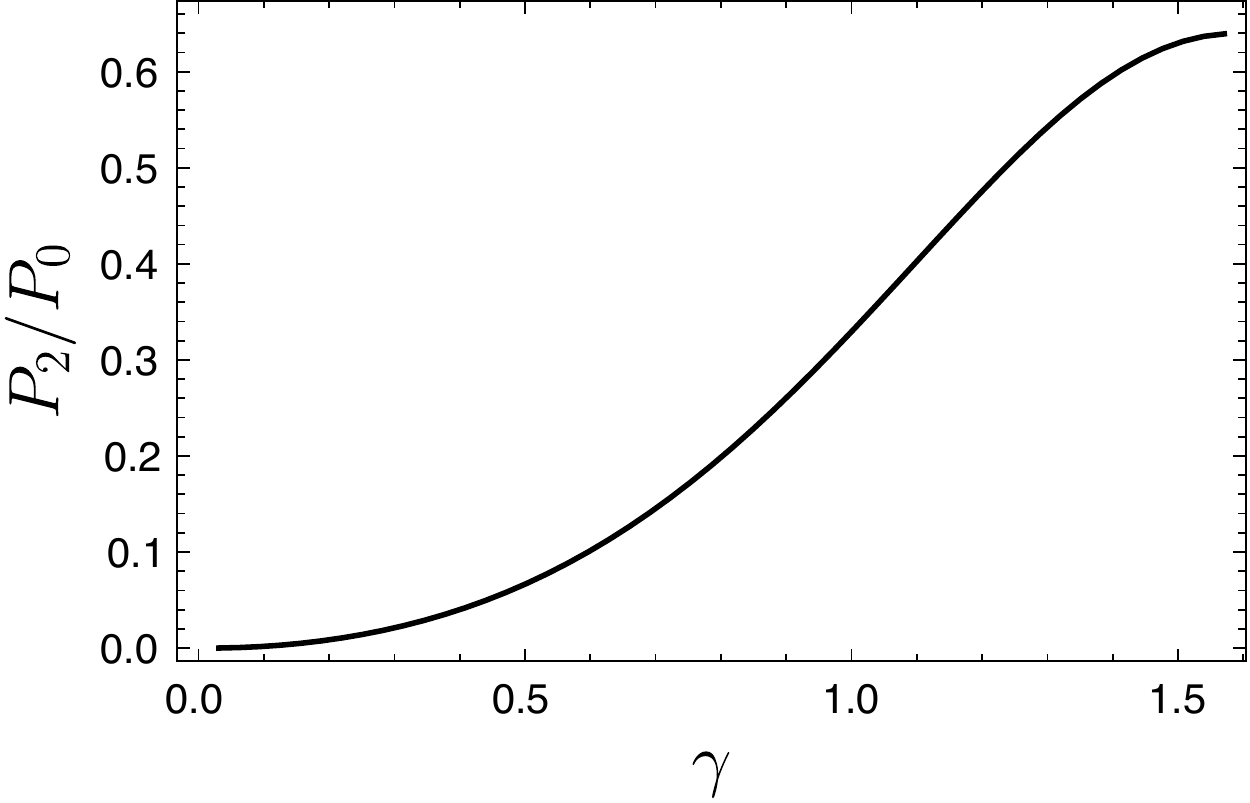}\hspace*{0.2cm}
\includegraphics[scale=0.5]{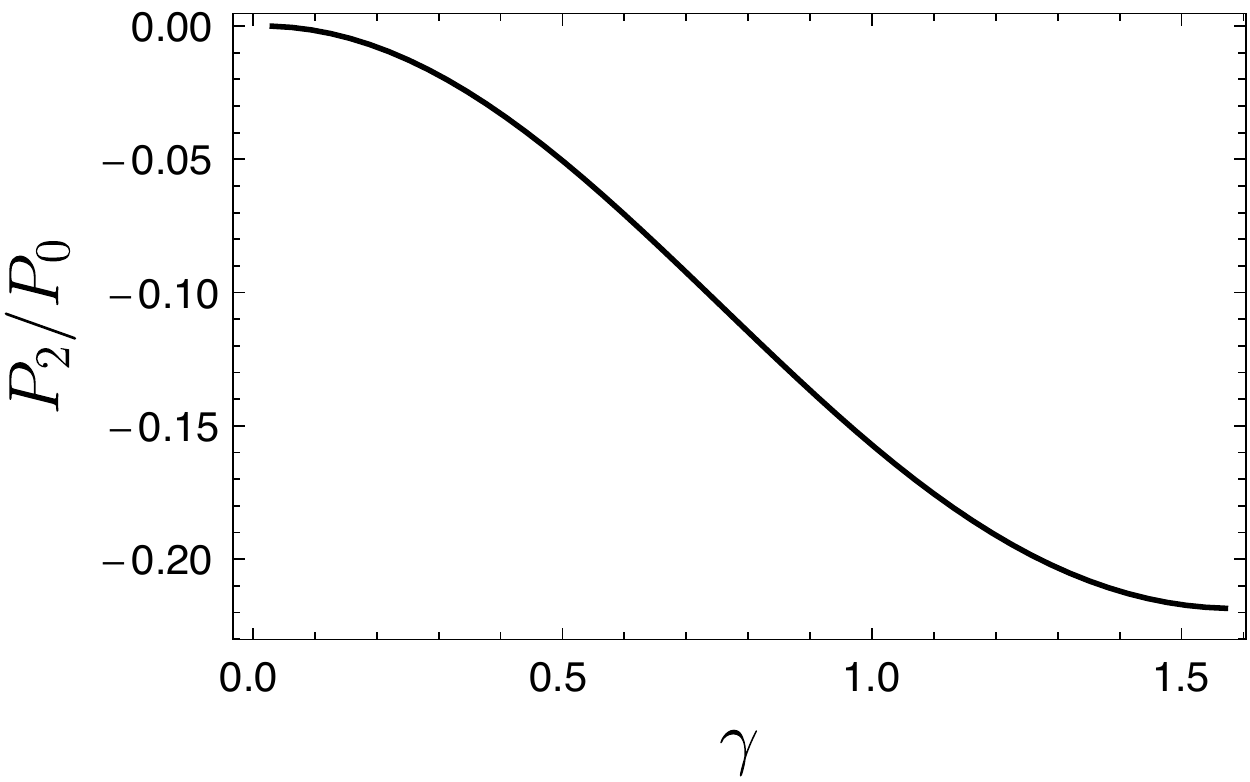}
\caption{Low-$\beta$ slow modes at $M_A=0.7$: quadrupole to monopole ratio of power spectrum of intensity fluctuations. Left-hand panel: the observed ratio in the case of negligible dust absorption or at scales $K\gtrsim \Delta_0^{-1}$. Right-hand panel: the observed ratio in the presence of dust absorption at scales $K\lesssim \Delta_0^{-1}$.}
\label{anisslowlb}
\end{figure*}

First important conclusion is that at large lags $R$, the level of anisotropy is independent of $R$. However, level of anisotropy is highly dependent on Alfv\'en Mach number $M_A$ and $\gamma$. Here, we use their result to predict the level of anisotropy of intensity fluctuations due to various MHD modes.  The quadrupole to monopole ratio of power spectrum of intensity fluctuation in the thin slice regime of VCA has been illustrated for various MHD modes in figures \ref{anisalfven}, \ref{anisfast}, \ref{anisslowhb} and \ref{anisslowlb}. These figures clearly show that the quadrupole to monopole ratio of each of these MHD modes change their sign from small wavenumber $K$ to large $K$. As Alfv\'en modes and low-$\beta$ fast modes do not contribute at $\gamma\sim 0$, while slow modes do not contribute at $\gamma\sim \upi/2$, in a realistic situation involving mixture of different modes, the anisotropy level decreases at large $K$ in comparison to that at small $K$. It is important to stress that the change in anisotropy level from small $K$ to large $K$ is simultaneously accompanied by the change is the spectral slope of the power spectrum. In the presence of self-absorption, it was shown in \citetalias{kandel2016extending} that the anisotropy level in the universal regime is similar to that of the thin slice regime. As our finding in Section \ref{sec:selfab} suggests, dust absorption restricts the range of universal regime, but the anisotropy level in this universal regime is still similar to that of the thin slice regime of negligible dust absorption.

\begin{figure}
\centering
\includegraphics[scale=0.5]{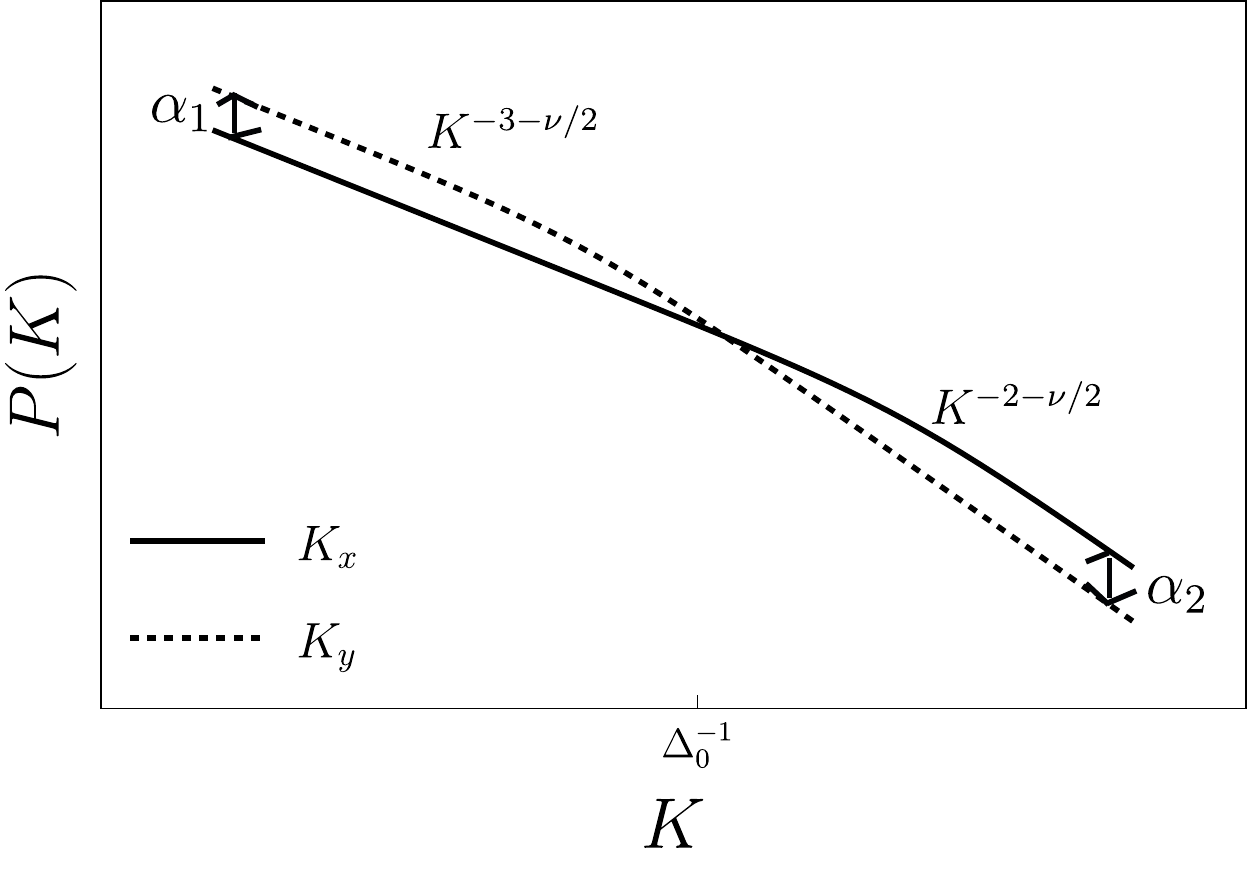}
\caption{Representation of both spectrum and anisotropy for Alfv\'en modes at $M_A=0.7$ and $\gamma=\upi/2$. $\alpha_1$ and $\alpha_2$ represent the isotropy degree in logarithmic scale.}
\label{anisalf}
\end{figure}

\begin{figure*}
\centering
\includegraphics[scale=0.5]{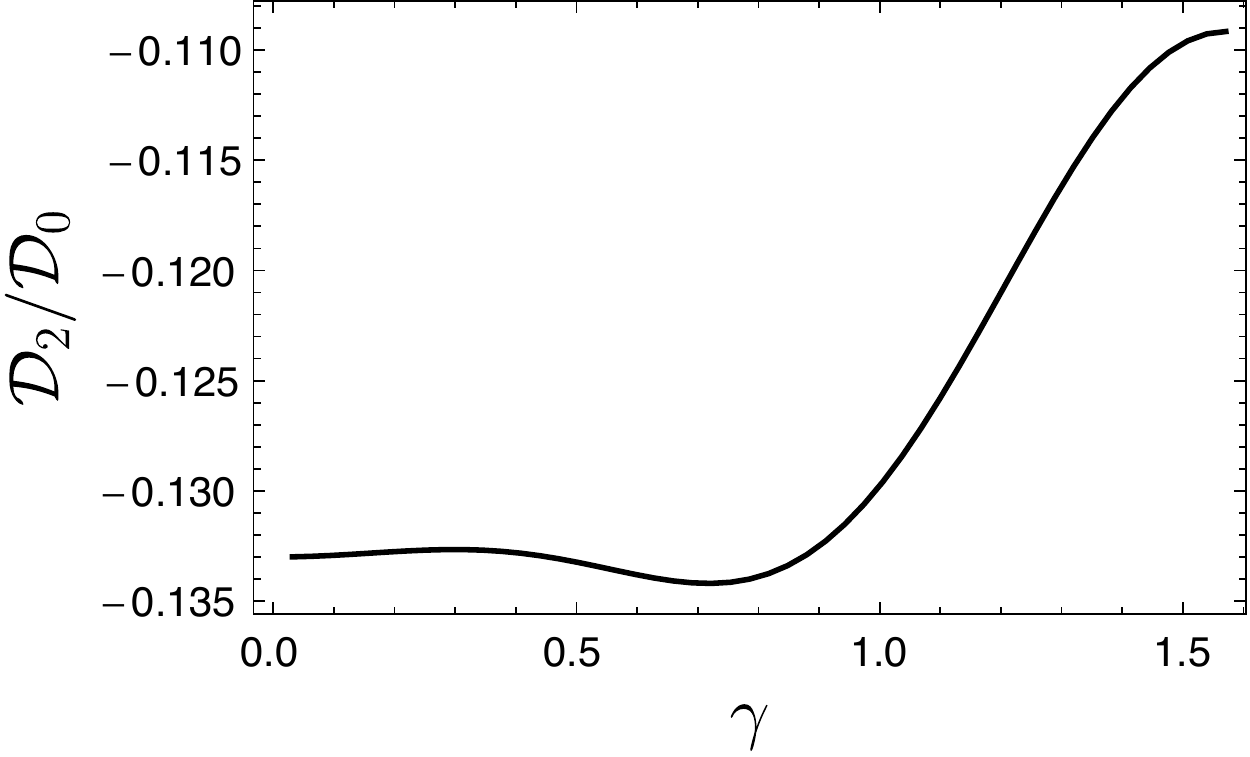}\hspace*{0.2cm}
\includegraphics[scale=0.5]{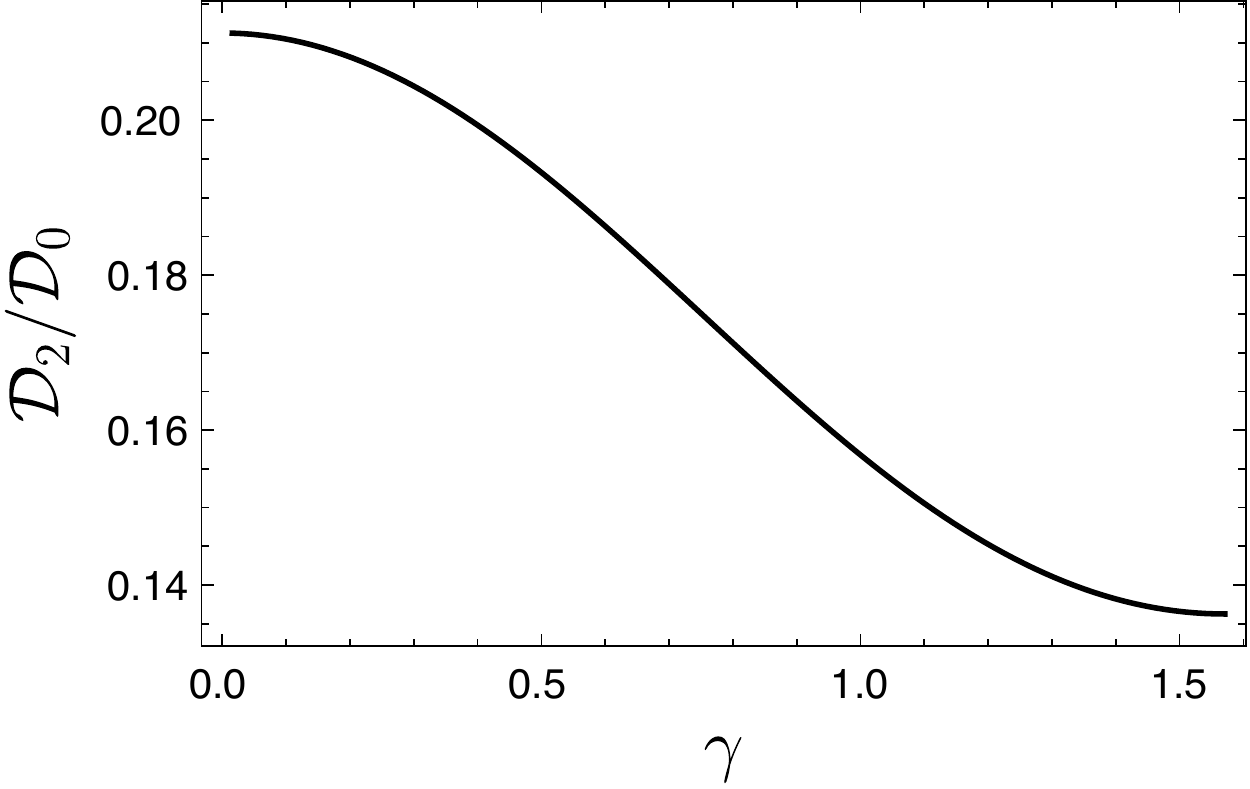}
\caption{Quadrupole to monopole ratio of centroid structure function at scales $R\gtrsim \Delta_0$ where dust absorption is significant. Left-hand panel: Alfv\'en mode at $M_a=0.7$. Right-hand panel: low-$\beta$ fast mode.}
\label{canisalfvenfast}
\end{figure*}

\begin{figure*}
\centering
\includegraphics[scale=0.5]{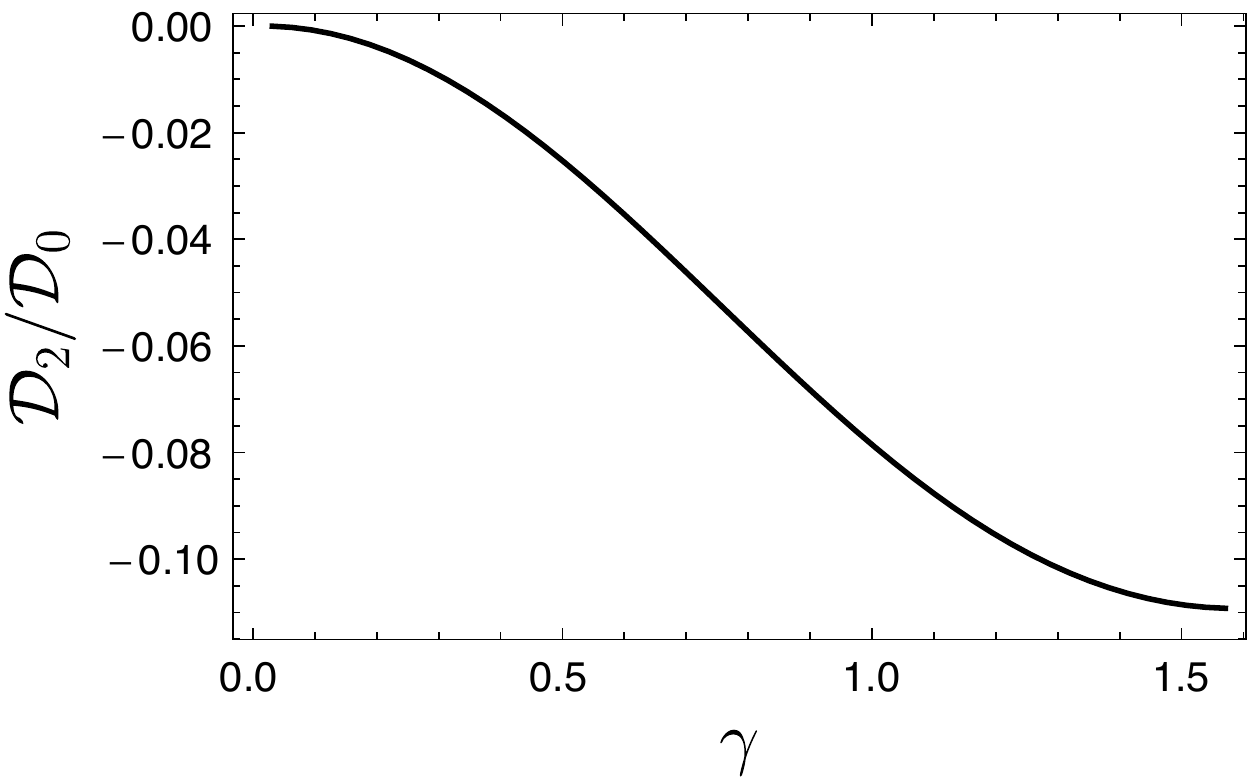}\hspace*{0.2cm}
\includegraphics[scale=0.5]{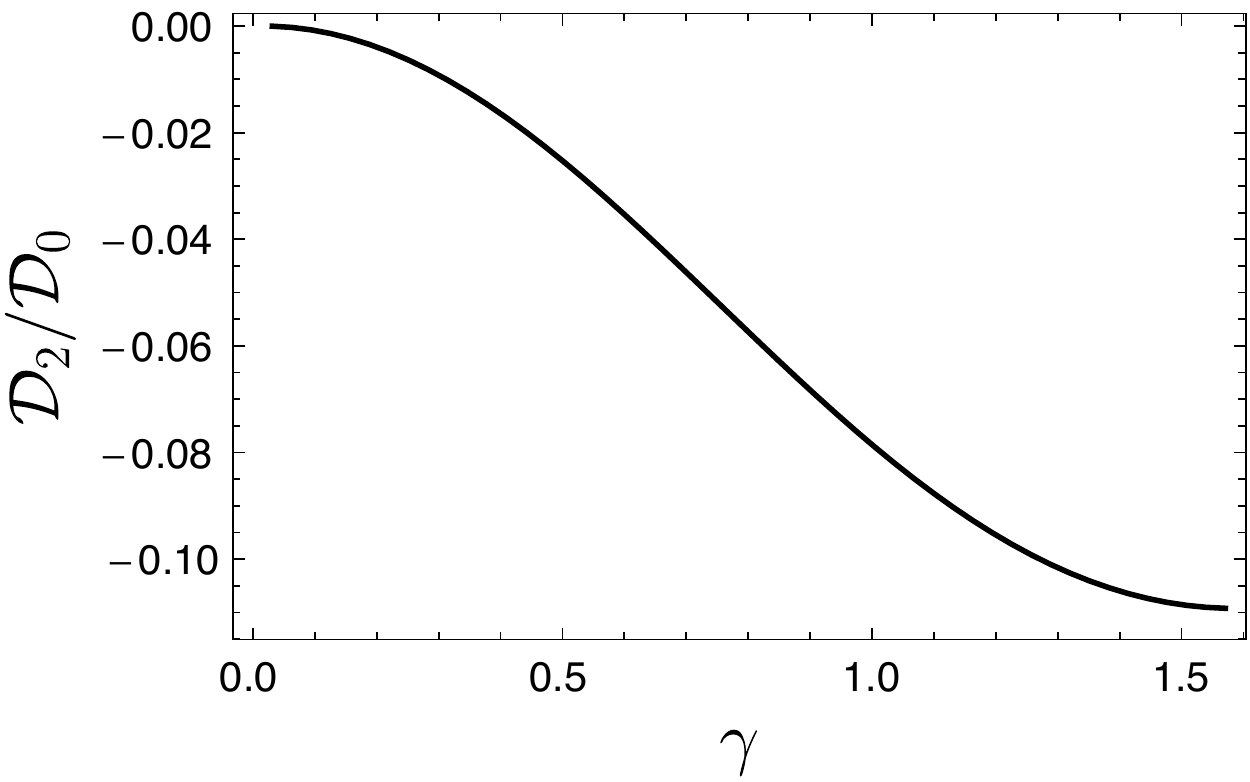}
\caption{Quadrupole to monopole ratio of centroid structure function for slow modes at $M_A=0.7$ at scales $R\gtrsim \Delta_0$ where dust absorption is significant. Left-hand panel: high-$\beta$ slow modes. Right-hand panel: low-$\beta$ slow modes.}
\label{canisslowhb}
\end{figure*}

The discussion carried above applies to anisotropy of centroids as well. At scales $R<\Delta_0$, the anisotropy level of centroids in the presence of dust-absorption is the same as that with the presence of dust-absorption. At $R>\Delta_0$, the centroid structure function is to a good approximation 
\begin{equation}
\mathcal{D}(\bm{R})\propto D_z(\bm{R},0)~,
\end{equation}
and combining this expression with the results of \citetalias{kandel2016extending}, one can easily obtain the anisotropy level of different MHD modes. The anisotropy level of centroid structure function of different MHD modes at $R\gtrsim \Delta_0$, where dust absorption is significant, is shown in figures \ref{canisalfvenfast} and \ref{canisslowhb}. If one compares these anisotropy levels with that in the absence of dust absorption (see \citetalias{Kandel21012017}), it is easy to see that the anisotropy level decreases at these scales in comparison to the one in the absence of dust or at lags $R\lesssim\Delta_0$. It is again important to note that the change of anisotropy level in going from $R\lesssim\Delta_0$ to $R\gtrsim \Delta_0$ is accompanied by the change in the spectral index of centroid structure function. 

\section{Effect of dust on collisionally excited emission lines}\label{sec:collext}
In some interstellar environments, like H\,{\small II} region, the ultra-violet, visible and infra-red spectra of emission lines are primarily collisionally excited lines of metal ions and recombination lines of Hydrogen and Helium. In this section, we study how collisionally excited lines can be used to obtain spectra of density and velocity field in a turbulent medium. 

\subsection{VCA for collisionally excited emission lines}\label{collvca}
The spectral intensity of a collisionally excited emission line in the presence of negligible dust absorption is given by
\begin{equation}
I_v(\bm{X})=\epsilon\int_0^{S}\mathrm{d}z\,\rho_g^2(\bm{x})\Phi_v(\bm{x})~,
\end{equation}
where $\rho_g$ is the density of emitters and $\rho_g^2$ characterises the collision rate. Under the assumption of zero correlation between density and velocity, all the major equations given by VCA is still applicable for collisionally excited emission lines, but the density correlation has to be properly modelled. The main equation for the correlation of intensity is given by
\begin{align}\label{collex}
\xi_I(\bm{R})=\int_{-\infty}^{\infty}\mathrm{d}v\,W(v)\int_0^S\mathrm{d}z\,\frac{\langle\rho_g^2(\bm{x}_1)\rho_g^2(\bm{x}_2)\rangle}{\sqrt{D_z(R,z)}}\nonumber\\
\times\exp\left[-\frac{v^2}{2D_z(R,z)}\right]~,
\end{align}
Eq. \eqref{collex} differs from the usual expression only through the structure of density correlation $\langle\rho_g^2(\bm{x}_1)\rho_g^2(\bm{x}_2)\rangle$. For sub-sonic turbulence, which has steep density spectra, density fluctuations are smaller in comparison to the mean density and therefore breaking density into mean and fluctuating part $\rho_g=\rho_0+\delta\rho_g$ gives
\begin{equation}
\langle\rho^2(\bm{x}_1)\rho^2(\bm{x}_2)\rangle\approx\rho_0^4+4\rho_0^2\xi_\rho(\bm{r})~.
\end{equation}
Thus, at this level of approximation, one returns to the usual predictions of VCA (with emissivity proportional to density of emitters rather than the square of their density). 

The subsequent part of section is focused specifically for supersonic turbulence, which has shallow density spectra, and thus has density fluctuations which are comparable to the mean density. For such density spectra, linear assumption is not reasonable. One way to proceed is assume a particular form of density PDF. Various studies have shown that the density PDF of molecular clouds (where turbulence is supersonic) is lognormal. Therefore, we model dust density $\rho$ as
\begin{equation}
\rho(\bm{x})=\rho_0\mathrm{e}^{-\sigma_{\mu}^2/2}\mathrm{e}^{\mu}~,
\end{equation}
where $\mu$ is a random Gaussian field with zero mean and variance $\sigma_{\mu}^2$ and is described by the two-point probability distribution function given by Eq. \eqref{d3}. Using Eqs. \eqref{eq:linden} and \eqref{eq:linden}, it can be shown that 
\begin{equation}\label{eq:sqdenc}
\langle\rho^2(\bm{x}_1)\rho^2(\bm{x}_2)\rangle=\frac{\langle\rho(\bm{x}_1)\rho(\bm{x}_2)\rangle^4}{\rho_0^4}~.
\end{equation}

We illustrate the above correlation with a specific model. For a shallow density, the correlation
\begin{equation}\label{eq:collexcorr}
\langle\rho^2(\bm{x}_1)\rho^2(\bm{x}_2)\rangle=\frac{\langle\rho(\bm{x}_1)\rho(\bm{x}_2)\rangle^4}{\rho_0^4}\approx \rho_0^2\left(\rho_0^2+4\sigma^2\left(\frac{r}{r_c}\right)^{-\nu_\rho}\right)~,
\end{equation}
for $r>r_c(\sigma/\bar{\rho})^{-2/\nu_\rho}$. Eq. \eqref{eq:collexcorr} represents a form of correlation that is similar to that of density correlation represented in Eq. \eqref{shallowdenx}, with an overall factor and different weight on amplitudes of correlation. Thus, for scales $R>r_c(\sigma/\bar{\rho})^{-2/\nu_\rho}$, all analytical expression derived in VCA are applicable to collisionally excited lines.

In the presence of dust absorption, the intensity of collisionally excited line is given by
\begin{equation}
I_v(\bm{x})=\epsilon\int_0^S\mathrm{d}z\,\rho^2(\bm{x})\Phi_v(\bm{x})\mathrm{e}^{-\kappa\int_0^z\mathrm{d}z'\,\rho}~.
\end{equation}

If density fluctuations are smaller than the mean, i.e. $\delta\rho<\bar{\rho}$, the factor $\rho^2$ can be approximated as $\rho^2\approx\bar{\rho}^2+2\bar{\rho}\delta\rho$. At this level of approximation, all the results that were derived in Sec. \ref{vcadust} are applicable to collisionally excited emission lines as well. Therefore, for sub-sonic turbulence (for which density fluctuations are smaller than the mean), the results of Sec. \ref{vcadust} are applicable. 

For supersonic turbulence, density fluctuations are usually comparable to the mean density. To obtain an appropriate model of intensity correlation can be worked out for strong dust absorption regime, we carry out integration by parts of the spectral intensity. We do this for thick and thin velocity slice regimes separately. 

In the thick velocity slice regime, by integrating over entire range of velocity, one obtains
\begin{equation}\label{eq:incorcol}
I(\bm{x})=\epsilon\int_0^S\mathrm{d}z\,\rho^2(\bm{x})\mathrm{e}^{-\kappa\int_0^z\mathrm{d}z'\,\rho}~.
\end{equation}
At this point, it is instructive to rewrite Eq. \eqref{eq:incorcol} as 
\begin{equation}
I=-\frac{\epsilon}{\kappa}\int_0^S\mathrm{d}z\,\rho(\bm{x})\frac{\mathrm{d}}{\mathrm{d}z}\mathrm{e}^{-\kappa\int_0^z\mathrm{d}z'\,\rho}~,
\end{equation}
which upon integration by parts gives
\begin{align}\label{intcorcol}
I&=\frac{\epsilon}{\kappa}\left(\rho(\bm{X},0)-\rho(\bm{X},S)\mathrm{e}^{-\kappa\int_0^S\mathrm{d}z'\,\rho(\bm{X},z')}\right)\nonumber\\
&\,\,\,\,+\frac{\epsilon}{\kappa}\int_0^S\mathrm{d}z\frac{\mathrm{d}\rho(\bm{X},z)}{\mathrm{d}z}\mathrm{e}^{-\kappa\int_0^z\mathrm{d}z'\,\rho(\bm{X},z')}\nonumber\\
&\approx \frac{\epsilon}{\kappa}\rho(\bm{X},0)+\frac{\epsilon}{\kappa}\int_0^\Delta\mathrm{d}z\frac{\mathrm{d}\rho(\bm{X},z)}{\mathrm{d}z}\nonumber\\
&=\frac{\epsilon}{\kappa}\rho(\bm{X},\Delta)~,
\end{align}
where $\Delta$ is the physical depth where optical depth reaches unity. The statistics of $\Delta$, including its mean $\bar{\Delta}$ and variance $\sigma_\Delta$, is presented in Appendix \ref{deltacutstat}. Comparing Eqs. \eqref{intcorcol} and \eqref{eq:Idensat}, one can see that even in the case of strong dust absorption, density effects do not get washed away for collisionally excited lines.

Using Eq. \eqref{intcorcol}, the correlation function of intensity correlation can be written as
\begin{equation}
\xi_I(\bm{R})=\frac{\epsilon^2}{\kappa^2}\langle\rho(\bm{X}_1,\Delta(\bm{X}_1))\rho(\bm{X}_2,\Delta(\bm{X}_2))\rangle.
\end{equation}
The evaluation of the above correlation is tricky due to the fact that $\Delta$ is itself a function of density. For fluctuating density, $\Delta$ across different lines of sight are different. The overall effect of fluctuating density is therefore smearing of the correlation. This effect can be captured by evaluating the average of density correlation function over a range of $\Delta$ given by $[\bar{\Delta}-\sigma_{\Delta}/2, \bar{\Delta}-\sigma_{\Delta}/2]$ such that
\begin{equation}
\xi_I(\bm{R})=\frac{\epsilon^2}{\kappa^2}\frac{1}{\sigma_{\Delta}}\int_{-\sigma_{\Delta}/2}^{\sigma_{\Delta}/2}\mathrm{d}z\,\left[\bar{\rho}^2+\xi_\rho(R, z)\right]~.
\end{equation}
Clearly, if $\sigma_\Delta$ is large, the correlation gets smeared out, while for small $\sigma_\Delta$, the intensity correlation function directly gives the density correlation function of approximately a two-dimensional thin slab.  In particular, one can obtain useful information about the correlation function of density by studying correlation function of intensity at lags $R$ given by
\begin{equation}
R>\sigma_{\Delta}~,
\end{equation}
while at shorter lags, the correlation is saturated.

Using similar idea as presented above, one can readily show that under the assumption of uncorrelated density and velocity field, the intensity correlation function in the thin-slice regime in the presence of strong dust absorption is
\begin{equation}
\xi_I(\bm{R})=\frac{\epsilon^2}{\kappa^2}\frac{1}{\sigma_{\Delta}}\int_{-\sigma_{\Delta}/2}^{\sigma_{\Delta}/2}\mathrm{d}z\,\frac{\bar{\rho}^2+\xi_\rho(R, z)}{\sqrt{D_z(R, z)}}~,
\end{equation}
where $D_z(R, z)$ is the velocity structure function. Unlike our previous discussion of the case which excluded collisionally excited lines, we see that in this case the density effects are not erased even at large lags. Again, for small $\sigma_\Delta$, one obtains correlation of intensity of approximately two-dimensional thin physical slab. 

\subsection{VCS for collisionally excited emission lines}
For collisionally excited emission lines, the power spectrum of intensity fluctuations along velocity coordinate in the presence of dust-absorption is given by 
\begin{align}\label{powerVCS1col}
P(k_v)=\mathrm{e}^{-\beta k_v^2}\int\mathrm{d}^2\bm{R}B(\bm{R})\int_0^{\Delta_{\text{av}}}\mathrm{d}z\,\xi_{\rho^2}(\bm{R}, z)\nonumber\\
\exp\left[-\frac{k_v^2D_z(\bm{R}, z)}{2}\right]~,
\end{align}
where $\xi_{\rho^2}(\bm{r})\equiv \langle\rho^2(\bm{x}_1)\rho^2(\bm{x}_2)\rangle$. The difference between this case and the usual VCS comes through the replacement of $\xi_\rho(\bm{r})$ by $\xi_{\rho^2}(\bm{r})$. However, as it was shown in the previous section, the form of $\xi_{\rho^2}(\bm{r})$ looks structurally similar to that of $\xi_{\rho}(\bm{r})$ for both steep and shallow spectra. Therefore, all the results of VCS from Sec. \ref{svcs} apply to collisionally excited emission lines. In particular, the extent of dust cut-off sets the range of wavenumbers where one can achieve narrow beam, and this range is given by Eq. \eqref{narrowrange}. For small dust cut-off scales, achieving narrow beam is difficult. All the relevant asymptotics are the same as in the usual VCS, and are given by Eqs. \ref{narrow} and \ref{wide}.

\subsection{Centroids for collisionally excited emission lines}
The centroids were shown to be useful for studies of subsonic turbulence. As it was mentioned earlier, density fluctuations are smaller than mean density for subsonic turbulence. Therefore, all the results derived in Sec. \ref{centroidsdust} is applicable even for collisionally excited emission lines.

\section{Comparison with observations}\label{obs}
Several studies (see \citet{munch1971distribution}) suggest that the Orion Nebula is dusty. The study of the reddening of the brightest region of the Orion Nebula as well as the behaviour of the He I lines $2^3P\rightarrow 2^3S$ and $3^3S\rightarrow 2^3P$ in \citet{munch1971distribution} suggests that the dust and gas in the Orion are well mixed. Thus, the results of this paper are applicable to this nebula. More importantly, ionised gas usually leads to collisionally excited emission lines. In this paper, we have discussed how VCA works with collisionally excited emission lines and shown that the previous analytical results are applicable even for the collisionally excited emission lines. The study of turbulence in ionised Orion Nebula has been carried out by \citet[hereafter \citetalias{arthur2016turbulence}]{arthur2016turbulence} by analysing observational data through VCA and centroids. Thus, our results validate the calculations performed by \citetalias{arthur2016turbulence} in order to find spectral index of velocity and density fields. However, one important point to note is that for strong dust absorption, density effects are suppressed. Therefore, this changes the result for shallow density spectra. 

\begin{figure*}
\includegraphics[scale=0.4]{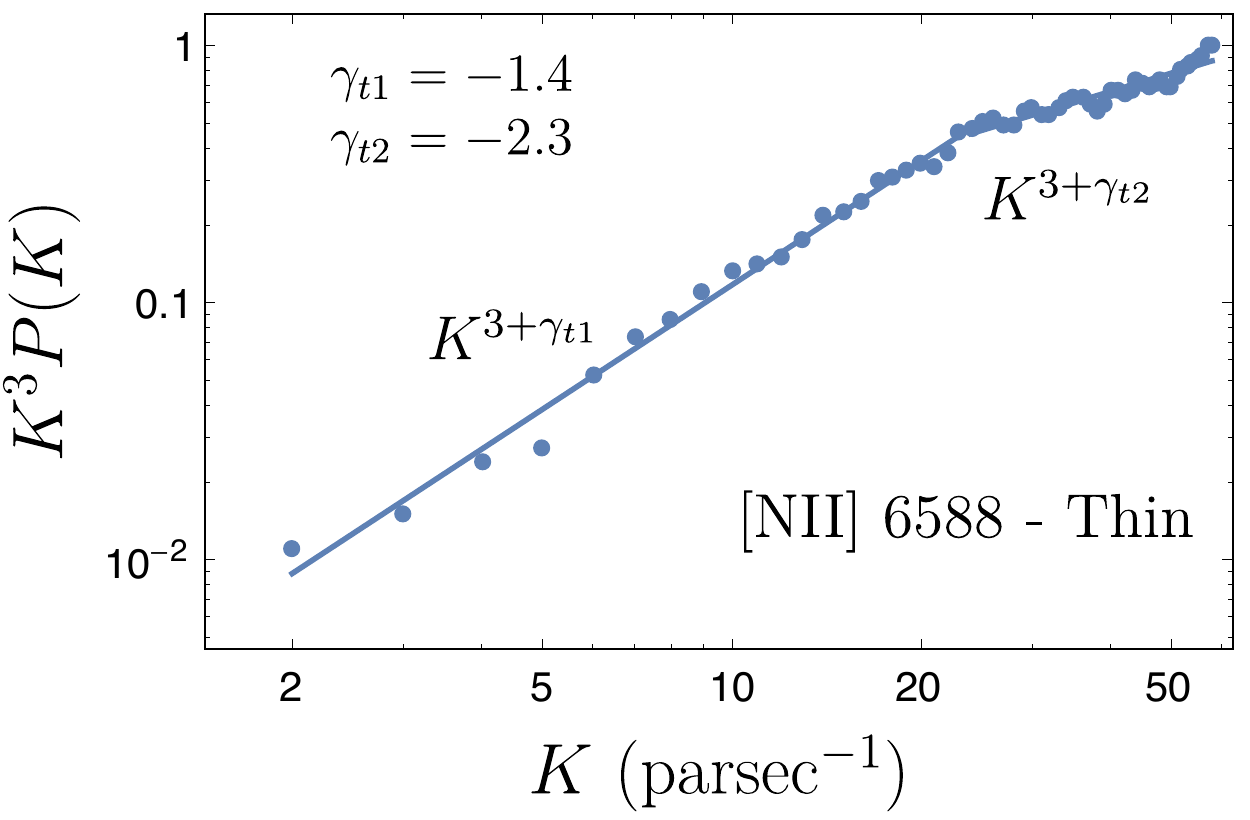}\hspace*{0.1cm}
\includegraphics[scale=0.4]{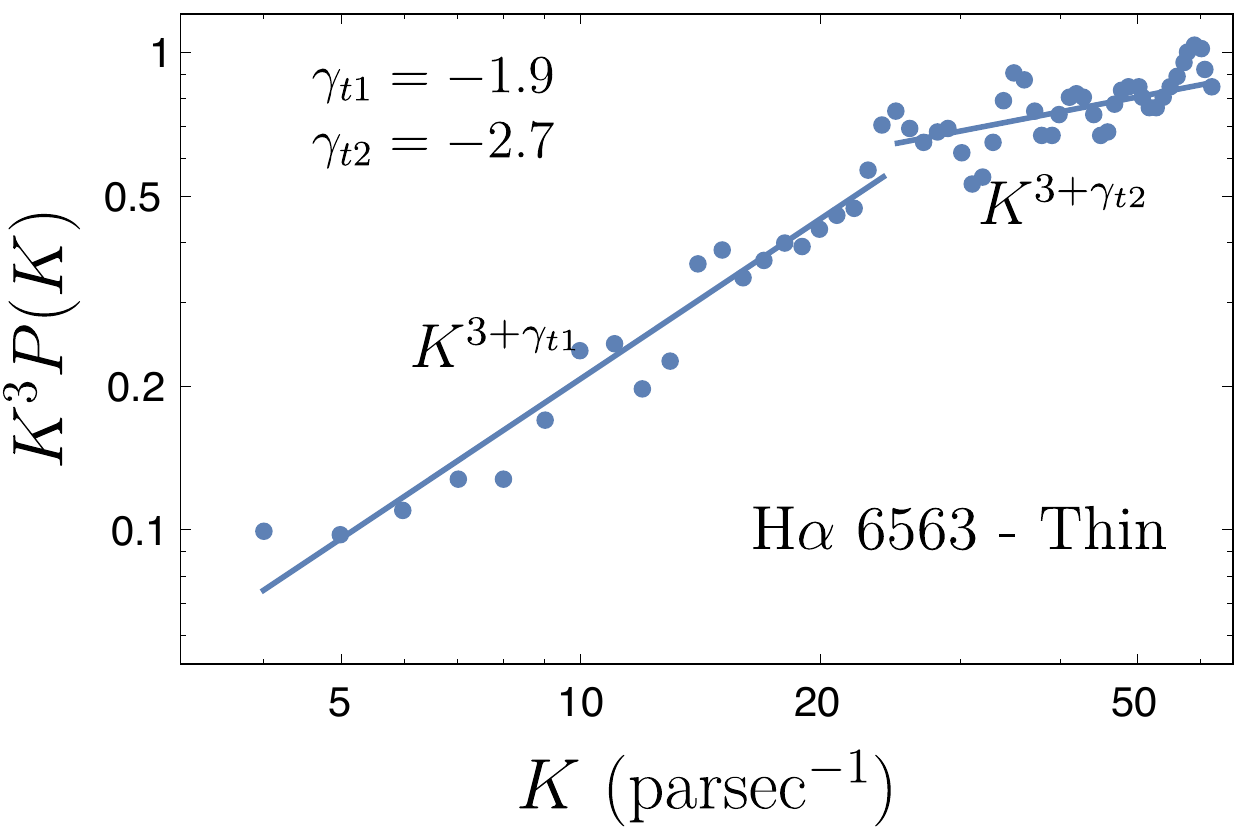}\hspace*{0.1cm}
\includegraphics[scale=0.4]{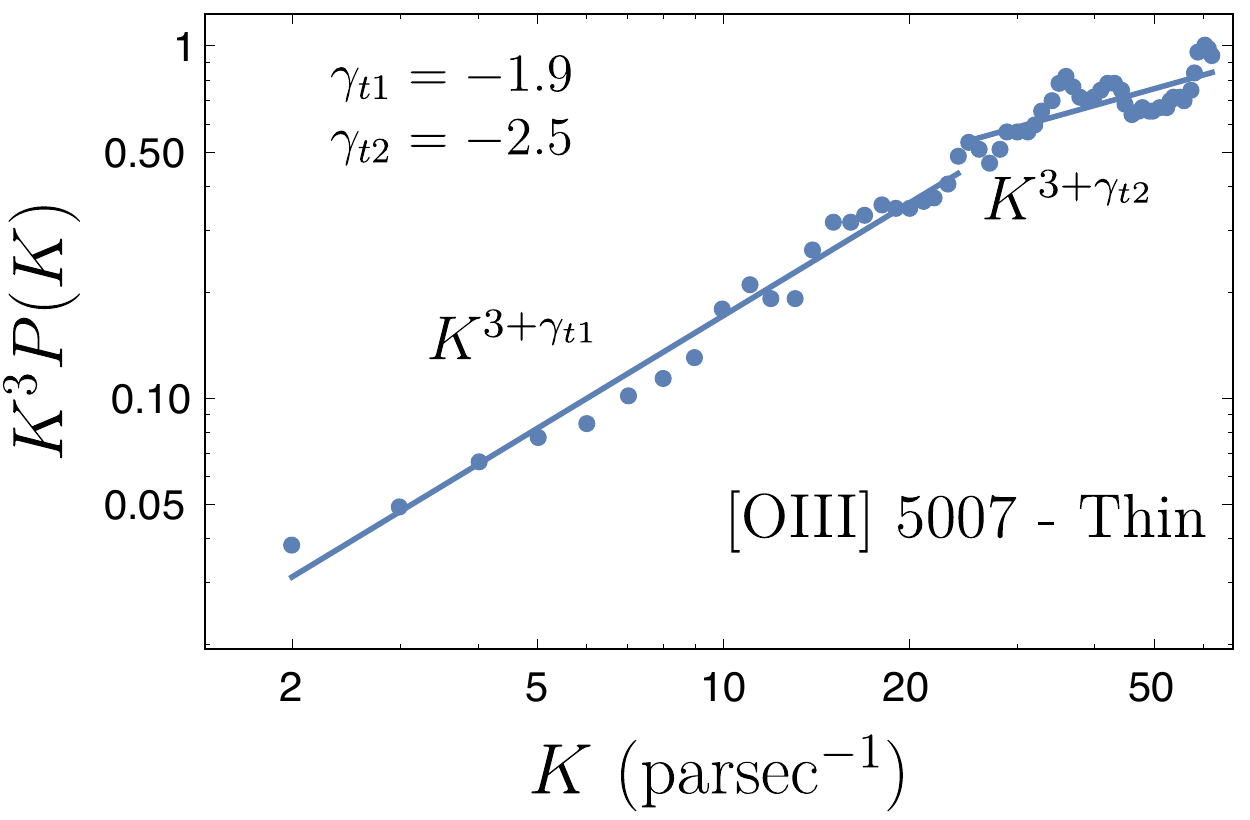}
\vspace*{0.2cm}
\includegraphics[scale=0.4]{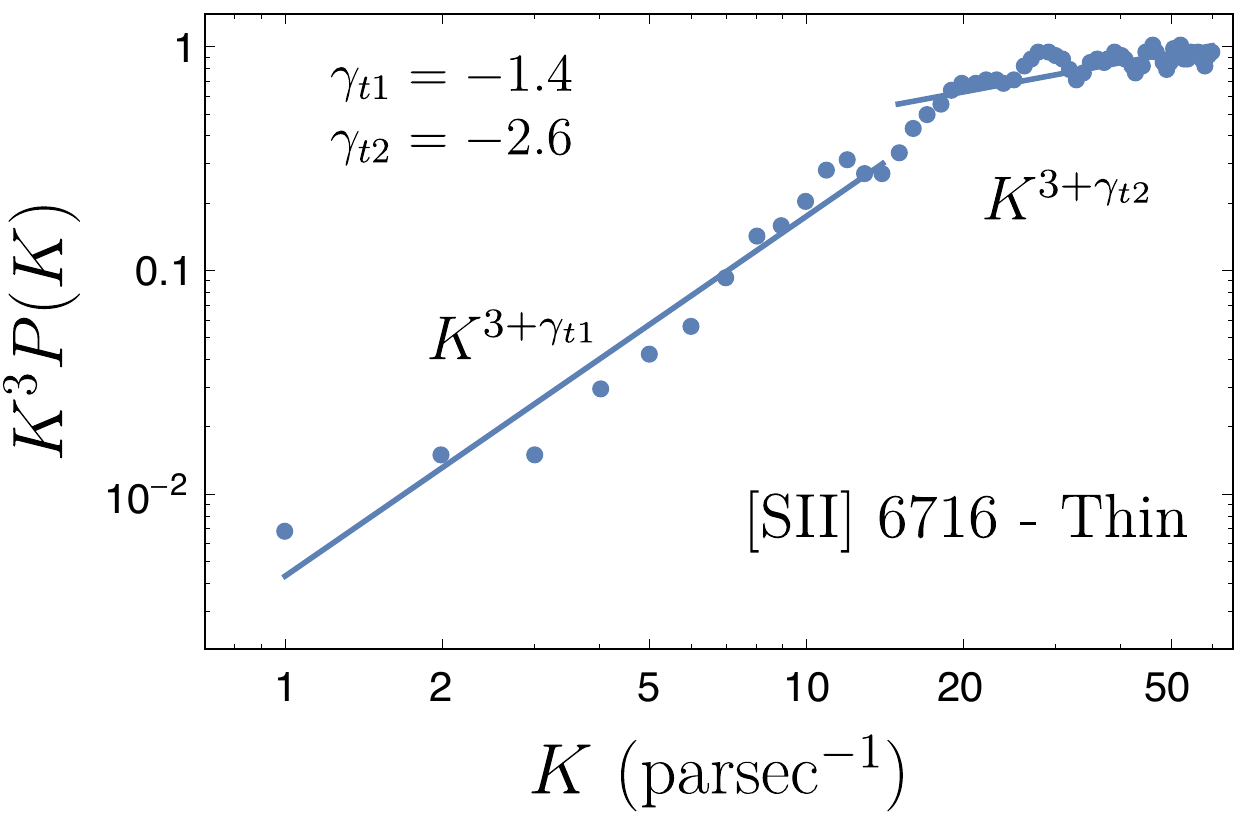}\hspace*{0.1cm}
\includegraphics[scale=0.4]{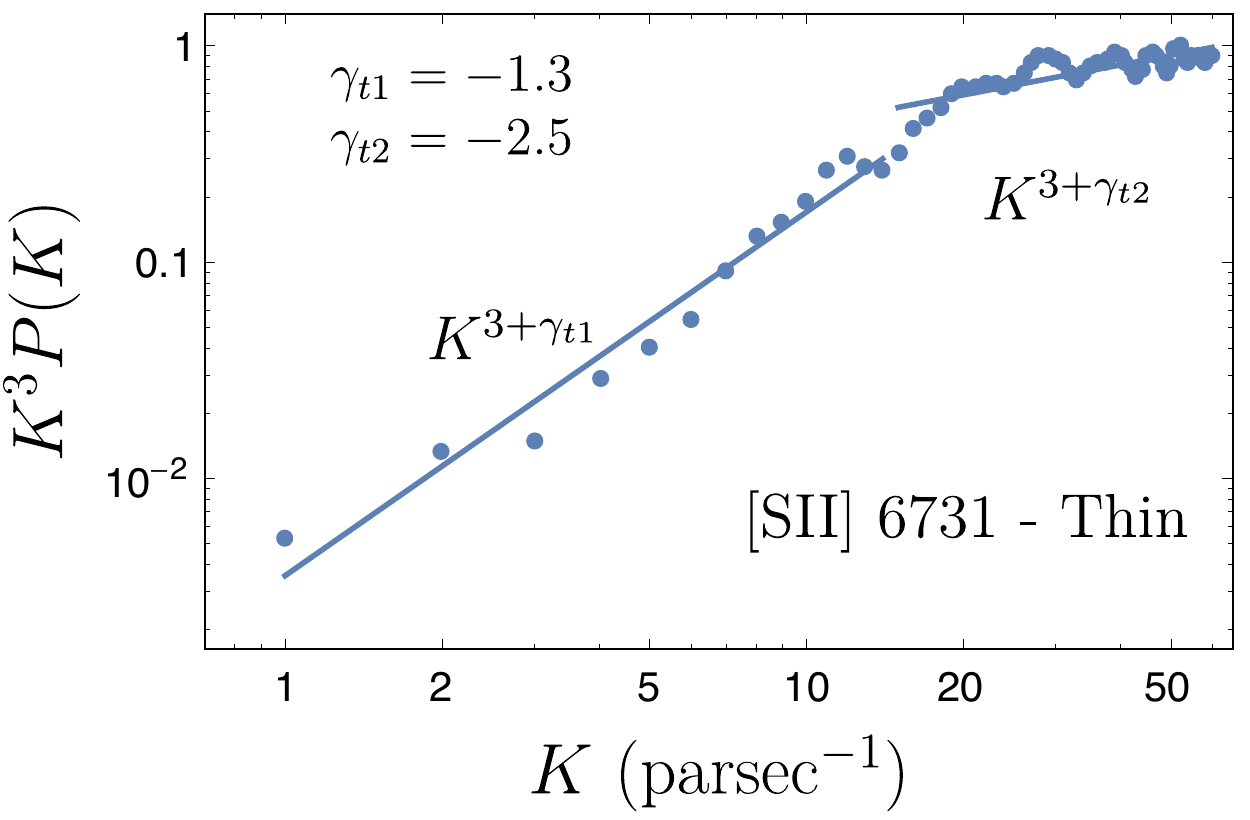}\hspace*{0.1cm}
\includegraphics[scale=0.4]{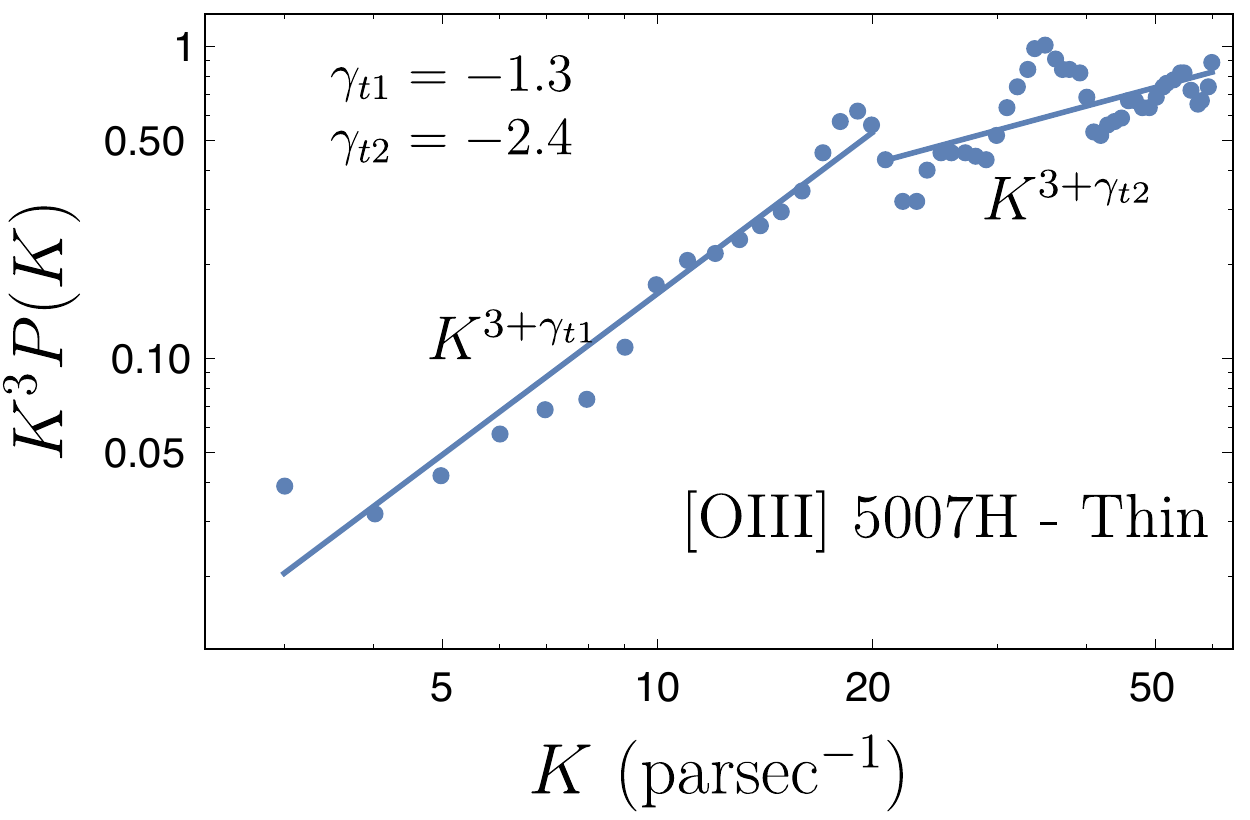}
\caption{Compensated power spectra $K^3P(K)$ of intensity fluctuations for the thin velocity slice case for different spectral lines. The two regimes with different slopes are shown, with corresponding best fit lines. One possible reason for the break in the slope is due to dust extinction. Data points are obtained from \citetalias{arthur2016turbulence}.}
\label{fig:obser1}
\end{figure*}

Of the results presented in \citetalias{arthur2016turbulence}, the most relevant ones for comparison with our prediction in the power spectrum of intensity fluctuations of  various spectral lines in the {\it thin velocity slice} regime. In this regime, our prediction was that the power spectrum of intensity fluctuations exhibits a break in the slope, and the breakpoint is characteristic to the mean strength of dust absorption. Moreover, the change in slope between the two regimes should be of unity.  We used data points obtained from \citetalias{arthur2016turbulence} to perform fit of the largest scales (smallest $K$), which was missing in \citetalias{arthur2016turbulence}. The so obtained plots for various spectral lines are shown in Fig. \ref{fig:obser1}. Note that we do not include any dispersion in the slopes of power spectrum, unlike in \citetalias{arthur2016turbulence}, where dispersion in slopes were from the different samplings of the PPV slits. Also notice the difference in Fig. \ref{fig:obser1}, where $K^3P(K)$ is plotted against $K$ and our schematic representation in Fig. \ref{powerspec}, where $P(K)$ is plotted against $K$.

As shown in Fig. \ref{fig:obser1}, clearly, there is a break in the spectral slope, and the difference in the slope is close to unity for all the spectral lines. Moreover, the characteristic break point for all the spectral lines is of order $\sim 20$ pc$^{-1}$, which corresponds to the dust-absorption cut-off $\Delta_0\sim 0.05$ pc. For a medium ionization zone of Orion nebula, whose markers are [O III], H II, He I and [Cl III], \citealt{2001RMxAC..10....1O} quote that the optical depths (which should be of similar to dust-absorption cut-off) is $\sim 0.06$ pc, totally consistent to our $\Delta_0\sim 0.05$ pc. Thus, one possible reason for this break is the dust extinction. Notice that, in Fig. \ref{fig:obser1}, different spectral lines have different break points, which is naturally expected as different spectral lines have different $\kappa$ values. Also notice that some of the markers in Fig. \ref{fig:obser1}, [N II] marks low ionization zone and [S II] marks ionization front. \citet{1994Ap&SS.216..267O, 2001RMxAC..10....1O};  \citet{1994ApJ...436..194O} find that the ionization front has an optical depth of $10^{-4}$ pc and low ionization zone $2\times 10^{-3}$ pc. Our $\Delta$ for these spectral lines seems to be different from the results of \citet{1994Ap&SS.216..267O, 2001RMxAC..10....1O}; and \citet{1994ApJ...436..194O}, thus it is difficult to assert if the break in the slopes for these spectral lines is due to dust-absorption.

Attribution of the break to dust absoprtion can also be affected by the presence of foreground extinction, and the finite thickness of the ionization layer (see \citetalias{arthur2016turbulence} for more details on this). In particular, if the thickness of the ionization layer is less than the physical depth at which optical depth reaches unity, then it would not be correct to conclude that the break is due to dust-absorption. 

\section{Discussion}\label{sdiscuss}

\subsection{Important results and their interpretation}
In this paper, we showed how dust absorption affects different statistical methods to study turbulence. Our results suggest that VCA, VCS and velocity centroids are able to trace velocity spectra even in the presence of dust absorption. However, it might not be possible to obtain density spectra with these techniques, especially if dust absorption is strong and the resolution of telescope is poor.  

In the presence of dust absorption, power spectrum of intensity fluctuations in the thin-slice regime has spectral index greater than -2 at small $K$, while at large $K$, it has spectral index less than -2. Thus it is appropriate to use correlation function at large $R$, while structure function is appropriate at small $R$. Our results suggest that density fluctuation is unimportant if dust absorption is strong as one effectively samples signals from only optical optical depth of unity, or equivalently from a fixed column-density. This allows one to obtain cleaner velocity statistics at larger scales. To study velocity and density spectra, one can also use power spectrum of intensity fluctuations. In the language of power spectrum of the intensity fluctuations, the spectral slope changes by unity\footnote{This is true for only if density spectra is steep or if dust absorption is strong. For shallow density in the weak absorption limit, spectral slope changes by $1-\nu_\rho$ (see Sec. \ref{sec:thin}).}, at the critical break point $K_c = \Delta_0^{-1}$. This is a manifestation of the fact that at the lags $R$ associated with these wavenumbers, the depth along LOS is much smaller than the separation between the lines of sight, thus statistics are effectively two dimensional. The break in the slope near dust cut-off scale is also exhibited by centroids. Thus, study of velocity spectra is possible using both centroids and thin velocity slice regime of VCA even in the presence of strong dust absorption, albeit intensity signal will be diminished if the dust absorption is strong. 

On the other hand, one can only study density spectra if absorption is not strong by  either using structure function at short scales $R$ or by using correlation function at  large scales $R$. Our calculations show that the correlation length set by dust absorption is short for shallow density and long for steep density. Thus, it is appropriate to study steep density using structure function, while correlation function is appropriate for shallow density. An important limitation is that the amplitude of correlation function is exponentially suppressed, and thus it is challenging to measure this correlation especially if instrumental noise is not properly factored out. We stress that in the thin slice regime of VCA, density effects are not important if dust absorption is strong, and only velocity statistics is manifested in the intensity statistics. This allows one to cleanly obtain velocity spectra even if the information of density spectra is not available.

Our results in this paper show that dust absorption {\it effectively} decreases the extent of self-absorption. This means that the universal regime $K^{-3}$ of intensity statistics, originally obtained in \citetalias{lazarian2004velocity}, is suppressed by strong dust absorption; in particular, the range of scale where one can see universal regime is decreased by a factor of $\Delta_0/S$. This is due to the fact that the cut-off in physical space introduced by dust absorption decreases the velocity window introduced by self-absorption, thus suppressing the universal regime. 

Our results also show that one can study intensity anisotropies even in the presence of dust absorption. However, in the context of VCA, dust absorption might lead to isotropization of statistics if proper statistical measure is not used. In particular, dust absorption leads to isotropic structure function at lags $R>\Delta_0$, where one needs to use correlation function in order to study both spectrum and anisotropy of intensity fluctuations. Similarly, one needs to study both spectrum and anisotropy at small $R$ for thick velocity slice limit of steep density. This ambiguity of choice of statistical measure is resolved if one studies anisotropy using power spectrum of intensity fluctuations. We stress that our description of anisotropy of interstellar turbulence uses the theory of strong non-linear MHD turbulence. The theory, including the issues of anisotropies, has been studied extensively both theoretically and numerically, in particular in the papers by \citet{cho2002compressible,  cho2003compressible} and \cite{kowal2010velocity}. In these papers the anisotropies of MHD turbulence were studied for a wide range of sonic and Alfv\'en Mach numbers. These anisotropies are not expected to depend on the ionization of the media, provided that the scales under study are larger than the scale of the decoupling of ions and neutrals (see \citealt{xu2016damping}). Therefore, our description of anisotropies applies to both turbulence within ionized gas in massive star forming regions and diffuse multiphase interstellar medium. Moreover, we study the anisotropies arising from the sub-Alfv\'enic and trans-Alfv\'enic turbulence.

In the context of VCS, we showed that both narrow beam and wide beam asymptote remain the same even in the presence of dust absorption. However, an important consequence of dust absorption is that the effective width of telescope beam $\Delta B$ increases, and one will be able to obtain narrow beam only if this width is less than the dust cut-off, i.e. $\Delta B<\Delta_0$. This might be difficult to achieve if dust absorption is strong.

Finally, an important piece of work that was carried out was understanding how different analytical techniques work with collisionally excited emission lines. Our results show that overall, all the analytical expressions and asymptotic predictions obtained in the previous studies are applicable to collisionally excited emission lines. However, an interesting regime is when dust absorption is present. In this case,  density effects are not suppressed for collisionally excited emission lines even when dust absorption is strong. 

In conclusion, the previous techniques of studying turbulence applied to mostly radio lines. Our study suggests that these techniques can be applied even to the optical and ultra-violet (UV) lines for which dust absorption can not be neglected.

\subsection{Model Assumptions}
In carrying out the analysis in this paper, we used several assumptions. Firstly, we considered the case when emission is proportional to the density of emitters, which excludes the case of collisionally excited emission, where emission proportional to the  square of density of emitters. Secondly, we assumed dust density to be proportional to the gas density. Correlation between HI and 100 micron (see \citealt{boulanger1996dust}) justifies  this assumption for the gas clouds in the ISM.  Thirdly, we assumed that the dust extinction coefficient $\kappa$ is independent of frequency within the frequency range of Doppler broadened spectral line. While this is not strictly true, it does not affect our analysis in any significant manner, especially in the case when thermal broadening of intensity profile is not significant. It is important to note that the dust extinction coefficient is different for different emission lines. 

Although density is better modelled as log-normal field, we argued that the column density can be approximated with reasonable accuracy to be Gaussian. For steep density, the density dispersion is small, therefore density and thus column-density can be modelled as Gaussian. For shallow density, we showed in Appendix \ref{appcolumn} that if the LOS depth is larger than the correlation length of the field, column-density tends to be Gaussian.

To study anisotropy of intensity correlations, we explicitly used models of different MHD modes, and obtained anisotropy for each of these modes. Description of turbulence in terms of MHD cascades is only applicable if these MHD modes approximately form  independent cascade. Numerical study carried out by \citet{cho2002compressible} shows that the coupling between various MHD modes is marginal as long as sonic Mach number is not too high, thus validating the applicability of this decomposition. Some possibilities of the validity the MHD modes in describing turbulence in the diffuse ISM has been questioned by \citet{caldwell2017dust} by considering the recent dust polarization result obtained by Planck observations. By assuming MHD model, \citet{caldwell2017dust} have shown that there is a very narrow range of parameters that could explain the Planck's result. Thus, they raise the possibility that at large scales ($\sim$ pc), turbulence could either be unimportant, or the ISM is not described by MHD model that is used. However, our ongoing work shows that the description of dust polarization using anisotropic MHD model is not inconsistent with the Planck result. 

\subsection{Other techniques to study turbulence}
In this paper, we mainly focused on how dust absorption affects VCA, VCS and centroids. However, there are other techniques to study turbulence, and these are also likely to be affected by dust absorption. An important technique to study turbulence using velocity slice of PPV space is the spectral correlation function (SCF, see \citealt{rosolowsky1999spectral}). SCF is very similar to VCA if one removes the adjustable parameters from it, as both of these technique measure correlations of intensity in velocity slices of PPV.  However, SCF treats outcomes empirically. The description of how intensity is affected by dust absorption that was presented in this paper is likely to help us understand the effect of dust absorption on SCF.

There also exist numerous techniques to identify and analyze clumps and shells in PPV (see \citealt{houlahan1992recognition}; \citealt{williams1994determining}; \citealt{stutzki1990high}; \citealt{pineda2006complete}; \citealt{ikeda2007survey}). Moreover, a more advanced technique to study a hierarchical structure of the PPV, namely dendrogram technique \citep{goodman2009role}, can provide a complementary insight to the values of sonic and Alfv\'en Mach numbers \citep{burkhart2013hierarchical}. Synergy of these different available techniques is very important to obtain a comprehensive picture of MHD turbulence.

\subsection{Spectroscopic and synchrotron studies of MHD turbulence}
In \citetalias{0004-637X-818-2-178}, a number of techniques were suggested aimed at obtaining the information
about magnetic field and the density of cosmic electrons using synchrotron intensity and polarization measurements. These techniques use position-position-frequency (PPF) cube (an analogue of PPV cube used in the VCA). Such studies provide a complimentary method to study the mean magnetic field direction, the degree of magnetization of the media and the contribution of Alfv\'en, slow and fast modes in a turbulent cloud. 

Understanding the properties of turbulent cascade in different interstellar phases is essential to understand the dynamics of the ISM as well as the transport process of heat and cosmic rays. While the turbulence in HI and molecular gas is well studied via spectroscopic measurements, synchrotron emission is useful in studying diffuse hot and warm turbulent media. The correspondence of the different important parameters characterizing MHD turbulence in different interstellar phases would testify a single turbulent cascade on the galactic scale, which can be a finding with important consequences for different branches of astrophysical research, e.g. for cosmic ray physics (see \citealt{2002cra..book.....S}). Since dust absorption is important at shorter wavelengths, it is less likely to affect synchrotron studies. Thus, synergy of synchrotron studies with spectroscopic studies can be beneficial to understand magnetic properties of a turbulent medium, even when it is significantly dusty.

\section{Summary}\label{ssumm}
On the basis of the analytical theory of the intensity fluctuations in the PPV space developed in the earlier works (\citetalias{lazarian2000velocity}; \citetalias{lazarian2004velocity};  \citetalias{lazarian2006studying}; \citetalias{kandel2016extending}) we provided the analytical description of how the results of those works are affected by dust absorption. The results of our study within the assumptions that were used can be summarized as follows: 
\begin{itemize}
\item Simple model of dust absorption was used to show how dust absorption affects study of VCA. In particular, it was shown that it is possible to obtain velocity spectra in the thin slice regime of VCA even if dust absorption is present. However, the spectral index of power spectrum of intensity changes by one at low wavenumbers to high wavenumbers. This change corresponds to studying the statistics at the scales comparable to the thickness of the dust layer with the optical depth of unity.

\item It was shown that the velocity spectra can be obtained using centroids in the presence of dust absorption. Similar to the VCA, the centroids, due to dust absorption, exhibit the break in the spectral slope of centroid structure function. The reason for that is the same as in the case of the VCA, namely the transfer to studying 2D surface layer of turbulence. 

\item The study of turbulence using VCS  in the presence of dust absorption shows that the asymptotes at different wavenumbers $k_v$ for both poor and high resolution beam remains the same as in the case of negligible dust absorption. One can state therefore, that the technique is robust in relation to the dust absorption.

\item Anisotropies measured by both VCA and centroids at small 2D lags $R$ are unaffected by dust absorption, whereas at larger lags one needs to use proper statistical measure (either correlation function or power spectrum) in order to study anisotropies. 

\item In the presence of both dust absorption and self-absorption of the gas, the statistics of the observed total intensities and velocity centroids change. In particular,
the universal regime of fluctuations $\sim K^{-3}$ that is independent from both underlying velocity and density fluctuations disappears when dust cut-off is much smaller than the LOS extent of the cloud. This allows study of velocity spectra even when self-absorption is present.

\item All the analytical formulations of VCA, VCS and centroids are valid even with collisionally excited emission lines. However, in the regime of strong dust absorption, density effects are not suppressed for collisionally excited emission lines. 

\item Our theoretical expectations correspond to the observations of turbulence in HII regions and explain the change of the spectral slope that is reported for studies of fluctuations at large scales.

\end{itemize}

\section{Acknowledgements}
D.~K. and D.~P. thank the Institut Lagrange de Paris, a LABEX funded by the ANR (under reference ANR-10-LABX-63) within the Investissements d'Avenir programme under reference ANR-11-IDEX-0004-02, where this work was started. A.~L. acknowledges the NSF grant AST 1212096 and Center for Magnetic Self Organization (CMSO). He also acknowledges Institut D'Astrophysique de Paris and Institut Lagrange de Paris for hospitality during his visit.

\footnotesize{
\nocite{*}
\bibliographystyle{mnras}
\bibliography{dust}
}

\appendix
\section{Full expression of intensity correlation}\label{fullexp}
We now focus on a simple case for which self-absorption is negligible. In this situation, Eq. \eqref{basic} gives
\begin{equation}
I_\nu=\epsilon\int_0^{S}\mathop{\mathrm{d}z}\rho(\bm{x})\Phi_\nu(\bm{x})\exp\left[-\int_0^z\mathop{\mathrm{d}z'}\kappa\rho(\bm{X}, z')\right]~,
\end{equation}
which can be equivalently written as
\begin{equation}
I_\nu=-\frac{\epsilon}{\kappa}\int_0^{S}\mathop{\mathrm{d}z}\Phi_\nu(\bm{x})\frac{\mathrm{d}}{\mathrm{d}z}\mathrm{e}^{-\int_0^z\mathop{\mathrm{d}z'}\kappa\rho(\bm{X}, z')}~.
\end{equation}
Assuming uncorrelated density and velocity field, the mean spectral intensity is given by
\begin{equation}\label{meanint}
\langle I_\nu\rangle=-\frac{\epsilon}{\kappa}\int_0^{S}\mathop{\mathrm{d}z}\langle\Phi_\nu(\bm{x})\rangle\frac{\mathrm{d}}{\mathrm{d}z}\left\langle\mathrm{e}^{-\int_0^z\mathop{\mathrm{d}z'}\kappa\rho(\bm{X}, z')}\right\rangle~.
\end{equation}
Assuming $\int_0^z\mathop{\mathrm{d}z'}\delta\rho(\bm{X}, z')$ to be Gaussian random quantity (which holds good if the density fluctuations are small), Eq. \eqref{meanint} can be written as
\begin{align}
\langle I_\nu\rangle=\epsilon\int_0^{S}\mathop{\mathrm{d}z}\langle\Phi_\nu(\bm{x})\rangle\left(\bar{\rho}-\kappa\int_0^z\mathrm{d}z'\xi_\rho(0,z-z')\right)\nonumber\\
\left\langle\mathrm{e}^{-\int_0^z\mathop{\mathrm{d}z'}\kappa\rho(\bm{X}, z')}\right\rangle
\end{align}
The correlation of spectral intensity is given by
\begin{align}\label{appdcor}
\langle I_{\nu_1}(\bm{X}_1)I_{\nu_2}(\bm{X}_2)\rangle=\frac{\epsilon^2}{\kappa^2}\int_0^{S}\mathop{\mathrm{d}z_1}\int_0^{S}\mathop{\mathrm{d}z_2}\langle\Phi_{\nu_1}(\bm{x}_1)\Phi_{\nu_2}(\bm{x}_2)\rangle\nonumber\\
\frac{\partial^2}{\partial z_1\partial z_2}\left\langle\mathrm{e}^{-\kappa\left[\int_0^{z_1}\mathop{\mathrm{d}z'}\rho(\bm{X}, z')+\int_0^{z_2}\mathop{\mathrm{d}z'}\rho(\bm{X}, z')\right]}\right\rangle~.
\end{align}
We assume fluctuations of column density to be Gaussian, so that
\begin{equation}\label{appdder}
\left\langle\mathrm{e}^{-\kappa\left[\int_0^{z_1}\mathop{\mathrm{d}z'}\rho(\bm{X}, z')+\int_0^{z_2}\mathop{\mathrm{d}z'}\rho(\bm{X}, z')\right]}\right\rangle=\mathrm{e}^{-\kappa\rho_0(z_1+z_2)}\mathrm{e}^{\kappa^2D_{\mathcal{A}}}~,
\end{equation}
where 
\begin{align}
D_{\mathcal{A}}(\bm{R}, z_1, z_2)
&=\frac{1}{2}\int_0^{z_1}\mathrm{d}z'\int_0^{z_1}\mathop{\mathrm{d}z''}\xi_\rho(0,z'-z'')\nonumber\\
&+\frac{1}{2}\int_0^{z_2}\mathrm{d}z'\int_0^{z_2}\mathop{\mathrm{d}z''}\xi_\rho(0,z'-z'')\nonumber\\
&+\int_0^{z_1}\mathrm{d}z'\int_0^{z_2}\mathop{\mathrm{d}z''}\xi_\rho(\bm{R},z'-z'').
\end{align}
With this, one can immediately write 
\begin{equation}
\frac{\partial^2}{\partial z_1\partial z_2}D_{\mathcal{A}}(\bm{R},z_1,z_2)=\xi_\rho(\bm{R},z_1-z_2)~,
\end{equation}
and
\begin{align}\label{eq:dustmeasureder}
&\frac{\partial^2}{\partial z_1\partial z_2}\mathrm{e}^{-\kappa\rho_0(z_1+z_2)}\mathrm{e}^{\kappa^2D_{\mathcal{A}}(\bm{R},z_1,z_2)}\nonumber\\
&=\kappa^2\mathrm{e}^{-\kappa\rho_0(z_1+z_2)}\mathrm{e}^{\kappa^2D_{\mathcal{A}}(\bm{R},z_1,z_2)}\bigg(\langle\rho(\bm{x}_1)\rho(\bm{x_2})\rangle-\kappa\left(\frac{\partial}{\partial z_1}+\frac{\partial}{\partial z_2}\right)\nonumber\\
& D_{\mathcal{A}}(\bm{R}, z_1, z_2)
+\kappa^2\frac{\partial D_{\mathcal{A}}(\bm{R},z_1,z_2)}{\partial z_1}\frac{\partial D_{\mathcal{A}}(\bm{R},z_1,z_2)}{\partial z_2}\bigg)~.
\end{align}
Using Eqs. \eqref{appdcor}, \eqref{appdder} and \eqref{eq:dustmeasureder}, one can immediately write
\begin{align}\label{mainformcor}
\langle I_{\nu_1}(\bm{X}_1)I_{\nu_2}(\bm{X}_2)\rangle=\epsilon^2\int_0^{S}\mathop{\mathrm{d}z_1}\int_0^{S}\mathop{\mathrm{d}z_2}\langle\Phi_{\nu_1}(\bm{x}_1)\Phi_{\nu_2}(\bm{x}_2)\rangle\nonumber\\
\mathrm{e}^{-\kappa\rho_0(z_1+z_2)}\mathrm{e}^{\kappa^2D_{\mathcal{A}}(\bm{R}, z_1, z_2)}
\bigg(\langle\rho(\bm{x}_1)\rho(\bm{x}_2)\rangle-\kappa\left(\frac{\partial}{\partial z_1}+\frac{\partial}{\partial z_2}\right)\nonumber\\
\times D_{\mathcal{A}}(\bm{R},z_1,z_2) +\kappa^2\frac{\partial D_{\mathcal{A}}(\bm{R},z_1,z_2)}{\partial z_1}\frac{\partial D_{\mathcal{A}}(\bm{R},z_1,z_2)}{\partial z_2}\bigg),
\end{align}
where $\langle\rho(\bm{x}_1)\rho(\bm{x_2})\rangle\equiv \rho_0^2+\xi_\rho(\bm{x}_1,\bm{x}_2)$.  

\section{Statistics of column-density}\label{appcolumn}
The column density 
\begin{equation}
\mathcal{N}(\bm{X})=\int_0^S\mathrm{d}z\rho(\bm{X},z)~,
\end{equation}
is an important parameter, especially at thick slice limit of the VCA, where it is proportional to the intensity of emission. We first write several important statistical descriptors of $\mathcal{N}$. The mean column-density is given by
\begin{equation}
\bar{\mathcal{N}}=\bar{\rho}S~,
\end{equation}
while the variance of the column-density {\it fluctuations}
$\mathcal{N-\bar N}$,
$\sigma_{\mathcal{N}}^2=\int_0^S\mathrm{d}z_1\int_0^S\mathrm{d}z_2\,\xi_{\rho}(0,z_1-z_2) $ is evaluated to
\begin{align}
\sigma_{\mathcal{N}}^2 &= \sigma_\rho^2 \; S \; r_c  ~,
\quad\quad\quad\quad \nu_\rho > 1 \\
\sigma_{\mathcal{N}}^2 &= \sigma_\rho^2 \; S^2 (r_c/S)^{\nu_\rho}
~, \quad \nu_\rho < 1 
\end{align}

The correlation function of column-density {\it fluctuations} $\mathcal{N-\bar N}$
is given by
\begin{equation}
\xi_{\mathcal{N}}(R)=\int_0^S\mathrm{d}z_1\int_0^S\mathrm{d}z_2\,\xi_{\rho}(\bm{R},z_1-z_2)~,
\end{equation}
which for {\it shallow} density at $R > r_c$ evaluates to
\begin{equation}\label{eq:xiNlong}
\xi_{\mathcal{N}}(\mathbf{R})\approx
\sigma_\rho^2Sr_c \left(\frac{r_c}{R}\right)^{\nu_\rho-1}
=\sigma_{\cal N}^2 \left(\frac{r_c}{R}\right)^{\nu_\rho-1}
~,\quad \nu_\rho>1~.
\end{equation}
The structure function of the column density, which can be computed as
\begin{equation}
d_{\mathcal{N}}(\mathbf{R})=2\!\int_0^S\!\!\!\mathrm{d}z_1
\!\int_0^S\!\!\!\mathrm{d}z_2
\, \left(\xi_{\rho}(0,z_1-z_2)-\xi_{\rho}(\bm{R},z_1-z_2)\right),
\end{equation}
for density with $\nu_\rho < 1$ behaves at $ R < r_c $ as
\begin{equation}
\label{eq:dNshort}
d_{\mathcal{N}}(R)\approx
\sigma_\rho^2 S r_c \left(\frac{R}{r_c}\right)^{1-\nu_\rho}
\!\!=\sigma_\mathcal{N}^2 \left(\frac{R}{S}\right)^{1-\nu_\rho} \!\!,
\quad -1 < \nu_\rho<1 \; .
\end{equation}
Note that since the column density is a projected 2D field, 
its behaviour reflects the shifted value $\nu_\rho-1$ of the 3D 
density index. Correspondingly, the
correlation function $\xi_\mathcal{N}(\mathbf{R})$
is applicable for large separation analysis only for $\nu_\rho > 1$ 
rather than $\nu_\rho > 0$ as for 3D density field. Similarly, one
should use the structure function $d_\mathcal{N}(\mathbf{R})$ for
$\nu_\rho < 1$ to study scaling regime through small $R$, while too steep
underlying 3D density with $\nu_\rho < -1$ will lead to saturated 
structure function statistics and needs to be studied separately.
For instance, Kolmogorov index for 3D density, $\nu_\rho=-2/3$ will result
in steep $5/3$ scaling of the column density structure function.

We can estimate the correlaton length
as the scale where $d_\mathcal{N}(R)$ or $\xi_\mathcal{N}(R)$ reaches
the value of the variance, extrapolating 
Eqs.~\eqref{eq:xiNlong} and \eqref{eq:dNshort}.
The second form of Eqs.~\eqref{eq:xiNlong} and \eqref{eq:dNshort} shows
that the correletion length of the column density is given by the 
underlying density scale $r_c$ for $\nu_\rho > 1$ but is increased to $S$ due
to accumulation of slow falling long wave correlations for $\nu_\rho < 1$.
Seemingly abrupt jump in column density correlation scale at $\nu_\rho=1$
is in reality a smooth transition which is not recovered by
our asymptotic analysis. While correlation lengths similar
to the size of the cloud $S$ are expected for $\nu_\rho < 0$, 
note that in the $0 < \nu_\rho < 1$ range, the column density
may have considerably larger correlation scale than underlying
3D density which is still shallow and is expected to have $r_c < S$.

\begin{figure*}
\centering
\includegraphics[scale=1]{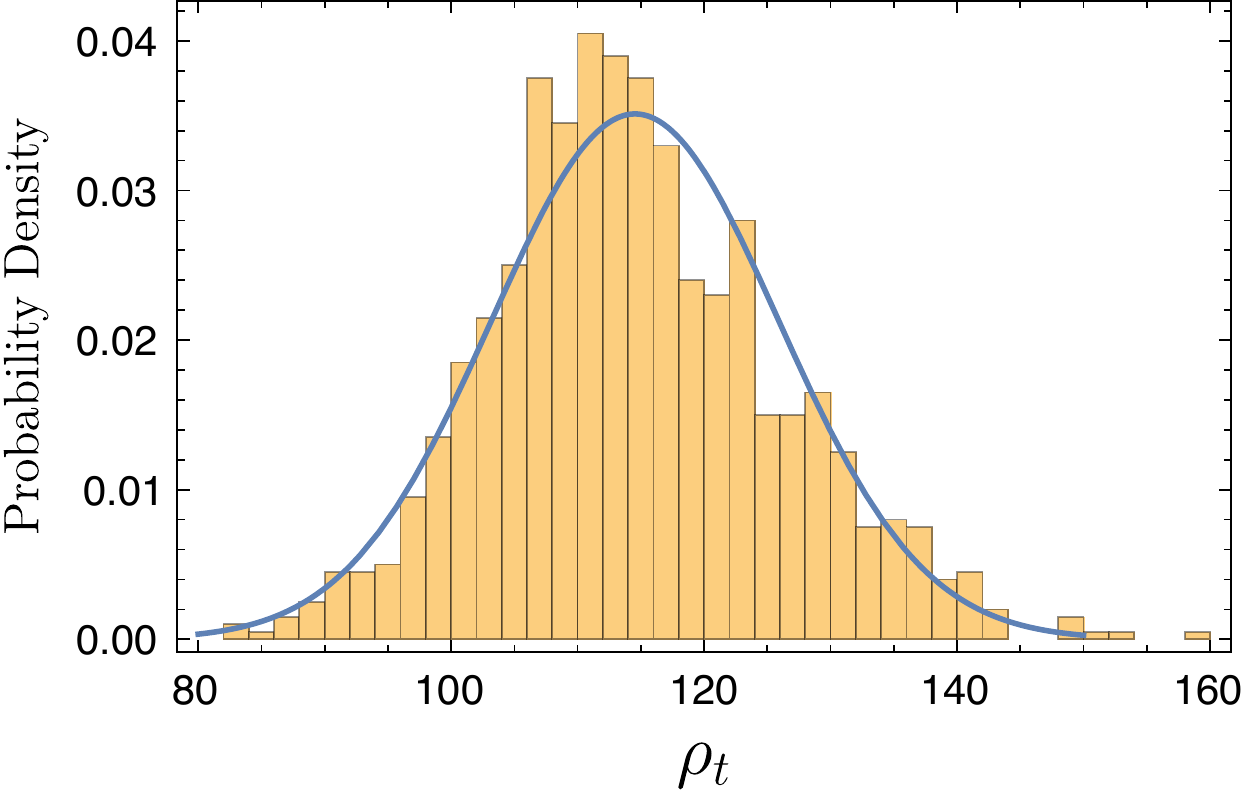}
\caption{Probability distribution of dust measure produced numerically, and overlayed Gaussian distribution. In the numerical calculation $\sigma_\rho/\bar{\rho}=2, S/r_c=1000$ and $\nu_\rho=1/3$.}
\label{pdf}
\end{figure*}

Let us turn to the kind of statistical model appropriate for column density.
For a steep density spectra, fluctuations are smaller than the mean density,
and Gaussian approximation for density fluctuations works well which translates
to Gaussian behaviour of the column densty.
However, for shallow density, fluctuations can be larger than the mean density, and thus density fluctuations follow lognormal distribution better than
the Gaussian distribution. 
If so, what kind of distribution does column density follow? We answer this question numerically. For that we  generate random correlated densities in one dimension. The most convenient way to do this is to start in spectral domain, and generate Gaussian random amplitudes $a(k)$ in spectral domain with dispersion given by their power spectra, which is related to the correlation function $\xi(\bm{r})$ as
\begin{equation}
P(k)=\int\mathrm{d}^n\bm{k}\,e^{i\bm{k}\cdot\bm{r}}\xi(\bm{r})~,
\end{equation}
where $n$ is the dimensionality of space, and in our test case $n=1$. We define $P(k)$ to satisfy power-law
\begin{equation}
P(k)=\sigma^2r_c(kr_c)^{-m}~,
\end{equation}
where $r_c$ is the correlation length of the density field, and is in general very small for a shallow density. The advantage of working in spectral domain is that these random amplitudes are uncorrelated. The Fourier transform of these random amplitudes corresponds to $\log\rho$ in real space:
\begin{equation}
\log\rho(r)=\int_0^L\mathrm{d}k\,\mathrm{e}^{ikr}a(k).
\end{equation}
With the obtained $\rho(r)$, we find out $\mathcal{N}$, and this is done for several realisation, so as to obtain distribution of $\mathcal{N}$. As shown in Fig. \ref{pdf}, our results show that $\mathcal{N}$ is to a very good extent Gaussian. One can understand this Gaussian behaviour by considering the fact that if the integration length $S$ is much larger than the correlation length $r_c$, then at LOS lengths larger than $r_c$, the fluctuations are essentially independent. Thus, from Central limit theorem, the sum of fluctuations should tend to be Gaussian. 

\section{Statistics of dust cutoff Delta}\label{deltacutstat}
The main effect of dust absorption is the introduction of cut-off scale $\Delta(\bm{X})$ beyond which signals are effectively cut out. This cut-off scale is given by
\begin{equation}\label{eqdustcut}
\kappa\int_0^{\Delta(\bm{X})}\mathrm{d}z\,\rho(\bm{r})=1~.
\end{equation}
$\Delta(\bm{X})$ is fluctuating random quantity, since $\rho(\bm{r})$ is fluctuating. 
In this appendix we derive several important statistical properties for $\Delta$ assuming lognormal distribution for $\rho$. We model dust density $\rho$ as
\begin{equation}\label{d2}
\rho(\bm x)  = \bar\rho e^{-\sigma_\mu^2/2} e^\mu~,
\end{equation}
where $\mu(r)$ is a random Gaussian field with zero mean and  variance $\sigma_\mu^2$, and is described by the two-point probability function
\begin{equation}\label{d3}
P(\mu_1,\mu_2) = \frac{1}{2 \pi \sqrt{\sigma_\mu^4 - \xi_\mu^2}}
\exp\left[ -\frac{1}{2} \frac{\sigma_\mu^2 \mu_1^2 + \sigma_\mu^2 \mu_2^2 -
2 \xi_\mu \mu_1 \mu_2}{\sigma_\mu^4 - \xi_\mu^2}\right]
\end{equation}
where $\mu_1 = \mu(\bm x_1)$, $\mu_2 = \mu(\bm x_2)$ and $\xi_\mu(|\bm x_1 - x_2|)$ is the correlation function of $\mu$ field.

Using the following relations 
\begin{eqnarray}
\langle e^\mu \rangle &=& \langle e^{-\mu} \rangle = e^{\sigma_\mu^2/2} ~,\\
\langle e^{2 \mu} \rangle &=& \langle e^{-2 \mu} \rangle = 
e^{2 \sigma_\mu^2} ~ , \\
\langle e^{ \pm n \mu} \rangle &=& e^{n^2 \sigma_\mu^2/2} ~.
\end{eqnarray}
as well as
\begin{equation}
\langle e^{\mu_1 + \mu_2} \rangle = e^{\sigma_\mu^2+\xi_\mu(\bm r)} ~.
\end{equation}
we obtain for the moments of the dust density distribution
\begin{equation}
\langle \rho \rangle = \bar\rho, \quad 
\langle \rho^2 \rangle = {\bar\rho}^2 e^{\sigma_\mu^2}~, ~\ldots ~ ,
\langle \rho^n \rangle = {\bar\rho}^n e^{n(n-1)\sigma_\mu^2/2}~,
\end{equation}
and for the density correlation function
\begin{equation}\label{eq:linden}
\langle \rho(\bm x_1) \rho(\bm x_2) \rangle = {\bar\rho}^2 e^{\xi_\mu(\bm r)} ~.
\end{equation}
Non-Gaussianity of the density field is reflected in (higher order moments).

It can also be shown that the correlation of square of density is 
\begin{equation}\label{eq:squareden}
\langle \rho^2(\bm x_1) \rho^2(\bm x_2) \rangle={\bar\rho}^4 \mathrm{e}^{4(\sigma_{\mu}^2+\xi_{\mu})}~.
\end{equation}

For dust absorption we shall consider two regimes, depending on comparison
of the absorption length $\Delta({\bm X})$ and the density 
perturbations correlation length $r_c$.  

First regime is $\Delta({\bm X}) < r_c$.  A weak, but explicit 
necessary condition for this to hold is is $\kappa \bar\rho r_c \gg 1 $. 
This regime is primarily relevant for the case of steep spectrum of density, when $r_c$ is associated with the largest scale of the inertial range. Since density can be thought as almost constant within $r_c$ length scale, we have
\begin{equation}\label{appdelta}
\Delta({\bm X}) \approx \frac{1}{\kappa \rho(\bm X,0)}, \quad  
\Delta({\bm X}) \ll r_c ~.
\end{equation}
With the help of Eqs. \eqref{d2}, \eqref{d3} and \eqref{appdelta}, statistical properties of the absorption length are immediately available. The mean and the 
second moment of $\Delta$ are
\begin{eqnarray}\label{meandelta}
\langle \Delta \rangle =\frac{e^{\sigma_\mu^2}}{\kappa \bar \rho} &=& 
\frac{ \langle \rho^2 \rangle }{\kappa {\bar\rho}^3}
= \frac{1}{\kappa \bar\rho}
\left(1+ \frac{\langle \delta \rho^2 \rangle}{\bar{\rho}^2} \right)\\
\langle \Delta^2 \rangle =\frac{e^{3 \sigma_\mu^2}}{\kappa^2 \bar{\rho}^2} &=&
\frac{ \langle \rho^2 \rangle^3 }{\kappa^2 {\bar\rho}^8}
= \frac{1}{\kappa^2 \bar{\rho}^2} 
\left(1+ \frac{\langle \delta \rho^2 \rangle}{\bar{\rho}^2} \right)^3~,
\label{eq:Deltasquare1}
\end{eqnarray}
therefore relative fluctuations $\delta\Delta/\langle\Delta\rangle$ have
the variance equal to that of density fluctuations
\begin{equation}
\frac{\langle (\delta \Delta)^2 \rangle}{\langle \Delta \rangle^2} 
\equiv \frac{\langle \Delta^2 \rangle -\langle \Delta \rangle^2}
{\langle \Delta \rangle^2}
= \frac{\langle \delta \rho^2 \rangle}{\bar{\rho}^2}
\end{equation}
The correlation function is 
\begin{equation}
\langle \Delta_1 \Delta_2 \rangle = \frac{1}{\kappa^2} 
\left\langle \frac{1}{\rho_1 \rho_2} \right\rangle = 
\frac{1}{\kappa^2 \bar{\rho}^2} 
\frac{\langle\rho^2\rangle^2}{\bar{\rho}^4} 
\frac{\langle \rho_1 \rho_2 \rangle}{\bar{\rho}^2} 
\end{equation}
and
\begin{equation}
\frac{\langle \delta \Delta_1 \delta \Delta_2 \rangle}{\langle\Delta\rangle^2}
= \frac{\langle \delta \rho_1 \delta \rho_2 \rangle}{\bar{\rho}^2}~.
\end{equation}
An interesting observation that can be made from Eq. \eqref{meandelta} is that the mean depth $\bar{\Delta}$ is longer than $1/(\kappa\bar{\rho})$.

Second regime is $\Delta(\bm{X})>r_c$, which is particularly relevant for shallow spectra, which has $r_c$ much smaller than the size of the cloud. For shallow spectra density is uncorrelated at scales larger than the correlation length. Formally, given a distribution of $\rho$, we are interested in the distribution of $\Delta$ such that the following holds
\begin{equation}
\tau\equiv \kappa\int_0^{\Delta}\mathrm{d}z\,\rho=1~.
\end{equation}
In this situation, the accumulation of optical depth along the LOS can be described by Brownian random walks with a net drift. 

A formal way to obtain the distribution of physical depths $\Delta$ is via diffusion approximation. Let $\Pi(\tau_n, \Delta_n)$ be the distribution of paths for which the threshold optical depth of 1 has not been met. In this picture $\delta \tau\equiv \tau_n-\tau_{n-1}=\kappa\rho r_c$ and $\delta\Delta=\Delta_n-\Delta_{n-1}=r_c$, where $r_c$ is the minimum length at which correlation 
of random jumps  at each step can be neglected. The case of constant density implies $r_c\rightarrow 0$. The distribution function $\Pi$ at any optical depth depends upon the distribution at the immediately prior optical depth, and hence satisfies Chapman-Kolmogorov equation
\begin{equation}\label{chapman}
\Pi(\tau_n, \Delta_n)=\int_{-\infty}^{\infty}d\tau_{n-1}P(\tau_n,\Delta_n|\tau_{n-1},\Delta_{n-1})\Pi(\tau_{n-1}, \Delta_{n-1})~,
\end{equation}
where $P(\tau_n,\Delta_n|\tau_{n-1},\Delta_{n-1})$ is the transition probability to have a field value of $\tau_n$ at $\Delta_n$, given the field value of $\tau_{n-1}$ at $\Delta_{n-1}$. If higher moments of $\Pi$ are ignored, Eq. \eqref{chapman} leads to Fokker-Plank equation 
\begin{equation}\label{fpeq}
\frac{\partial\Pi}{\partial \Delta}=\frac{\kappa^2\langle\rho^2\rangle r_c}{2}\frac{\partial^2 \Pi}{\partial \tau^2}-\kappa\bar{\rho}\frac{\partial\Pi}{\partial \tau}~.
\end{equation}
The initial condition that $\Pi$ satisfies is $\Pi(\tau, 0)=\delta_D(\tau)$, where $\delta_D(\tau)$ is the Dirac delta function, and the boundary  condition satisfied by $\Pi$ is $\Pi(1,\Delta)=0$.

Instead of solving for complete form of probability distribution $\Pi$, it is convenient to introduce {\it first passage length} (an analogue of first passage time in random walks), which is the physical length required to reach optical depth of unity for the first time. We denote this length to be $\Delta(\tau)$, and the probability density associated with it to be $f(\tau, \Delta)$, which gives the probability that optical depth $\tau$ is reached at physical depth $\Delta$. In order to calculate $f(\tau, \Delta)$, we need to solve Eq. \eqref{fpeq} with proper boundary conditions. 

We denote the first passage distance from $\tau_0$ to $\tau=1$ by $\Delta(\tau_0)$. In this scenario, it is useful to introduce the notion of {\it survival probability} $S(\tau_0, \Delta)$, which is the probability that at physical depth $\Delta$, optical depth of unity has not been reached yet, assuming that the motion started at optical depth of $\tau_0$. The survival probability satisfies the following relations
\begin{align}
S(\tau_0,\Delta)=\int\mathrm{d}\tau\Pi(\tau,\Delta|\tau_0, 0)~,\\
f(\tau_0, \Delta)=-\frac{\partial S(\tau_0, \Delta)}{\partial \Delta}~.
\end{align}

We are now interested in moments of physical depths, i.e. $\langle \Delta^n\rangle$, which is given by
\begin{align}\label{spdf}
\Delta_n\equiv\langle \Delta^n\rangle &=\int_0^\infty\mathrm{d}\Delta \Delta^nf(\tau_0,\Delta)\nonumber\\
&=n\int_0^\infty\mathrm{d}\Delta \Delta^{n-1}f(\tau_0,\Delta)~,
\end{align}
which was obtained by integration by parts and assuming $S(\infty)-S(0)=0$ for a well-defined probability distribution function $S(\Delta)$. Eq. \eqref{spdf} can also be written as
\begin{equation}\label{nthmom}
\langle \Delta^n\rangle=n\int \mathrm{d}\tau g_{n-1}(\tau;\tau_0)~,
\end{equation}  
where 
\begin{equation}
g_{n}(\tau;\tau_0)=\int_0^\infty\mathrm{d}\Delta \Delta^n\Pi(\tau, \Delta;\tau_0,0)~.
\end{equation}
Integrating by parts Eq. \eqref{nthmom} $n$ times and using Fokker Plank equation (Eq. \eqref{fpeq}) finally gives differential equation governing moments of $\Delta$
\begin{equation}\label{mainmoment}
\frac{\kappa^2\langle\rho^2\rangle r_c}{2}\frac{\partial^2 \Delta_n(\tau)}{\partial \tau^2}-\kappa\bar{\rho}\frac{\partial\Delta_n(\tau)}{\partial \tau}=-n\Delta_{n-1}~,
\end{equation}
satisfying the initial and boundary conditions
The moments satisfies the following boundary condition
\begin{equation}\label{b1}
\left(\frac{\partial \Delta_n(\tau)}{\partial \tau}\right)_{\tau=0}=0~,
\end{equation}
\begin{equation}\label{b2}
\Delta_n(\tau=1)=0~.
\end{equation}
Eq. \eqref{mainmoment} can be solved for $\Delta_1(0)$ and $\Delta_2(0)$, to obtain $\langle \Delta\rangle$ and $\langle\Delta^2\rangle$. Here, we quote the final result
\begin{equation}
\langle\Delta\rangle=\frac{1}{\kappa\bar{\rho}}+\frac{\langle\rho^2\rangle r_c}{2\bar{\rho}^2}\left[\exp\left(-\frac{2\bar{\rho}}{\kappa\langle\rho^2\rangle r_c}\right)-1\right]~,
\end{equation}
and 
\begin{align}
\langle\Delta^2\rangle&=\frac{1}{\kappa^2\bar{\rho}^2}-3\frac{\langle\rho^2\rangle r_c}{\kappa\bar{\rho}^3}\exp\left(-\frac{2\bar{\rho}}{\kappa\langle\rho^2\rangle r_c}\right)\nonumber\\
&+\frac{\langle\rho^2\rangle^2r_c^2}{2\bar{\rho}^4}\left[\exp\left(-\frac{2\bar{\rho}}{\kappa\langle\rho^2\rangle r_c}\right)-1\right] \left[\exp\left(-\frac{2\bar{\rho}}{\kappa\langle\rho^2\rangle r_c}\right)+2\right]~.
\label{eq:Deltasquare2}
\end{align}
In the above expression, $r_c$ is the minimal step size at which correlation
between subsequent steps can be neglected. Clearly in the case when $r_c\rightarrow 0$, the variance of $\Delta$ is square of its mean.

\bsp	
\label{lastpage}
\end{document}